\PassOptionsToPackage{usenames,dvipsnames}{xcolor}
\documentclass[a4paper,11pt,1p]{elsarticle}
\usepackage[titletoc,toc,title]{appendix}

\usepackage[T1]{fontenc}
\usepackage[utf8]{inputenc}
\usepackage[british]{babel}
\usepackage{geometry}
\usepackage{graphicx} 
\usepackage{amssymb} 
\usepackage{amsmath,pdftexcmds}
\usepackage{fancyvrb}
\usepackage{multirow}
\usepackage{mathtools}
\usepackage{multicol}
\usepackage{xcolor}
\usepackage{cuted}
\usepackage{etoolbox}
\usepackage{bbm}
\usepackage{comment}
\usepackage{tabularx}
\usepackage{hyperref}
\usepackage{adjustbox}  % for resizebox, scalebox
\usepackage{url}
\usepackage{booktabs}
\usepackage{mathrsfs}
\usepackage{textcomp}
\usepackage[version=3]{mhchem}
\usepackage{framed}
\usepackage[all]{xy}
\usepackage{amsfonts}
\usepackage{anysize}
\marginsize{2.4cm}{2.4cm}{1.2cm}{2.6cm}
\DeclareFontFamily{U}{cbgreek}{}
\DeclareFontShape{U}{cbgreek}{m}{n}{
        <-6>    grmn0500
        <6-7>   grmn0600
        <7-8>   grmn0700
        <8-9>   grmn0800
        <9-10>  grmn0900
        <10-12> grmn1000
        <12-17> grmn1200
        <17->   grmn1728
      }{}
\DeclareFontShape{U}{cbgreek}{bx}{n}{
        <-6>    grxn0500
        <6-7>   grxn0600
        <7-8>   grxn0700
        <8-9>   grxn0800
        <9-10>  grxn0900
        <10-12> grxn1000
        <12-17> grxn1200
        <17->   grxn1728
      }{}

\DeclareRobustCommand{\Qoppa}{%
  \text{\usefont{U}{cbgreek}{\normalorbold}{n}\symbol{21}}%
}
\DeclareRobustCommand{\sampi}{%
  \text{\usefont{U}{cbgreek}{\normalorbold}{n}\symbol{27}}%
}
\DeclareRobustCommand{\Sampi}{%
  \text{\usefont{U}{cbgreek}{\normalorbold}{n}\symbol{23}}%
}

\DeclareRobustCommand{\digamma}{%
  \text{\usefont{U}{cbgreek}{\normalorbold}{n}\symbol{147}}%
}

\makeatletter
\newcommand{\normalorbold}{%
  \ifnum\pdf@strcmp{\math@version}{bold}=\z@ bx\else m\fi
}
\makeatother

\newcommand{\overbar}[1]{\mkern 1.5mu\overline{\mkern-1.5mu#1\mkern-1.5mu}\mkern 1.5mu}

\makeatletter
\def\ps@pprintTitle{
   \let\@oddhead\@empty
   \let\@evenhead\@empty
   \def\@oddfoot{\reset@font\hfil\thepage\hfil}
   \let\@evenfoot\@oddfoot
}
\makeatother

\newsavebox{\leftbox}
\newsavebox{\rightbox}

\definecolor{DarkMidnightBlue}{rgb}{0.0, 0.09, 0.42}

\hypersetup{hidelinks,
backref=true,
pagebackref=true,
hyperindex=true,
breaklinks=true,
colorlinks=true,
linkcolor=blue, % default= red
citecolor=blue, % default= green
urlcolor=DarkMidnightBlue,
bookmarks=true,
bookmarksopen=false,
pdftitle={Title},
pdfauthor={Author}}

\VerbatimFootnotes

\title{\textbf{\textsf{P-Wave Two-Particle Bound and  Scattering States in a Finite Volume including QED}}}

\author[hiskp,bctp]{Gianluca~Stellin} 
\author[hiskp,bctp,fzj,tsu]{Ulf-G.~Mei\ss{}ner}
\address[hiskp]{Helmholtz Institut für Strahlen- und Kernphysik, Universität Bonn, Nu\ss{}allee 14-16, 53115 Bonn, Germany}
\address[bctp]{Bethe Center for Theoretical Physics, Universität Bonn, Nu\ss{}allee 12, 53115 Bonn, Germany}
\address[fzj]{Insitute for Advanced Simulation, Institut f\"ur Kernphysik and J\"ulich Center for Hadron Physics, \\
Forschungszentrum J\"ulich, 52425 J\"ulich, Germany}
\address[tsu]{Ivane Javakhishvili Tbilisi State University,  0186 Tbilisi, Georgia}

\date{\today}

\begin{document}

%--------------------------------------------------------------------------------------------------------------------------------------------------------------------

\begin{abstract}
\begin{small}
The mass shifts for two-fermion bound and scattering P-wave states subject to the long-range interactions
due to QED
in the non-relativistic regime are derived. Introducing a short range force coupling the spinless fermions
to one
unit of angular momentum in the framework of pionless EFT, we first calculate both perturbatively and
non-perturbatively
the Coulomb corrections to fermion-fermion scattering in the continuum and infinite volume context.
Motivated by the research on particle-antiparticle bound states, we extend the results to fermions
of identical mass
and opposite charge. Second, we transpose the system onto a cubic lattice with periodic boundary
conditions and
we calculate the finite volume corrections to the energy of the lowest bound and unbound $T_1^{\pm}$
eigenstates.
In particular, power law corrections proportional to the fine structure constant and resembling the recent
results for S-wave states are found. Higher order contributions in $\alpha$ are neglected, since the gapped
nature of the momentum operator in the lattice environnement allows for a perturbative treatment of the QED
interactions. 
\end{small}
\end{abstract}


\maketitle

\clearpage

%--------------------------------------------------------------------------------------------------------------------------------------------------------------------

\tableofcontents

\clearpage

%--------------------------------------------------------------------------------------------------------------------------------------------------------------------

\section{\textsf{Preamble}}\label{S-1.0}

Effective Field Theories \cite{BiM96,Kol99,Epe06,DLe09,EHM09,MaE11,LaM19,HKK20}
nowadays play a fundamental role in the
description of many-body systems in nuclear and subnuclear physics, employing the quantum fields which
can be excited in a given regime of energy. Once the breakdown scale $\Lambda$ of the EFT is set,
the scattering
amplitudes are usually expressed in power series of $p/\Lambda$, where $p$ represents the characteristic
momentum of the processes under consideration. The Lagrangian density is typically written in terms
of local operators of increasing
dimensions obeying pertinent symmetry constraints. Moreover, power counting rules establish a hierarchy among
the interaction terms to include in the Lagrangian, thus permitting to filter out the contributions
that become relevant only at higher energy scales \cite{LaM19}.

In the case of systems of stable baryons at energies lower than the pion mass, the Lagrangian density
contains only the nucleon fields and their Hermitian conjugates, often combined toghether with differential
operators.
The corresponding theory, the so-called pionless EFT \cite{Kol99,KSW96,KSW98-01,BeK98,BHK99-01}
counts a number of successes in the description of nucleon-nucleon scattering and structure properties
of few-nucleon systems.
Despite the original difficulties in the reproduction of S-wave scattering lengths, that were solved
via the introduction of the Power Divergence Subtraction (PDS) as a regularization scheme
\cite{KSW98-01,Geg98,KSW98-02},
the theory has permitted so far to reproduce the ${}^1S_0$ $np$ phase shift \cite{Kap97,CRS99},
structure properties of the triton as a $dn$ S-wave compound \cite{BHK99-01,BHK99-02,BHK00} and
the scattering length \cite{AfP03,Gri04} and the phase shift \cite{GBG00,BlG01,BRG03} of the
elastic $dn$ scattering process.

In the first applications of QED in pionless EFT, the electromagnetic interactions were treated
perturbatively,
as in the case of the electromagnetic form factor \cite{KSW99} and electromagnetic polarizability \cite{CGS98}
for the deuteron or the inelastic process of radiative neutron capture on protons \cite{SSW99}. Afterwards,
a non-perturbative treatment of electromagnetic (Coulomb) interactions on top of the same EFT was
set up, in the context of proton-proton S-wave elastic \cite{KoR00} and inelastic \cite{KoR99,KoR01}
scattering.

Inspired by the P-wave interactions presented in refs.~\cite{BHK02,GLR08,Fur14}, we generalize in
the first part of the present paper the analysis in ref.~\cite{KoR00} to fermion-fermion low-energy
elastic scattering ruled by the interplay between the Coulomb and the strong forces transforming as
the $\ell=1$ representation of the rotation group (cf. ref.~\cite{BCH82} for the empirical
S- and P-wave phase shifts in the $pp$ case).
As in ref.~\cite{KoR00}, we treat the Coulomb photon exchanges both in a perturbative and in a
non-perturbative fashion. During the derivation of the T-matrix elements, we observe that at
sufficiently low energy the repulsion
effects from the Coulomb ladders become comparable to the ones of the strong forces, leading to
the breakdown of the perturbative regime of non-relativistic QED. In the determination of the
closed expressions for the scattering parameters in terms of the coupling constants, we take
advantage of the separation of the Coulomb
interaction from the strong forces, considered first in refs.~\cite{LaS44,LaS47,Bet49} and eventually
generalized to strong couplings of arbitrary angular momentum $\ell$ in ref.~\cite{KMB82}. The
importance of particle-antiparticle systems led us to the applicaton of the formalism to
fermion-antifermion scattering, where the attractive Coulomb force gives rise to bound states.
This case provides a laboratory for the study of $p\bar{p}$ bound \cite{KBM02} 
and unbound states \cite{CDP91}, also referenced as \textit{protonium}. 

Of fundamental importance for the study of few- and many-particle systems with QED
are Lattice Effective Field Theories and Lattice Quantum Chromodynamics (LQCD). 
The latter has matured to the point where basic properties of light mesons and baryons are
being calculated at or close to the
physical pion mass \cite{DaS14}. In particular, in the case of the lowest-lying mesons, their
properties are attaining
a level of accuracy where it is necessary to embed the strong interactions within the
full standard model \cite{DFF08,AII12,DFH14,AAB17,Tot19,AAB20}. Despite the open computational
challenges represented by the inclusion of the full QED in LQCD simulations, in the last decade
quenched QED \cite{UnH08} together with flavour-symmetry
violating terms have been included in the Lagrangian, with the aim of reproducing some features
of the observed hadron spectrum \cite{PDF12,DFL13,BDF13,DBH13,BDF14}. Conversely, the
perspective to add QED interactions in LQCD simulations for light nuclei appears still futuristic,
due to the limitations in the computational resources.
Nevertheless, in two-body scattering processes like $\pi^{\pm} \pi^{\pm}$ scattering \cite{Met16,AGM17},
the time is  ripe for the introduction of QED interactions in the present LQCD calculations.

It is exactly in this context that, in the second part of the paper, we immerse our fermion-fermion
EFT into a cubic box with periodic boundary conditions (PBC). The lattice environment has a
number of consequences, the most glaring of them are the breaking of rotational symmetry
\cite{Joh82,BNL14,BNL15,SEM18} and the discretization of the spectrum of the operators representing
physical observables \cite{LeT07,KLH12,PWH19}.
Concerning the Hamiltonian, its spectrum consists of levels that in the infinite-volume limit become
part of the continuum (scattering states) and in others that are continuously transformed into
the bound states.
For two- and three-body systems governed by strong interactions, the shifts of the bound
energy levels with respect to the counterparts at infinite volume depend on the lattice size $L$
through negative exponentials,
often multiplied by nontrivial polynomials in $L$. Apart the pioneering work on two-bosons subject
to hard-sphere potentials in ref.~\cite{HuY57}, these effects for two-body systems
have been extensively analyzed by L\"uscher in refs.~\cite{Lue86-01,Lue86-02} (\cite{Lue91}), where
the energy of the lowest unbound (bound) states has been expressed in terms of the scattering
parameters and the lattice size.

In the last three decades, L\"uscher formulas for the energy shifts have been extended in several
directions including non-zero angular momenta \cite{KLH11,KLH12,LuS11,DMO12} moving frames
\cite{KSS05,RuG95,DaS11,BKL11,DMO12,GHL12,REK14},
generalized boundary conditions \cite{SaV05,DMO11,BDL14,KoL16,SNR16,CSW17} and particles
with intrinsic spin \cite{BLM08,BrH15}. Moreover, considerable advances have been made in
the derivation of analogous formulas for the energy corrections of bound states of three-body
\cite{PWH19,PoR12,MRR15} and N-body systems \cite{KoL18}. See also the review~\cite{BDY18}.

However, the presence of the long-range interactions induced by QED leads to significant
modifications in the form of the corrections associated to the finite volume energy levels. Irrespective
on whether a state is bound or unbound, in fact, the energy shifts take the form of polynomials in
the reciprocal of the lattice size
\cite{DaS14} and the exponential damping factors disappear. Moreover, the gapped nature of the
momentum of the particles on the lattice allows for a perturbative treatment of the QED contributions,
even at low energies
\cite{DaS14,UnH08,DET96,BeS14}. In this regime, composite particles receive corrections of the same kind
both in their mass \cite{DaS14} and in the energies of the two-body states that they can form \cite{BeS14}.

As shown in ref.~\cite{BeS14}, the leading-order energy shift for the lowest S-wave bound state is
proportional
to the fine-structure constant and has the same sign of the counterpart in absence of QED, presented in
refs.~\cite{KLH12,KLH11}. In the second part of this work, we demonstrate that the same relation holds
for the lowest bound P-wave state, whose finite volume correction is negative as the one for the counterpart
without electromagnetic interactions. Additionally, we prove that the QED energy-shifts for S- and P-wave
eigenstates have the same magnitude if order $1/L^3$ terms are neglected, a fact that remains valid in
the absence
of interactions of electromagnetic nature. At least for the $\ell=0$ and $1$ two-body bound eigenstates,
in fact, the sign of the correction depends directly on the parity of the wavefunction associated to the
energy state, whose tails are truncated at the boundaries of the cubic box, as observed in ref.~\cite{KLH12}.

Concerning scattering states, the energy shift formula for the lowest P-wave state that we present
in this paper has close similarities with the one in ref.~\cite{BeS14}, despite an overall
$\xi/M \equiv 4\pi^2/M L^2$ factor,
owing to the fact that the energy of the lowest unbound state with analogous transformation properties under
discrete rotations ($T_1$ irrep\footnote{Throughout, we use the abbreviation ``irrep'' for an irreducible
representation.} of the cubic group) is different from zero. Additionally, further scattering
parameters appear in the expression for the $\ell = 1$ finite volume energy correction, even as
coefficients of the smallest powers of $1/L$.

The present article is structured into two parts and its content can be summarized as
follows. After this preamble, the theoretical framework that is the basis for both the infinite and the
finite volume treatment is introduced, by starting from the Lagrangian with the strong P-wave interactions
alone. Next, in the end of sec.~2.0, the T-matrix for two-body fermion-fermion scattering to all orders in
the strength parameter of the potential is computed. Subsequently, in sec.~2.1, non-relativistic QED is
presented in the same fashion of refs.~\cite{CaL86,KiN96} and the Lagrangian is reduced to the
case of spinless fermions and electrostatic interactions. After displaying the amplitudes corresponding
to tree-level and one-loop diagrams with one Coulomb photon exchange, the non-perturbative treatment
of the Coulomb interaction is implemented. To this aim, we resort to the formalism of ref.~\cite{KoR00},
that we recapitulate in the end of sec.~2.1. As in the introductory section, the T-matrix matrix
element accounting for
both Coulomb and strong interactions is derived to all orders in $\alpha$, thanks to the Dyson-like
identities that hold among the free, the Coulomb and the full two-body Green's functions.
As in ref.~\cite{KoR00}, section~2.2 closes with the expressions of the scattering length and the
effective range in terms of the
physical constants of our EFT Lagrangian, that are obtained from the effective-range expansion. The first
part of the analysis is concluded in sec.~2.3 with the calculation of the same amplitude for the
fermion-antifermion scattering case.

Then, the two fermion-system is transposed onto a cubic lattice of size $L$ and the distortions induced by
the new environment in the laws of electrodynamics \cite{DaS14,HiP83} and in the masses of possibly
composite particles \cite{DaS14} are briefly summarized in sec.~3.0. Next, the quantization
conditions, that give access to the energy spectrum in finite volume though the expression of the
T-matrix elements \cite{BeS14}, are displayed and discussed (sec.~3.1) in the perturbative regime of QED.
Next, the finite volume counterpart
of the $\ell=1$ effective range expansion is presented, together with the expressions of the new L\"uscher
functions, shown in the end of sec.~3.2. Subsequently, the energy eigenvalues of the lowest bound and
scattering states are presented along with the details of the whole derivation, which can be skipped
by an experienced reader.
The pivotal results of the calculation are indeed given by the concluding formulas of secs.~3.3.1 and 3.3.2.
In the section that follows some hints are given regarding the consequences of the addition of transverse
photon interactions within our EFT for Coulomb and strong forces coupled to one unit of angular momentum.

The general conclusions of our work are drawn in sec.~5, where the main results are qualitatively
recapitulated.
The appendices provide supplemental material to the reader interested in the derivation of the
scattering amplitudes in sec.~2.2 and 2.3 and/or in the three-dimensional Riemann sums arising from the
approximations of the L\"uscher functions in secs.~3.1, 3.2 and 3.3. 

%--------------------------------------------------------------------------------------------------------------------------------------------------------------------

\section{\textsf{Effective Field Theory for non-relativistic fermions}}\label{S-2.0}

Our analysis of two-particle scattering and bound states in the infinite- and finite-volume context
is based on pionless Effective Field Theory \cite{KSW96,KSW98-01,Geg98,KSW98-02,BeK02,HaP10,GGP12,KGH16}.
The theory, developed more than two decades ago \cite{KSW96}, describes the strong interactions
between nucleons at energy scales smaller than the pion mass, $M_{\pi}$  \cite{LaM19,KoR00,BeK02}. The action
is non-relativistic and is constructed by including all the possible potential terms made of nucleon
fields and their derivatives, fulfilling the symmetry requirements of the strong interactions at low energies
(i.e. parity, time reversal and Galilean invariance) \cite{Aga17}. The importance of the various interaction
terms decreases with their canonical dimension while approaching the zero energy limit. Besides, even the
dominant contribution at low energies for local contact interactions between four-nucleon fields is of
dimension six, thus making the theory non-renorma\-lizable \cite{KoR00} in the classical sense.

Analogously to ref.~\cite{BeS14}, we begin by extending pionless EFT to spinless fermions of mass $M$
and charge $e$, and we assume that the theory is valid below an upper energy $\Lambda_{E^*}$
in the center-of-mass
frame (CoM). More specifically, if  the fermions represent hadrons, the latter energy cutoff
can be chosen to coincide with the pion mass. Second, we construct the interactions in terms
of four-fermion operators, selecting the ones that transform explicilty as the $2\ell +1$-dimensional
irreducible representation of $\mathrm{SO(3)}$,
 \begin{equation}
V^{(\ell)}(\mathbf{p},\mathbf{q}) \equiv \langle \mathbf{q}, -\mathbf{q} | \hat{\mathcal{V}}^{(\ell)}|
\mathbf{p},
-\mathbf{p} \rangle  =  \left(c_0^{(\ell)} + c_2^{(\ell)}\mathbf{p}^2  + c_4^{(\ell)}\mathbf{p}^4
+ \dots\right) \mathcal{P}_{\ell}\left(\mathbf{p}\cdot\mathbf{q}\right)\label{E-2.0-01}
\end{equation}
where $\mathcal{P}_{\ell}$ is a Legendre polynomial, $\pm \mathbf{p}$ ($\pm \mathbf{q}$) are the
three-momenta of two
incoming (outcoming) particles in the CoM frame, such that $|\mathbf{p}| = |\mathbf{q}|$,
$\hat{\mathcal{V}}^{(\ell)}$
is the potential in terms of second quantized operators and the  $c_{2j}^{(\ell)}$ are low-energy
(LECs) constants,
whose importance at low-energy scales diminishes for increasing values of $j$ . In particular, for the three
lowest angular momentum couplings ($\ell \leq 2$), the interaction potentials take the form  
\begin{equation}
V^{(0)}(\mathbf{p},\mathbf{q}) = C_0 + C_2\mathbf{p}^2+C_4\mathbf{p}^4 + \dots~, \label{E-2.0-02}
\end{equation}
\begin{equation}
V^{(1)}(\mathbf{p},\mathbf{q}) = \left(D_0 + D_2\mathbf{p}^2+D_4\mathbf{p}^4 + \dots\right)\mathbf{p}\cdot\mathbf{q}~,
\label{E-2.0-03}
\end{equation}
and 
\begin{equation}
V^{(2)}(\mathbf{p},\mathbf{q}) = \left(F_0 + F_2\mathbf{p}^2 + F_4\mathbf{p}^4 + \dots\right)
\left[3(\mathbf{p}\cdot\mathbf{q})^2-1\right]~. \label{E-2.0-04}
\end{equation}
As shown in sec.~II of ref.~\cite{BeS14}, the terms in eq.~\eqref{E-2.0-02} (eq.~\eqref{E-2.0-03}
and~\eqref{E-2.0-04})
proportional to even powers of the momentum (a gradient expansion in configuration space), can be
encoded by a single interaction with energy-dependent coefficient $C(E^*)$ ($D(E^*)$ and $F(E^*)$)
for S-waves (P- and D-waves),
where $E^*$ represents the CoM energy of the colliding particles, equal to $2M + \mathbf{p}^2/M$.
While the case
of fermions coupled to zero angular momentum via a single contact interaction proportional to $C(E^*)$ is the
starting-point of the analysis in ref.~\cite{BeS14}, the fundamental P-wave interaction in
eq.~\eqref{E-2.0-03}
with energy-dependent coefficient $D(E^*)$ becomes the key tool of the present investigation.
Although interactions
of the same form have been already adopted  in pionless  EFT for nucleons (cf. eq.~(4) in
ref.~\cite{BHK02}) and
in EFT with dimeron fields (cf. eq.~(2) in ref.~\cite{BHK03}), the P-wave counterpart of Kong and Ravndal's
analysis on fermion-fermion scattering in ref.~\cite{KoR00} is not available in the literature. Adopting
 the conventions of ref.~\cite{BHK02} for the coupling constants (cf. the Feynman rules in
app.~\ref{S-A-1.0}), the Lagrangian density assumes the form
\begin{equation}
\mathcal{L} = \psi^{\dagger}\left[i\hbar\partial_t + \frac{\hbar^2\nabla^2}{2M}\right]\psi
+ \frac{D(E^*)}{8}(\psi \overleftrightarrow{\nabla}_i \psi)^{\dagger}(\psi \overleftrightarrow{\nabla}_i \psi)~,
\label{E-2.0-05}
\end{equation} 
where $\overleftrightarrow{\nabla} = \overleftarrow{\nabla} - \overrightarrow{\nabla}$ denotes the Galilean
invariant derivative for fermions.Recalling the Feynman rules in app.~\ref{S-A-1.0}, two-body elastic
scattering processes without QED are represented by chains of bubbles, analogous to the ones in
Fig.~3 in ref.~\cite{KoR00}.

\begin{figure} [hb]
\includegraphics[width=0.49\columnwidth]{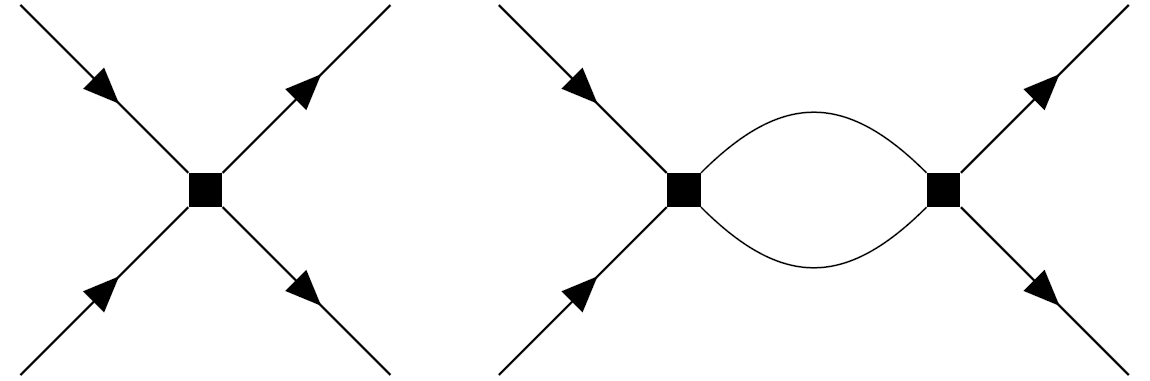}
\includegraphics[width=0.49\columnwidth]{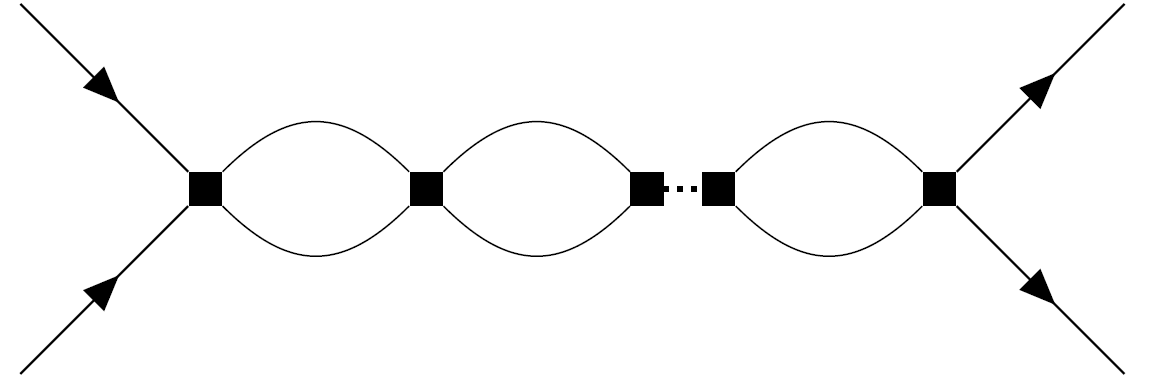}
\caption{Tree-level (upper line, left), 1-loop (upper line, right) and n-loops diagrams
(lower line) representing fermion-fermion
elastic scattering with the strong P-wave potential in eq.~\eqref{E-2.0-06}.}
\label{F-2-01}
\end{figure}

In particular, the tree-level diagram, consisting of a single four-fermion vertex,
leads to an amplitude equal to
$-\mathrm{i}D(E^*) \mathbf{p}\cdot \mathbf{p}'$ (cf. ref.~\cite{Fur14}) where $\pm\mathbf{p}$
and $\pm\mathbf{p}'$
are, respectively, the momenta of the incoming and outcoming particles in the CoM frame. As a consequence,
the two-body $\ell=1$ (pseudo)potential in momentum space takes the form
\begin{equation}
V^{(1)}(\mathbf{p},\mathbf{q}) \equiv \langle \mathbf{q}, -\mathbf{q} \lvert \hat{\mathcal{V}}^{(1)} \lvert \mathbf{p}, -\mathbf{p}\rangle = D(E^*)~\mathbf{p}\cdot \mathbf{q}~,\label{E-2.0-06}
\end{equation}
which coincides with the tree-level diagram multiplied by the imaginary unit. Considering also 
the other possible diagrams in momentum space with amputated legs in fig.~\ref{F-2-01}, the
expression for the full scattering amplitude due to strong interactions can be written as,
\begin{equation}
\mathrm{i}T_{\mathrm{S}}(\mathbf{p},\mathbf{q}) = \mathrm{i} \langle \mathbf{q}, -\mathbf{q} \lvert
\left[\hat{\mathcal{V}}^{(1)}+ \hat{\mathcal{V}}^{(1)}\hat{G}_0^{E}\hat{\mathcal{V}}^{(1)}
+ \hat{\mathcal{V}}^{(1)}\hat{G}_0^{E}\hat{\mathcal{V}}^{(1)}\hat{G}_0^{E}\hat{\mathcal{V}}^{(1)} + ...
\right]\lvert \mathbf{p}, -\mathbf{p}\rangle~.\label{E-2.0-07}
\end{equation}
where $\hat{G}_0^{E} \equiv \hat{G}_0^{(+)}(E)$ is the two-body unperturbed retarded $(+)$ Green's
function operator,
\begin{equation}
\hat{G}_0^{(\pm)}(E) = \frac{1}{E-\hat{H}_0\pm \mathrm{i}\varepsilon}~,\label{E-2.0-08}
\end{equation}
with $\hat{H}_0$ the two-body free Hamiltonian in relative coordinates $\hat{H}_0 =
\hat{\mathbf{p}}^2/M$, and $M/2$
is the reduced mass of a system of identical fermions. Inserting a complete set of plane wave eigenstates
$|\mathbf{q},-\mathbf{q}\rangle \equiv |\mathbf{q}\rangle$ in the numerator, the latter expression becomes
\begin{equation}
\hat{G}_0^{(\pm)}(E) = M\int_{\mathbb{R}^3}\frac{\mathrm{d}^3 q}{(2\pi)^3}
\frac{|\mathbf{q}\rangle\langle\mathbf{q}|}{\mathbf{p}^2 - \mathbf{q}^2 \pm \mathrm{i}\varepsilon}~,
\label{E-2.0-09}
\end{equation}
where $E = \mathbf{p}^2/M$ is the energy eigenvalue at which the retarded $(+)$ and advanced $(-)$ Green's
functions are evaluated. In configuration space the latter take the form
\begin{equation}
\langle \mathbf{q}, -\mathbf{q} \lvert \hat{G}_0^{E} \lvert \mathbf{p}, -\mathbf{p}\rangle
= \frac{(2\pi)^3\delta(\mathbf{p}-\mathbf{q})}{E-\mathbf{p}^2/M + \mathrm{i}\varepsilon}~,\label{E-2.0-10}
\end{equation}
that is diagrammatically depicted by two propagation lines. The explicit computation of the three lowest order
contributions to the sum in eq.~\eqref{E-2.0-07} yields
\begin{equation}
\mathrm{i}\langle \mathbf{q}, -\mathbf{q} \lvert \hat{\mathcal{V}}^{(1)} \lvert \mathbf{p}, -\mathbf{p}\rangle
= \mathrm{i} D(E^*)\mathbf{p}\cdot \mathbf{q},\label{E-2.0-11}
\end{equation}
\begin{equation}
\mathrm{i}\langle \mathbf{q}, -\mathbf{q} \lvert \hat{\mathcal{V}}^{(1)}\hat{G}_0^{E}\hat{\mathcal{V}}^{(1)} \lvert
\mathbf{p}, -\mathbf{p}\rangle = \mathrm{i} D(E^*)^2~p_i q_j~\partial_{i}\partial_{j}'G_0^E(\mathbf{r},\mathbf{r}')
\Big\lvert_{\substack{\mathbf{r} = \mathbf{0} \\ \mathbf{r}'= \mathbf{0}}} \equiv \mathrm{i}D(E^*)^2~\mathbf{q}\cdot \mathbb{J}_{0}
\mathbf{p},\label{E-2.0-12}
\end{equation}
and
\begin{equation}
\begin{split}
\mathrm{i}\langle \mathbf{q}, -\mathbf{q} \lvert \hat{\mathcal{V}}^{(1)}\hat{G}_0^{E}\hat{\mathcal{V}}^{(1)}\hat{G}_0^{E}
\hat{\mathcal{V}}^{(1)} \lvert \mathbf{p}, -\mathbf{p}\rangle = \mathrm{i} D(E^*)^3~q_i p_k~\partial_{i}\partial_{j}'
 G_0^E(\mathbf{r},\mathbf{r}')\Big\lvert_{\substack{\mathbf{r} = \mathbf{0} \\ \mathbf{r}'= \mathbf{0}}} \\ \cdot \partial_{j}'\partial_{k}''
G_0^E(\mathbf{r}',\mathbf{r}'')\Big\lvert_{\substack{\mathbf{r}' = \mathbf{0} \\ \mathbf{r}''= \mathbf{0}}}  & = \mathrm{i}
D(E^*)^3 \mathbf{q} \cdot \mathbb{J}_{0}^2\mathbf{p}~,\label{E-2.0-13}
\end{split}
\end{equation}
where $\partial_i \equiv \partial/\partial r_i$, $\partial_i' \equiv \partial/\partial r_i'$ and
$\partial_i'' \equiv \partial/\partial r_i''$, while $\mathbb{J}_{0}$ is a symmetric matrix whose elements
are given by 
\begin{equation}
(\mathbb{J}_{0})_{ij} =  \partial_{i}\partial_{j}'G_0^E(\mathbf{r},\mathbf{r}')\Big
\lvert_{\substack{\mathbf{r} = \mathbf{0} \\ \mathbf{r}'= \mathbf{0}}}~,\label{E-2.0-14}
\end{equation}
and Einstein's index convention is henceforth understood. Extending the computation to higher orders, it is
evident that the infinite superposition of chains of bubbles translates into a geometric series in the
total scattering amplitude, as in the $\ell=0$ case, and a formula analogous to eq.~(2) in ref.~\cite{KoR00}
is obtained,
\begin{equation}
\mathrm{i}T_{\mathrm{S}}(\mathbf{p},\mathbf{q}) =\mathrm{i}D(E^*)~\mathbf{q}\cdot \left(\mathbbm{1}
+ D(E^*)\mathbb{J}_{0} + D(E^*)^2\mathbb{J}_{0}^2 + D(E^*)^3\mathbb{J}_{0}^3 +\ldots
\right)\mathbf{p}
= \mathbf{q}\cdot\frac{D(E*)}{\mathbbm{1}-D(E^*)\mathbb{J}_{0}}\mathbf{p}~.\label{E-2.0-15}
\end{equation}
Furthermore, performing the Fourier transform of the potential in eq.~\eqref{E-2.0-06} into configuration
space,
\begin{equation}
V^{(1)}(\mathbf{r},\mathbf{r}') \equiv \langle \mathbf{r'} \lvert \hat{\mathcal{V}}^{(1)} \lvert \mathbf{r}\rangle
= D(E^*)~\nabla\delta(\mathbf{r})\cdot\nabla'\delta(\mathbf{r}')~,\label{E-2.0-16}
\end{equation}
the full scattering amplitude can be recovered independently in position space by means of
partial integrations
and cancellations of surface integrals at infinity,
\begin{equation}
T_{\mathrm{S}}(\mathbf{r},\mathbf{r}') = \nabla \delta(\mathbf{r}) \cdot \frac{D(E^*)}{\mathbbm{1}
-D(E^*)\mathbb{J}_{0}}\nabla'\delta(\mathbf{r}')~.\label{E-2.0-17}
\end{equation}
The matrix elements of $\mathbb{J}_{0}$ can be, now, evaluated by dimensional regularization. Applying the
formula in eq.~(B18) of ref.~\cite{Ram97} for d-dimensional integration, eq.~\eqref{E-2.0-14} in arbitrary
d-dimensions becomes
\begin{equation}
\begin{split}
(\mathbb{J}_{0})_{ij}(d) = \partial_{i}\partial_{j}'G_0^E(d;\mathbf{r},\mathbf{r}') \Big\lvert_{\substack{\mathbf{r}
= \mathbf{0} \\ \mathbf{r}'= \mathbf{0}}} = \left(\frac{\mu}{2}\right)^{3-d}\int_{\mathbb{R}^d} \frac{\mathrm{d}^d
\mathbf{k}}{(2\pi)^d}  \frac{k_i k_j}{E-\mathbf{k}^2/M+\mathrm{i}\varepsilon}  \\ = -\delta_{ij}
\frac{M^2(E+\mathrm{i}\varepsilon)}{d (4\pi)^{d/2}} \left(\frac{\mu}{2}\right)^{3-d} & [-M(E+\mathrm{i}\varepsilon)]^{d/2-1}
\Gamma \textstyle{\left(\frac{2-d}{2}\right)}
\label{E-2.0-18}
\end{split}
\end{equation}
where $\mu$ is the renormalization scale introduced by the minimal subtraction (MS) scheme.
Like the S-wave counterpart,
the integral proves to be finite in three dimensions and, within this limit is given by
\begin{equation}
\lim_{d\rightarrow 3} \partial_{i}\partial_{j}'G_0^E(\mathbf{r},\mathbf{r}';d)\Big\lvert_{\mathbf{r},\mathbf{r}'=0} 
= \partial_{i}\partial_{j}'G_0^E(\mathbf{r},\mathbf{r}')\Big\lvert_{\mathbf{r},\mathbf{r}'=0} =  -\delta_{ij}\frac{M}{4\pi}
\frac{\mathrm{i}|\mathbf{p}|^3}{3}~,\label{E-2.0-19}
\end{equation}
where the energy $E$ in the CoM frame has been eventually expressed as $\mathbf{p}^2/M$.
For the sake of completeness, we derive the contribution to $(\mathbb{J}_{0})_{ij}$  from the power
divergence subtraction (PDS) regularization scheme, in which the power counting of the EFT is
manifest \cite{KSW98-01,KSW98-02}.
To this aim, the eventual poles of the regularized integral for $d\rightarrow 2$ should be taken into account.
In this limit, it turns out from eq.~\eqref{E-2.0-18} that the Euler's Gamma has a pole singularity of the
kind $2/(2-d)$. As a consequence, the original dimensional regularization result in eq.~\eqref{E-2.0-18}
acquires a finite PDS contribution, transforming into
\begin{equation}
(\mathbb{J}_0)_{ij}^{\mathrm{PDS}} = \partial_{i}\partial_{j}'G_0^E(3;\mathbf{r},\mathbf{r}')\Big\lvert_{\substack{\mathbf{r}
= \mathbf{0} \\\mathbf{r}'= \mathbf{0}}}^{\mathrm{PDS}}  = -\delta_{ij}\frac{M}{4\pi}
\left(\frac{\mathrm{i}|\mathbf{p}|^3}{3} +\mu \frac{\mathbf{p}^2}{2}\right)~.\label{E-2.0-20}
\end{equation}
This can be compared with the one in eq.~(4) in ref.~\cite{KoR00} for the S-wave interactions. Since
the $\mathbb{J}_0$ matrix is diagonal  (eq.~\eqref{E-2.0-20}), few efforts are needed for the computation
of the fermion-fermion scattering amplitude,
\begin{equation}
T_{\mathrm{S}}(\mathbf{p},\mathbf{q}) = \frac{12\pi}{M}\frac{D(E^*)\mathbf{p}\cdot\mathbf{q}}{\frac{12\pi}{M}
+\mathrm{i}D(E^*)|\mathbf{p}|^3}~.\label{E-2.0-21}
\end{equation}
With reference to scattering theory \cite{Omn70}, the $T_{\mathrm{S}}$ matrix for P-wave elastic scattering with
phase shift $\delta_1$ can be written as
\begin{equation}
T_{\mathrm{S}}(\mathbf{p},\mathbf{q}) = -\frac{4\pi}{M} \frac{e^{\mathrm{i}2\delta_1}-1}{2\mathrm{i}|\mathbf{p}|} 3\cos\theta
= -\frac{12\pi}{M}\frac{\mathbf{p}\cdot\mathbf{q}}{|\mathbf{p}|^3\cot\delta_1-\mathrm{i}|\mathbf{p}|^3}  ~,
\label{E-2.0-22}
\end{equation}
where $\theta$ is the angle between the incoming and outcoming direction of particles in the CoM frame. Recalling
the effective-range expansion (ERE) for $\ell=1$ scattering \cite{Omn70},  
\begin{equation}
|\mathbf{p}|^3\cot\delta_1 = -\frac{1}{a} +\frac{1}{2}r_0\mathbf{p}^2+r_1\mathbf{p}^4 + r_2\mathbf{p}^6
+\ldots~,\label{E-2.0-23}
\end{equation}
an expression for the scattering parameters in terms of the momenta of the particles, the coupling
constant and
the mass $M$ can be drawn. In particular, a formula for the scattering length analogous to eq.~(2.16) of
ref.~\cite{KSW98-02} can be recovered,
\begin{equation}
a = \frac{M}{4\pi}\frac{D(E^*)}{3}~.\label{E-2.0-24}
\end{equation}
Furthermore, the effective range parameter $r_0$ vanishes, as in the zero angular momentum case.
Plugging the PDS-regularized expression of $\mathbb{J}_{0}$ in eq.~\eqref{E-2.0-20} into
eq.~\eqref{E-2.0-15} and exploiting
the ERE again, finally, the renormalized form of the coupling constant $D(E^*)$ is obtained, 
\begin{equation}
D(E^*,\mu) = 3a\left(\frac{4\pi}{M}+\mu\frac{\mathbf{p}^2}{2}\right)~.\label{E-2.0-25}
\end{equation}
Unlike in the $\ell=0$ case, we note that the $\mu$-dependent version of $D(E^*) = 12\pi a/M$ is
quadratic in the momentum of the incoming fermions.

%--------------------------------------------------------------------------------------------------------------------------------------------------------------------

\subsection{\textsf{Coulomb corrections}}\label{S-2.1}

We introduce the interactions of electromagnetic nature in the non Lorentz-covariant fashion
of ref.~\cite{KiN96}
and \cite{CaL86}. The formalism of non-relativistic quantum electrodynamics (NRQED), introduced
in ref.~\cite{CaL86}, is designed to reproduce the low-momentum behaviour of QED to any desired
accuracy. Besides,
only non-relativistic momenta are allowed in the loops and in the external legs of the diagrams.
The contributions
arising from relativistic momenta in the QED loops, in fact, are absorbed as renormalizations of the coupling
constants of the local interactions in the non-relativistic counterpart of QED \cite{CaL86}. The Lagrangian
is determined by the particle content and by the symmetries of the theory, such as gauge invariance,
locality, hermiticity,
parity conservation, time reversal symmetry and Galilean invariance. The particles are fermionic,
characterized
by mass $M$ and unit charge $e$, and are represented by two-component non-relativistic Pauli spinor fields
$\Psi$. In compliance to these prescriptions, the NRQED Lagrangian density in ref.~\cite{CaL86} assumes the form,

\begin{equation}
\begin{split}
\mathcal{L}^{\mathrm{NRQED}} = -\frac{1}{2}\left(\mathbf{E}^2-\mathbf{B}^2\right) + \Psi^{\dagger}\left(i\partial_t -e\phi
+ \frac{\mathbf{D}^2}{2M}\right)\Psi  + \Psi^{\dagger}\left[c_1\frac{\mathbf{D}^4}{8M^3}  + c_2\frac{e}{2M}
\boldsymbol{\sigma}\cdot\mathbf{B} \right. \\  \left. + c_3\frac{e}{8M^2}\nabla\cdot\mathbf{E} + c_4\frac{e}{8M^2}
\mathrm{i}\mathbf{D}\times\boldsymbol{\sigma}\right]\Psi   + \Psi^{\dagger}\left[d_1\frac{e}{8M^3} \{\mathbf{D}^2, \right. & \left. 
\boldsymbol{\sigma}\cdot\mathbf{B}\} \right]\Psi + \ldots~,
\label{E-2.1-01}
\end{split}
\end{equation}

\noindent
where $\mathbf{D} = \nabla + \mathrm{i}e\mathbf{A}$ is the covariant derivative, while
$\mathbf{E} = -\nabla\phi -\partial_t \mathbf{A}$ and $\mathbf{B} = \nabla \times \mathbf{A}$ denote the
electric
and magnetic fields, respectively. The terms in the first row encode the leading ones of
$\mathcal{L}^{\mathrm{NRQED}}$,
containing the minimal coupling of the fermionic fields with the vector potential $\mathbf{A}$ and
the scalar potential,
$\phi$. The interactions proportional to the constants $c_1$-$c_4$ and $d_1$ in eq.~\eqref{E-2.1-01} are
next-to-leading-order terms, corresponding to corrections of order $v^4/c^4$ and $v^6/c^6$,
respectively \cite{KiN96},
whereas the ellipses represent contributions containing higher order covariant derivatives,
$\mathcal{O}(v^8/c^8)$.

\begin{figure} [h]
\begin{center}
\includegraphics[width=0.50\columnwidth]{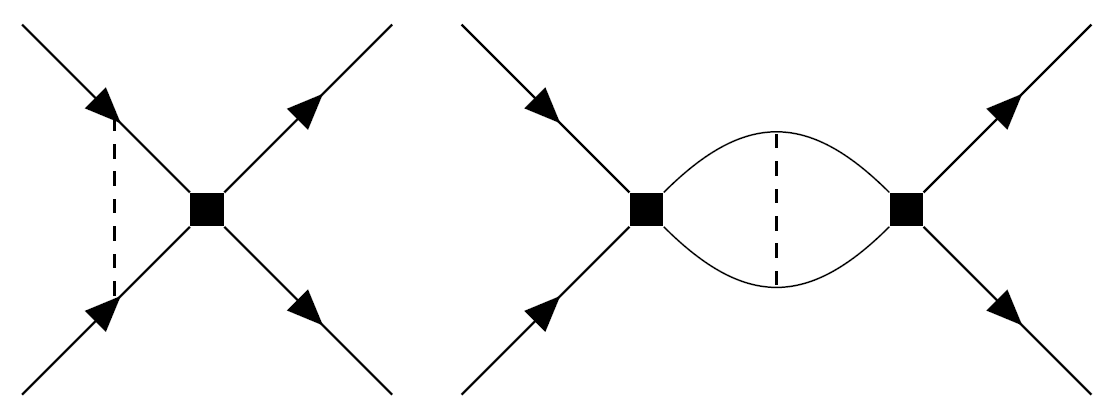}
\end{center}
\caption{The tree-level (left) and one-loop (right) fermion-fermion scattering diagram for
strong $\ell=1$ interactions with one Coulomb photon insertion (dashed lines).}\label{F-2-02}
\end{figure}

Since the Coulomb force dominates at very low energies and transverse photons couple proportionally
to the fermion
momenta, in the present treatment we choose to retain in the Lagrangian only the scalar field and
its lowest order
coupling to the fermionic fields as in ref.~\cite{KoR00}. Moreover, we reduce the latter to spinless
fields $\psi$,
consistently with sec.~\ref{S-2.0} and with ref.~\cite{BeS14}. As a consequence, the full
Lagrangian density of
the system becomes the superposition of the one in eq.~\eqref{E-2.0-05} with the one involving
the electrostatic
potential and its leading order coupling to the spinless fermions, namely
\begin{equation}
\mathcal{L}^{\mathrm{NRQED}~\mathrm{corr}} = -\frac{1}{2} \nabla\phi \cdot \nabla\phi - e\phi~\psi^{\dagger}\psi~.
\label{E-2.1-02}
\end{equation}
Alternatively, on top of the P-wave interaction in eq.~\eqref{E-2.0-06} the Coulomb force, that in
momentum space regulated by an IR cutoff $\lambda$, reads 
\begin{equation}
V_{\mathrm{C}} (\mathbf{p},\mathbf{q}) \equiv \langle \mathbf{q}, -\mathbf{q}|\hat{V}_{\mathrm{C}}|\mathbf{p},-\mathbf{p}\rangle
= \frac{e^2}{(\mathbf{p}-\mathbf{q})^2 +\lambda^2}~,\label{E-2.1-03}
\end{equation}
has been added. The introduction of the electrostatic potential generates the additional Feynman rules listed
in app.~\ref{S-A-1.0}. Consequently, the T-matrix is enriched by new classes of diagrams (cf.
fig.~\ref{F-2-02}),
in which the Colulomb photon insertions either between the external legs and within the loops begin to appear.
Unlike transverse photons, the scalar ones do not propagate between different bubbles.

From the Feynman rules, the amplitude for the tree-level diagram with one photon insertion in the
left part of fig.~\ref{F-2-02} gives,
\begin{equation}
-\mathrm{i}T_{\mathrm{SC}}^{\mathrm{tree}}(\mathbf{p},\mathbf{p}') =-D(E^*)\int_{\mathbb{R}^4}\frac{\mathrm{d}^4l}{(2\pi)^4}
\frac{\mathrm{i}~e}{\frac{E}{2}+l_0-\frac{(\mathbf{p}-\mathbf{l})^2}{2M}+\mathrm{i}\varepsilon}  \frac{ (\mathbf{p}
-\mathbf{l})\cdot\mathbf{p}' }{\mathbf{l}^2+\lambda^2} \frac{\mathrm{i}~e}{\frac{E}{2}-l_0-\frac{(\mathbf{l}
-\mathbf{p})^2}{2M}+\mathrm{i}\varepsilon}~.\label{E-2.1-04}
\end{equation}
After integrating over the free energy $l_0$, the tree-level amplitude in eq.~\eqref{E-2.1-04} can
be decomposed as follows
 \begin{equation}
-\mathrm{i}T_{\mathrm{SC}}^{\mathrm{tree}}(\mathbf{p},\mathbf{p}') = \int_{\mathbb{R}^3}\frac{\mathrm{d}^3\mathbf{l}}{(2\pi)^3}
\frac{-\mathrm{i}e^2}{\mathbf{l}^2+\lambda^2}\frac{M D(E^*)~\mathbf{p}\cdot\mathbf{p}'}{\mathbf{p}^2-(\mathbf{p}-\mathbf{l})^2
+\mathrm{i}\varepsilon}   - \mathrm{i}  \int_{\mathbb{R}^3}\frac{\mathrm{d}^3\mathbf{l}}{(2\pi)^3} \frac{e^2}{\mathbf{l}^2
+\lambda^2}\frac{D(E^*) M~\mathbf{l}\cdot\mathbf{p}'}{\mathbf{l}^2-2\mathbf{p}\cdot\mathbf{l}-\mathrm{i}\varepsilon}~.
\label{E-2.1-05}
\end{equation}
In particular, in the last rewriting the first integral on the r.h.s. turns out to be identical to the
one in eq.~(9) of
ref.~\cite{KoR00} except for the factor $\mathbf{p}\cdot\mathbf{p}' = \mathbf{p}^2 \cos\theta$, therefore
it can be immediately integrated. Conversely, the last integral in eq.~\eqref{E-2.1-05} represents a
new contribution,
whose evaluation in dimensional regularization is carried out in app.~\ref{S-A-2.0}. Adding the two
contributions
together, the tree level amplitude with Coulomb photon insertion in eq.~\eqref{E-2.1-05} becomes
\begin{equation}
T_{\mathrm{SC}}^{\mathrm{tree}}(\mathbf{p},\mathbf{p}') = \cos\theta D(E^*)\frac{\alpha M}{2} \left[|\mathbf{p}|
\left(\mathrm{i} - \frac{\pi}{2}\right) - \mathrm{i}|\mathbf{p}|\log\frac{2|\mathbf{p}|}{\lambda} \right]
+ \mathcal{O}(\lambda)~,\label{E-2.1-06}
\end{equation}
where the limit $\lambda \rightarrow 0$ for the $\mathcal{O}(\lambda)$ terms is understood. From the
last equation
we infer that, due to the linear dependence in the momenta of the incoming particles $\mathbf{p}$ in the
CoM frame,
P-wave fermion-fermion scattering is suppressed with respect to the S-wave one in the low-$\mathbf{p}$
limit. However,
both the $\ell=0$ and $\ell=1$ tree level amplitudes with one Coulomb photon insertion are divergent in the
$\lambda \rightarrow 0$ limit.

Nevertheless, due to the fact that the latter logarithmic contribution is imaginary, the infinite
term does not
contribute to the $\mathcal{O}(\alpha)$ corrections of the strong cross section, which is proportional
to $|C(E^*) + T_{\mathrm{SC}}^{\mathrm{tree}}|^2 \approx C(E^*)$ $\cdot[C(E^*) + 2
\mathfrak{Re}T_{\mathrm{SC}}^{\mathrm{tree}}]$ up to first order
in $\alpha$. The corrected cross section to that order turns out to be IR finite and proportional
to $1-\pi\eta$,
where $\eta = \alpha M/2|\mathbf{p}|$ for particles with equal unit charge. As observed in
ref.~\cite{KoR00}, the
inclusion of $n$ Coulomb photon exchanges leads to corrections proportional to $\eta^n$ in the cross section.
Therefore, the feasibility of a perturbative treatment  for the Coulomb force is regulated by the smallness
of the parameter $\eta$, i.e. by a constraint on the momenta of the incoming particles, $|\mathbf{p}|
\gg \alpha M/2$.
As a consequence, if the momenta of the incoming particles are too small, the Coulomb force is expected
to have a strong influence on the cross-section of the elastic process and a non-perturbative
treatment becomes necessary.

Furthermore, the Feynman rules for the one-loop diagram with one photon insertion on the right part of
fig.~\ref{F-2-02} yield
\begin{equation}
\begin{split}
-\mathrm{i}T_{\mathrm{SC}}^{\mathrm{1-loop}}(\mathbf{p},\mathbf{p}') = \int_{\mathbb{R}^4}\frac{\mathrm{d}^4 l}{(2\pi)^4}
\int_{\mathbb{R}^4}\frac{\mathrm{d}^4 k}{(2\pi)^4} \frac{-\mathrm{i}e^2}{(\mathbf{l}-\mathbf{k})^2+\lambda^2} 
 \frac{D(E^*)}{\frac{E}{2}+l_0-\frac{\mathbf{l}^2}{2M}+\mathrm{i}\varepsilon} \\ \cdot \frac{\mathrm{i}
\mathbf{p}\cdot\mathbf{l}  }{\frac{E}{2}-l_0-\frac{\mathbf{l}^2}{2M}+\mathrm{i}\varepsilon} 
\frac{ \mathrm{i} \mathbf{k}\cdot\mathbf{p}'}{\frac{E}{2}+k_0-\frac{\mathbf{k}^2}{2M}+\mathrm{i}\varepsilon}
& \frac{D(E^*)}{\frac{E}{2}-k_0-\frac{\mathbf{k}^2}{2M}+\mathrm{i}\varepsilon}~.\label{E-2.1-07}
\end{split}
\end{equation}
Similarly, the contour integration with respect to the free energies $k_0$ and $l_0$, followed by the
momentum translation $\mathbf{l}\mapsto\mathbf{q}\equiv \mathbf{l}-\mathbf{k}$, leads to
\begin{equation}
T_{\mathrm{SC}}^{\mathrm{1-loop}}(\mathbf{p},\mathbf{p}') = -  M^2 \int_{\mathbb{R}^3}\frac{\mathrm{d}^3 \mathbf{k}}{(2\pi)^3}
\int_{\mathbb{R}^3}\frac{\mathrm{d}^3 \mathbf{q}}{(2\pi)^3}\frac{ e^2  [D(E^*)]^2   }{\mathbf{q}^2+\lambda^2} 
\frac{\mathrm{i} \mathbf{k}\cdot \mathbf{p}' }{\mathbf{p}^2-(\mathbf{q}+\mathbf{k})^2+\mathrm{i}\varepsilon}
\frac{\mathrm{i} \mathbf{p}\cdot (\mathbf{q}+\mathbf{k}) }{\mathbf{p}^2-\mathbf{k}^2+\mathrm{i}\varepsilon}~.
\label{E-2.1-08}
\end{equation}
The remaining momentum integrations are performed in dimensional regularization (cf. sec.~\ref{S-A-2.0})
and give
\begin{equation}
T_{\mathrm{SC}}^{\mathrm{1-loop}}(\mathbf{p},\mathbf{p}') = \cos\theta [D(E^*)]^2 
\frac{\alpha M^2}{4\pi}\frac{\mathbf{p}^4}{6} \left[\frac{1}{\epsilon} 
 -\gamma_E +\frac{7}{3} + \mathrm{i}\pi -\log\left(\frac{2\mathbf{p}^2}{\pi\mu}\right)  \right]~,\label{E-2.1-09}
\end{equation}
where $\epsilon \equiv d -3$ and $\gamma_E \approx 0.5772$ denotes the Euler-Mascheroni constant.
The amplitude in
eq.~\eqref{E-2.1-09} displays a pole at $d = 3$ as the one in eq.~(15) of ref.~\cite{KoR00}, an ultraviolet
divergence that can be reabsorbed with a redefinition of the strength parameter $D(E^*)$ via the
renormalization
process. However, $T_{\mathrm{SC}}^{\mathrm{1-loop}}$ is devoid of the logarithmic divergence in the
zero-momentum limit,
due to the multiplication by a factor $\mathbf{p}^2$. Consequently, in comparison with the zero
angular momentum
counterpart, the $\ell=1$ one-loop scattering amplitude with one-photon insertion is suppressed
in the limit of
zero momentum $\mathbf{p}$ of the incoming particles in the CoM frame. Since $T_{\mathrm{SC}}^{\mathrm{1-loop}}$
possesses also a pole in the $d\rightarrow 2$ limit, the implementation of the PDS scheme results into an
additional term proportional to the renormalization scale (or mass), 
\begin{equation}
T_{\mathrm{SC}}^{\mathrm{1-loop}}(\mathbf{p},\mathbf{p}')\Big\lvert^{\mathrm{PDS}}
=   \cos\theta [D(E^*)]^2 \frac{\alpha  M^2 }{4\pi} \frac{\mathbf{p}^{2}}{2} \left[\frac{\mathbf{p}^{2}}{3}
\left(\frac{1}{\epsilon} -\gamma_E+\frac{7}{3}-\log\frac{2\mathbf{p}^2}{\pi\mu} + \mathrm{i}\pi \right) 
 + \frac{\mu}{2}\right]~.\label{E-2.1-10}
\end{equation}
Differently from the $\ell=0$ counterpart in eq.~(14) of ref.~\cite{KoR00}, the logaritmic term in
the CoM momentum of the colliding fermions does not give rise to a divergence in the zero momentum limit,
due to the $\mathbf{p}^4$
prefactor. Nevertheless, the dressing of the one-bubble diagram with two or more Coulomb photon
insertions results
in the multiplication of $T_{\mathrm{SC}}^{\mathrm{1-loop}}$ by one or more powers of $\eta = \alpha M /2
|\mathbf{p}|^2$,
so that, at order higher than four in $\alpha$, the amplitude becomes singular in the limit
$|\mathbf{p}| \rightarrow 0$.
It follows that the perturbative approach breaks down and the effects of Coulomb repulsion need
to be treated to all orders in $\alpha$.

\begin{figure}[h]
\includegraphics[width=1.00\columnwidth]{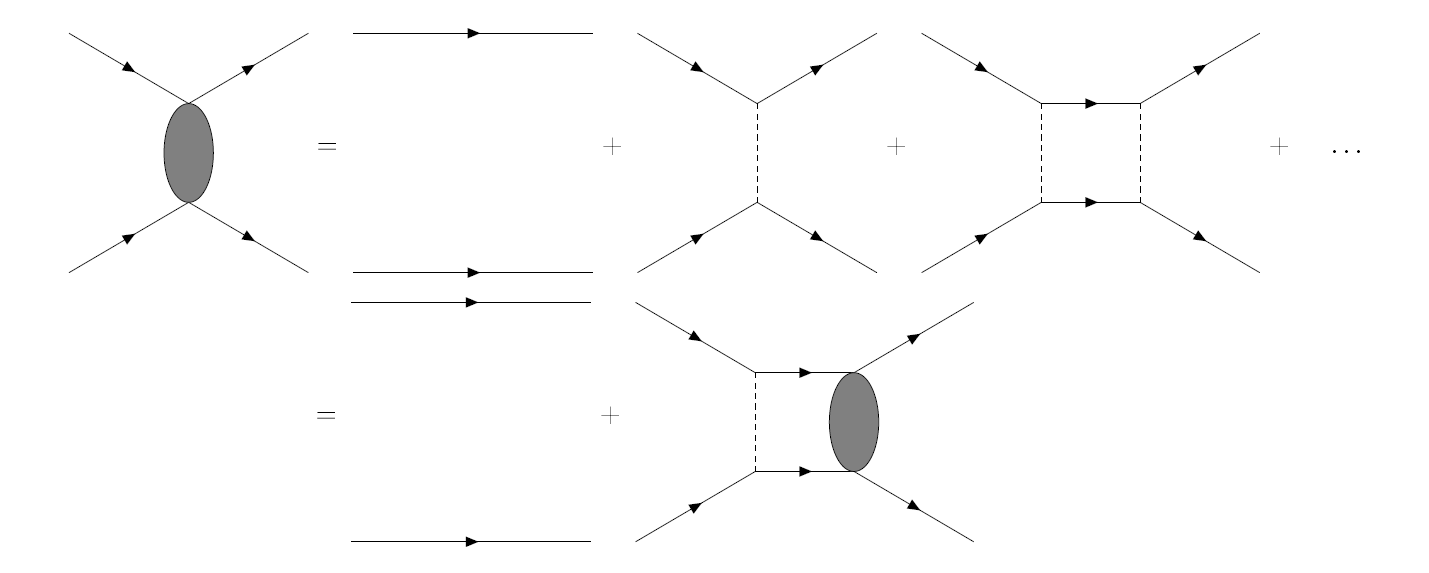}
\caption{The Coulomb propagator $G_{\mathrm{C}}$ as an infinite superposition of ladder diagrams (upper row),
which can be compactly incorporated in a self-consistent identity (lower row).}
\label{F-2-03}
\end{figure}

Since our interest resides in the low-momentum sector of fermion-fermion elastic scattering, we
incorporate the
Coulomb ladders in the amplitude of the process to all orders in the fine structure constant.
For scalar photons,
this amounts to replacing the free-fermion propagators in the bubble diagrams of fig.~\ref{F-2-01}
with the Coulomb
propagators in fig.~\ref{F-2-03}). To this aim, we follow the procedure outlined in ref.~\cite{KoR00}
and introduce
the Coulomb Green's functions. 
The inclusion of the Coulomb potential (cf. eq.~\eqref{E-2.1-03}) in the Hamiltonian yields
the Coulomb Green's function operator,
\begin{equation}
\hat{G}_{\mathrm{C}}^{(\pm)}(E) = \frac{1}{E-\hat{H}_0-\hat{V}_{\mathrm{C}} \pm \mathrm{i}\varepsilon}~,\label{E-2.1-11}
\end{equation}
an expression that, together with eq.~\eqref{E-2.0-08}, admits a self-consistent rewriting
\textit{\`a la Dyson} \cite{FeW71}, 
\begin{equation}
\hat{G}_{\mathrm{C}}^{(\pm)} = \hat{G}_0^{(\pm)} + \hat{G}_0^{(\pm)}\hat{V}_{\mathrm{C}}\hat{G}_{\mathrm{C}}^{(\pm)}~,\label{E-2.1-12}
\end{equation}
that can be diagrammatically represented as in fig.~\ref{F-2-03}.

Moreover, the solutions of Schr\"odinger equation with a repulsive Coulomb potential, $(\hat{H}_0
+ V_{\mathrm{C}}-E)|\Psi_{\mathbf{p}}^{(\pm)}\rangle$, can be formally expressed in terms of the free ones as
\begin{equation}
|\psi_{\mathbf{p}}^{(\pm)}\rangle = \hat{G}_{\mathrm{C}}^{(\pm)}\hat{G}_0^{-1}|\mathbf{p}\rangle
=  \left[1+\hat{G}_{\mathrm{C}}^{(\pm)}\hat{V}_{\mathrm{C}}\right]|\mathrm{p}\rangle~, \label{E-2.1-13}
\end{equation}
see eq.~(18) in ref.~\cite{KoR00}. The above eigenstates share with the plane waves the generalized
normalization property, i.e. $\langle \psi_{\mathbf{p}}^{(\pm)}|\psi_{\mathbf{q}}^{(\pm)}\rangle =
(2\pi)^3\delta(\mathrm{q}-\mathrm{p})$. If the potential is repulsive, the solution with outgoing spherical
waves in the future is given by 
\begin{equation}
\psi_{\mathbf{p}}^{(+)}(\mathbf{r}) = e^{-\frac{1}{2}\pi\eta}\Gamma(1+\mathrm{i}\eta)M(-\mathrm{i}\eta,1;\mathrm{i}pr
-\mathrm{i}\mathbf{p}\cdot\mathbf{r})e^{\mathrm{i}\mathbf{p}\cdot\mathbf{r}}~,\label{E-2.1-14}
\end{equation}
while the state with incoming spherical waves in the distant past coincides with
\begin{equation}
\psi_{\mathbf{p}}^{(-)}(\mathbf{r}) = e^{-\frac{1}{2}\pi\eta}\Gamma(1-\mathrm{i}\eta)M(\mathrm{i}\eta,1;-\mathrm{i}pr
-\mathrm{i}\mathbf{p}\cdot\mathbf{r})e^{\mathrm{i}\mathbf{p}\cdot\mathbf{r}}~,\label{E-2.1-15}
\end{equation}
where $M(a,b;c)$ is a Kummer function. In particular, the squared modulus of the two given spherical waves
evaluated in the origin, i.e. the probability of finding the two fermions at zero separation, is equal to 
\begin{equation}
C_{\eta}^2 \equiv |\psi_{\mathbf{p}}^{(\pm)}(0)|^2 = e^{-\pi\eta}\Gamma(1+\mathrm{i}\eta)\Gamma(1-\mathrm{i}\eta)
= \frac{2\pi\eta}{e^{2\pi\eta}-1}~,\label{E-2.1-16}
\end{equation}
known as the \textit{Sommerfeld factor} \cite{BeS57,HuB59}. Since the scattering eigenfunctions of the
repulsive
Coulomb Hamiltonian form a complete set of wavefunctions, they can be employed in an operatorial
definition of the Coulomb Green's functions analogous to eq.~\eqref{E-2.0-09}, 
\begin{equation}
\hat{G}_{\mathrm{C}}^{(\pm)} = M\int_{\mathbb{R}^3}\frac{\mathrm{d}^3\mathbf{q}}{(2\pi)^3}
\frac{|\psi_{\mathbf{q}}^{(\pm)}\rangle\langle\psi_{\mathbf{q}}^{(\pm)}|}{\mathbf{p}^2-\mathbf{q}^2\pm \mathrm{i}\varepsilon}~.
\label{E-2.1-17}
\end{equation}
In a way fully analogous to the one with which we have defined the Coulomb Green's functions in
eq.~\eqref{E-2.1-11},
we introduce the full Green's function, including both the strong and the electrostatic interactions.
Therefore,
we add the operator $\hat{V}_{\mathrm{S}} \equiv \hat{\mathcal{V}}^{(1)}$ to the kinetic and Coulomb potential in
eq.~\eqref{E-2.1-11}, so that
\begin{equation}
\hat{G}_{\mathrm{SC}}^{(\pm)}(E) = \frac{1}{E-\hat{H}_0-\hat{V}_{\mathrm{C}} - \hat{V}_{\mathrm{S}} \pm \mathrm{i}\varepsilon}~.
\label{E-2.1-18}
\end{equation}
Then, we define the incoming and outcoming wavefunctions as in ref.~\cite{KoR00},
\begin{equation}
|\chi_{\mathrm{p}}^{(\pm)}\rangle = \left[1+\hat{G}_{\mathrm{SC}}^{(\pm)}\left(\hat{V}_{\mathrm{S}}+\hat{V}_{\mathrm{C}}\right)\right]|
\mathrm{p}\rangle~,\label{E-2.1-19}
\end{equation} 
similar to the eq.~\eqref{E-2.1-13}. Exploiting the operator relation $A^{-1}-B^{-1}=B^{-1}(B-A)A^{-1}$ with
$A~=~ \hat{G}_{\mathrm{SC}}^{(\pm)}(E)$ and $B~=~\hat{G}_{\mathrm{C}}^{(\pm)}(E)$ we find the self-consistent
Dyson-like identity
\begin{equation}
\hat{G}_{\mathrm{SC}}^{(\pm)} - \hat{G}_{\mathrm{C}}^{(\pm)} = \hat{G}_C^{(\pm)}\hat{V}_{\mathrm{S}}\hat{G}_{\mathrm{SC}}^{(\pm)}~,
\label{E-2.1-20}
\end{equation}
that permits to rewrite the eigenstates of the full Hamiltonian in terms of the Coulomb states,
\begin{equation}
|\chi_{\mathbf{p}}^{(\pm)}\rangle = \left[1+\sum_{n=1}^{+\infty}(\hat{G}_{\mathrm{C}}^{(\pm)}\hat{V}_{\mathrm{S}})^n\right]|
\psi_{\mathbf{p}}^{(\pm)}\rangle~.\label{E-2.1-21}
\end{equation}
Subsequently, the scattering amplitude can be computed via the S-matrix element, given by the overlap between
an incoming state with momentum $\mathbf{p}$ and an outcoming state $\mathbf{p}'$, 
\begin{equation}
S(\mathbf{p}',\mathbf{p}) = \langle \chi_{\mathbf{p}'}^{(-)} |\chi_{\mathbf{p}}^{(+)} \rangle = (2\pi)^3\delta(\mathbf{p}'
-\mathbf{p}) - 2\pi \mathrm{i}~\delta(E'-E)  T(\mathbf{p}',\mathbf{p})\label{E-2.1-22}
\end{equation}
where $T(\mathbf{p}',\mathbf{p}) = T_{\mathrm{C}}(\mathbf{p}',\mathbf{p}) + T_{\mathrm{SC}}(\mathbf{p}',\mathbf{p})$ as
in eq.~(4) in ref.~\cite{Har65} (for the complete derivation of eq.~\eqref{E-2.1-22} we refer to chap.~5 of
ref.~\cite{GoW64}). In particular $T_{\mathrm{C}}(\mathbf{p}',\mathbf{p}) = \langle \mathbf{p}'|\hat{V}_{\mathrm{C}}|
\psi_{\mathbf{p}}^{(+)}\rangle$ is the pure electrostatic scattering amplitude and $T_{\mathrm{SC}}(\mathbf{p}',\mathbf{p})
= \langle \psi_{\mathbf{p}'}^{(-)}|\hat{V}_{\mathrm{S}}|\chi_{\mathbf{p}}^{(+)}\rangle$ is the strong scattering amplitude
modified by Coulomb corrections. Since the eigenstates $\psi_{\mathbf{p}}$ of the former are known, the scattering
amplitude due only to the Coulomb interaction can be computed in closed form and admits the following partial
wave expansion \cite{KoR00},
\begin{equation}
T_{\mathrm{C}}(\mathbf{p}',\mathbf{p}) = - \frac{4\pi}{M}\sum_{\ell =0}^{+\infty}(2\ell +1)\left[\frac{e^{2\mathrm{i}\sigma_{\ell}}
-1}{2\mathrm{i}|\mathbf{p}|}\right]\mathcal{P}_{\ell}(\cos\theta)~,\label{E-2.1-23}
\end{equation}
where $\theta$ is the angle between $\mathbf{p}$ and $\mathbf{p}'$ and $\sigma_{\ell} = \arg \Gamma (1+\ell
+\mathrm{i}\eta)$ is the Coulomb phase shift. In particular, the strong scattering amplitude $T(\mathbf{p},\mathbf{p}')$
possesses a phase shift $\sigma_{\ell}+\delta_{\ell}$. Furthermore, the Coulomb corrected version of $T_{\mathrm{S}}$
can be expanded in terms of the Legendre polynomials $\mathcal{P}_{\ell}$ as
\begin{equation}
T_{\mathrm{SC}}(\mathbf{p}',\mathbf{p}) = - \frac{4\pi}{M}\sum_{\ell =0}^{+\infty}(2\ell +1)e^{2\mathrm{i}\sigma_{\ell}}
\left[\frac{e^{2\mathrm{i}\delta_{\ell}}-1}{2\mathrm{i}|\mathbf{p}|}\right]\mathcal{P}_{\ell}(\cos\theta)\label{E-2.1-24}
\end{equation}
where $\delta_{\ell}$ is the strong contribution to the total phase shift. After expressing the eigenstates of the
full Hamiltonian in terms of the Coulomb eigenstates (cf. eq.~\eqref{E-2.1-21}), we concentrate on the P-wave amplitude.
Since the strong interaction couples the fermions to one unit of angular momentum and Coulomb
forces are central, the only nonzero component of $T_{\mathrm{SC}}$ of the expansion in eq.~\eqref{E-2.1-24} is the one
with $\ell=1$. Analogously to eq.~(31) in ref.~\cite{KoR00}, we can, thus, write
\begin{equation}
|\mathbf{p}|^3 (\cot\delta_1 - \mathrm{i}) = - \cos\theta \frac{12\pi\mathbf{p}^2}{M}
\frac{e^{2\mathrm{i}\sigma_1}}{T_{\mathrm{SC}}(\mathbf{p}',\mathbf{p})}~,\label{E-2.1-25}
\end{equation}
and we replace the ERE of the l.h.s. of the last equation with the $\ell = 1$ version (cf. ref.~\cite{KMB82})
of the generalized effective-range expansion formulated in ref.~\cite{Bet49} for the repulsive Coulomb interaction,
\begin{equation}
\mathbf{p}^2\left(1+\eta^2\right)\left[C_{\eta}^2|\mathbf{p}|(\cot\delta_1-\mathrm{i}) + \alpha M H(\eta)\right] 
 = -\frac{1}{a_{\mathrm{C}}^{(1)}} +\frac{1}{2}r_0^{(1)}\mathbf{p}^2 + r_1^{(1)}\mathbf{p}^4 + \ldots~,\label{E-2.1-26}
\end{equation}
where $a_{\mathrm{C}}^{(1)}$, $r_0^{(1)}$ and $r_1^{(1)}$ are the scattering length, the effective range and the shape parameter,
respectively. By comparison with the S-wave counterpart in eq.~(32) of ref.~\cite{KoR00}, we can observe that, apart from
the different power of the momentum of the incoming particles in front of the $\cot\delta_1 - \mathrm{i}$ term, the
most significant difference is provided by the polynomial on the l.h.s. of the eq.~\eqref{E-2.1-26}, containing all
the even powers of $\eta$ from zero to $2\ell$, as shown in eq.~(10.10) in ref.~\cite{KMB82}. Besides, the
function $H(\eta)$, that represents the effects of Coulomb force on the strong interactions at short distances,
is given by
\begin{equation}
H(\eta) = \psi(\mathrm{i}\eta)+\frac{1}{2i\eta}-\log(\mathrm{i}\eta)~,\label{E-2.1-27}
\end{equation}
where $\psi(z) = \Gamma'(z)/\Gamma(z)$ id the Digamma function. Despite the appearance, the generalized ERE is
real, since the imaginary parts arising from $H(\eta)$ cancel exactly with the imaginary part in the l.h.s. of
eq.~\eqref{E-2.1-26}. Due to the following identity on the logarithmic derivative of the Gamma function, 
\begin{equation}
\mathfrak{Im} \psi(\mathrm{i}\eta) = \frac{1}{2\eta} + \frac{\pi}{2}\coth\pi\eta~,\label{E-2.1-28}
\end{equation}
in fact, the imaginary part of $H(\eta)$ proves to coincide with $C_{\eta}^2/2\eta$. For the sake of completeness,
in the case of fermion-antifermion scattering the Coulomb potential is attractive and $H(\eta)$ in the effective
range expansion (cf. eq.~\eqref{E-2.1-26}) should be replaced by
\begin{equation}
\bar{H}(\eta) = \psi(\mathrm{i}\eta) + \frac{1}{2\mathrm{i}\eta}-\log(-\mathrm{i}\eta)~,\label{E-2.1-29}
\end{equation}
where $\eta = -\alpha M/2p$ is defined as a negative real parameter.

%--------------------------------------------------------------------------------------------------------------------------------------------------------------------

\subsection{\textsf{Repulsive channel}}\label{S-2.2}

Considering the results of the previous section, all the elements for the derivation of the Coulomb-corrected
strong scattering amplitude, $T_{\mathrm{SC}}(\mathbf{p},\mathbf{p}')$, are available. Recalling the definition
and eq.~\eqref{E-2.1-21}, the amplitude can be computed by evaluating each of the terms in the expansion,
whose insertions are given by retarded Coulomb propagators $\hat{G}_{\mathrm{C}}^{(+)}$ followed by $\ell=1$
four-point vertices, $\hat{V}_{\mathrm{S}}=\hat{\mathcal{V}}^{(1)}$. In particular, the lowest order contribution
to the T-matrix reads
\begin{equation}
\begin{split}
\langle \psi^{(-)}_{\mathbf{p}'}|\hat{\mathcal{V}}^{(1)}| \psi_{\mathbf{p}}^{(+)}\rangle  = \int_{\mathbb{R}^3}\mathrm{d}^3 r'
\int_{\mathbb{R}^3} \mathrm{d}^3 r \langle~\psi^{(-)}_{\mathbf{p}'}| \mathbf{r}'\rangle \langle \mathbf{r}' |
\hat{\mathcal{V}}^{(1)}| \mathbf{r}\rangle \\ \cdot \langle \mathbf{r} | \psi_{\mathbf{p}}^{(+)}\rangle
=  D(E^*)~\nabla' \psi^{(-)*}_{\mathbf{p}'}(\mathbf{r}') \Big\lvert_{\mathbf{r}'=\mathbf{0}} &  \cdot \nabla
\psi_{\mathbf{p}}^{(+)}(\mathbf{r}) \Big\lvert_{\mathbf{r}=\mathbf{0}}~,\label{E-2.2-01}
\end{split}
\end{equation}
where eqs.~\eqref{E-2.0-16} and \eqref{E-2.1-14}-\eqref{E-2.1-15} have been exploited, partial integration for
the two variables has been performed and the vanishing surface terms dropped. The explicit computation of
the two integrals over the free Coulomb wavefunctions (cf. eqs.~\eqref{E-2.1-14}-\eqref{E-2.1-15}) in the last
row is carried out in app.~\ref{S-A-3.0}, and yields
\begin{equation}
D(E^*)~\nabla' \psi^{(-)*}_{\mathbf{p}'}(\mathbf{r}') \Big\lvert_{\mathbf{r}'=\mathbf{0}} \cdot \nabla
\psi_{\mathbf{p}}^{(+)}(\mathbf{r}) \Big\lvert_{\mathbf{r}=\mathbf{0}} = D(E^*)~(1+\eta^2) C_{\eta}^2 e^{\mathrm{i}2\sigma_1}
\mathbf{p}'\cdot\mathbf{p}~,\label{E-2.2-02}
\end{equation}
where $C_\eta^2$ is the Sommerfeld factor, a function of $\eta = \alpha M /2|\mathbf{p}|$. As it can be
observed, the polynomial $1+\eta^2$ in the l.h.s. of the generalized effective range expansion appears,
see eq.~\eqref{E-2.1-26}. As shown in app.~\ref{S-A-3.0}, the P-wave strong vertex \emph{projects out} of
the integral all the components of the Coulomb wavefunctions $\psi_{p}^{(\pm)}$  with $\ell \neq 1$ appearing
in the angular momentum expansion
\begin{equation}
\psi_{\mathbf{p}}^{(+)}(\mathbf{r}) = \frac{4\pi}{|\mathbf{p}| r} \sum_{\ell=0}^{+\infty}\sum_{m=-\ell}^{\ell}
\mathrm{i}^{\ell} e^{i\sigma_{\ell}} Y_{\ell}^{m*}(\hat{\mathbf{p}}) Y_{\ell}^{m}(\hat{\mathbf{r}})
F_{\ell}(\eta,|\mathbf{p}|r)~,\label{E-2.2-03}
\end{equation}
where $\hat{\mathbf{p}}$ and $\hat{\mathbf{r}}$ are unit vectors parallel to $\mathbf{p}$ and $\mathbf{r}
\equiv r~\hat{\mathbf{r}}$ respectively and $F_{\ell}(\eta, |\mathbf{p}|r)$ is the regular free Coulomb wavefunction.
The latter functions, derived by Yost, Wheeler and Breit in ref.~\cite{YWB36}, display a regular behaviour in
the vicinity of the origin, in contrast with the $G_{\ell}(\eta, |\mathbf{p}|r)$, linearly independent solutions
of the Whittaker equation for a repulsive Coulomb potential which are irregular for $r\rightarrow 0$.
Explicitly, $F_{\ell}(\eta, |\mathbf{p}|r)$ has the form given in ref.~\cite{HuB59},
\begin{equation}
F_{\ell}(\eta,|\mathbf{p}|r) = \frac{2^{\ell}e^{-\pi\eta/2}|\Gamma(1+\ell+\mathrm{i}\eta)|}{(2\ell+1)!}  (|\mathbf{p}|r)^{\ell+1} 
e^{\mathrm{i}|\mathbf{p}|r} M(1+\ell+\mathrm{i}\eta, 2\ell+2,-2\mathrm{i} |\mathbf{p}|r)~,\label{E-2.2-04}
\end{equation}
whereas the expansion for the incoming waves is obtained via the complex-conjugation property
$\psi_{\mathbf{p}}^{(-)}(\mathbf{r}) = \psi_{-\mathbf{p}}^{(+)*}(\mathbf{r})$. Next, we proceed with derivation of
the next-to-leading order contribution to the scattering amplitude,
\begin{equation}
\begin{split}
\langle \psi_{\mathbf{p}'}^{(-)}|\hat{\mathcal{V}}_1\hat{G}_{\mathrm{C}}^{(+)}\hat{\mathcal{V}}_1|\psi_{\mathbf{p}}^{(+)} \rangle
=  M[D(E^*)]^2 \int_{\mathbb{R}^3} \mathrm{d}^3 \mathbf{r}'~\delta(\mathbf{r}')  \int_{\mathbb{R}^3} \mathrm{d}^3
\mathbf{r}''~\delta(\mathbf{r}'') \int_{\mathbb{R}^3} \mathrm{d}^3 \mathbf{r}'''~\delta(\mathbf{r}''') \\ \cdot \int_{\mathbb{R}^3}
\mathrm{d}^3 \mathbf{r}''''~\delta(\mathbf{r}'''') \partial_i'\psi_{\mathbf{p}}^{(-)*}(\mathbf{r'})  \partial_i''
\partial_j'''G_{\mathrm{C}}^{(+)}(\mathbf{r}'',\mathbf{r}''') & \partial_j'''' \psi_{\mathbf{p}}^{(+)}(\mathbf{r''''})~,
\label{E-2.2-05}
\end{split}
\end{equation}
where partial integration has been exploited and Einstein summation convention over repeated indices is understood.
More succintly, the last equation can be recast as
\begin{equation}
\begin{split}
\langle \psi_{\mathbf{p}'}^{(-)}|\hat{\mathcal{V}}_1\hat{G}_{\mathrm{C}}^{(+)}\hat{\mathcal{V}}_1|\psi_{\mathbf{p}}^{(+)} \rangle
= M [D(E^*)]^2~\partial'_i\psi_{\mathbf{p}'}^{(-)*}(\mathbf{r}') \Big\lvert_{\mathbf{r}'=0}  \partial_i'' \partial_j'''
G_{\mathrm{C}}^{(+)}(\mathbf{r}'',\mathbf{r}''')\Big\lvert_{\mathbf{r}'',\mathbf{r}'''=0} \\ \cdot \partial'''_i\psi_{\mathbf{p}'}^{(+)}
(\mathbf{r}'''') \Big\lvert_{\mathbf{r}''''=0} \equiv [D(E^*)]^2 \nabla' \psi_{\mathbf{p}'}^{(-)*}(\mathbf{r}') \Big
\lvert_{\mathbf{r}'=0} & \cdot \mathbb{J}_{\mathrm{C}} \nabla\psi_{\mathbf{p}'}^{(+)}(\mathbf{r}) \Big\lvert_{\mathbf{r}=0} ~.
\label{E-2.2-06}
\end{split}
\end{equation}
where in the second row, the Coulomb-corrected counterpart of the $\mathbb{J}_{0}$ matrix defined in eq.~\eqref{E-2.0-14}
has been introduced,
\begin{equation}
(\mathbb{J}_{\mathrm{C}})_{ij} =  \partial_{i}\partial_{j}'G_{\mathrm{C}}^{(+)}(\mathbf{r},\mathbf{r}')|_{\mathbf{r},\mathbf{r}'=0}~.
\label{E-2.2-07}
\end{equation}
Analogously to the $\ell=0$ case, the higher order contributions to the T-matrix possess the same structure of
eqs.~\eqref{E-2.2-02} and~\eqref{E-2.2-06} differ from the latter only in the powers of $\mathbb{J}_{\mathrm{C}}$ and
the coupling constant $D(E^*)$. Therefore we can again write
\begin{equation}
\mathrm{i}T_{\mathrm{SC}}(\mathbf{p}',\mathbf{p}) =\mathrm{i}D(E^*)~\nabla'\psi_{\mathbf{p}'}^{(-)*}(\mathbf{r}') \Big
\lvert_{\mathbf{r}'=0} \left[\mathbbm{1}  + D(E^*)\mathbb{J}_{\mathrm{C}} + D(E^*)^2\mathbb{J}_{\mathrm{C}}^2
+ D(E^*)^3\mathbb{J}_{\mathrm{C}}^3 +\ldots \right] \nabla\psi_{\mathbf{p}}^{(+)}(\mathbf{r}) \Big\lvert_{\mathbf{r}=0}~,
\label{E-2.2-08}
\end{equation}
and we can treat the terms enclosed by the round barckets as a geometric series,
\begin{equation}
T_{\mathrm{SC}}(\mathbf{p}',\mathbf{p}) = \nabla'\psi_{\mathbf{p}'}^{(-)*}(\mathbf{r}') \Big\lvert_{\mathbf{r}'=0} \cdot
\frac{D(E^*)}{\mathbbm{1}-D(E^*)\mathbb{J}_{\mathrm{C}}}\nabla\psi_{\mathbf{p}}^{(+)}(\mathbf{r}) \Big\lvert_{\mathbf{r}=0}~.
\label{E-2.2-09}
\end{equation} 
Recalling the tensor product between vectors and the definition of the Coulomb Green's function operators in
eq.~\eqref{E-2.1-17}, it is convenient to rewrite the overall matrix as
\begin{equation}
\mathbb{J}_{\mathrm{C}}  = M\int_{\mathbb{R}^3} \mathrm{d}^3 r~ \delta(\mathbf{r})  \int_{\mathbb{R}^3} \mathrm{d}^3 \mathbf{r}'
~\delta(\mathbf{r}')  \cdot\int_{\mathbb{R}^3} \frac{\mathrm{d}^3\mathbf{s}}{(2\pi)^3} \frac{\nabla \psi_{\mathbf{s}}^{(+)}
(\mathbf{r})\otimes \nabla' \psi_{\mathbf{s}}^{(+)*}(\mathbf{r'})}{\mathbf{p}^2-\mathbf{s}^2+\mathrm{i}\varepsilon} ~,
\label{E-2.2-10}
\end{equation}
and we observe that the numerator can be considerably simplified by means of the results of app.~\ref{S-A-3.0}.
In particular, eq.~\eqref{E-A-3.0-18} can be applied twice, yielding 
\begin{equation}
\int_{\mathbb{R}^3} \mathrm{d}^3 \mathbf{r}~\delta(\mathbf{r})  \int_{\mathbb{R}^3} \mathrm{d}^3 r'~\delta(\mathbf{r}')\nabla
\psi_{\mathbf{s}}^{(+)}(\mathbf{r})\otimes \nabla' \psi_{\mathbf{s}}^{(+)*}(\mathbf{r}')   = C_{\eta}^2 (1+\eta^2)~ \mathbf{s}
\otimes \mathbf{s}~.\label{E-2.2-11}
\end{equation}
Equipped with the last result together with eq.~\eqref{E-2.1-16}, we recast the components of the $\mathbb{J}_{\mathrm{C}}$
matrix as
\begin{equation}
\begin{split}
(\mathbb{J}_{\mathrm{C}})_{ij} = M\int_{\mathbb{R}^3} \frac{\mathrm{d}^3\mathbf{s}}{(2\pi)^3} \frac{2\pi
\eta(\mathbf{s})~s_i s_j}{e^{2\pi\eta(\mathbf{s})}-1} \frac{1+\eta(\mathbf{s})^2}{\mathbf{p}^2-\mathbf{s}^2+\mathrm{i}
\varepsilon} ~,\label{E-2.2-12}
\end{split}
\end{equation}
where the dependence of $\eta$ on the integrated momentum $\mathbf{s}$ has been made explicit. As the $\ell=0$ counterpart
in eq.~(43) of ref.~\cite{KoR00}, the integral is ultraviolet divergent. Additionally, all the off-diagonal matrix
elements of $\mathbb{J}_{\mathrm{C}}$ vanish, as the integrand is manifestly rotationally symmetric in three
dimensions except for the components $s_i s_j$, that are integrated over a symmetric interval around zero, see eq.~(4.3.4) 
in ref.~\cite{Col84}. In  dimensional regularization, eq.~\eqref{E-2.2-12} can be rewritten as
\begin{equation}
\left(\mathbb{J}_{\mathrm{C}}\right)_{ij}(d)  = M\frac{\delta_{ij}}{d}\int_{\mathbb{R}^d} \frac{\mathrm{d}^ds}{(2\pi)^d}
\frac{2\pi \eta(s)~s^2}{e^{2\pi\eta(s)}-1} \frac{1+\eta(s)^2}{\mathbf{p}^2-\mathbf{s}^2+\mathrm{i}\varepsilon}
\equiv  \mathbbm{j}_{\mathrm{C}}(d)\delta_{ij}~,\label{E-2.2-13}
\end{equation}
an expression that in three dimensions, combined with the results in app.~\ref{S-A-3.0}, allows to simplify
the Coulomb-corrected strong scattering amplitude as 
\begin{equation}
T_{\mathrm{SC}}(\mathbf{p}',\mathbf{p}) = (1+\eta^2) C_{\eta}^2 \frac{D(E^*) ~e^{2\mathrm{i}\sigma_1} \mathbf{p}\cdot\mathbf{p}'}
{1-D(E^*)~\mathbbm{j}_{\mathrm{C}}}~,\label{E-2.2-14}
\end{equation} 
in momentum space. 
Returning to eq.~\eqref{E-2.2-09} and ignoring the Feynman prescription in the denominator, we first exploit the
aforementioned trick and split the integral into three parts,
\begin{equation}
\begin{split}
\mathbbm{j}_{\mathrm{C}}(d) = \frac{M}{d} \int_{\mathbb{R}^d}\frac{\mathrm{d}^ds}{(2\pi)^d} \frac{2\pi\eta}{e^{2\pi\eta}-1}
\frac{\mathbf{p}^2}{\mathbf{s}^2}\frac{\mathbf{p}^2}{\mathbf{p}^2-\mathbf{s}^2}(1+\eta^2) \\ - \frac{M}{d}\int_{\mathbb{R}^d}
\frac{\mathrm{d}^ds}{(2\pi)^d}\frac{2\pi\eta}{e^{2\pi\eta}-1}\frac{\mathbf{p}^2}{\mathbf{s}^2}(1+\eta^2) \\
- \frac{M}{d}\int_{\mathbb{R}^d}\frac{\mathrm{d}^ds}{(2\pi)^d} \frac{2\pi\eta}{e^{2\pi\eta}-1} (1+\eta^2) \\
\equiv \mathbbm{j}_{\mathrm{C}}^{\mathrm{fin}}(d;\mathbf{p}) + \mathbbm{j}_{\mathrm{C}}^{\mathrm{div}, 1}(d;\mathbf{p})
+ \mathbbm{j}_{\mathrm{C}}^{\mathrm{div}, 2}(d;\mathbf{p})~.\label{E-2.2-14}
\end{split}
\end{equation}
While the first one proves to be finite, the other two display a pole for $d\rightarrow 3$ and the PDS regularization
scheme has to be implemented. We begin with integral in the first row of eq.~\eqref{E-2.2-14},
$\mathbbm{j}_{\mathrm{C}}^{\mathrm{fin}}$.  The numerator of the latter can be split into two parts, according
to the terms of the polynomial in $\eta$ inside the round brackets. Taking the limit $d \rightarrow 3$, we
observe that one of the two parts coincides with $J_0^{\mathrm{fin}}$ in eq.~(45) of ref.~\cite{KoR00}, up to a
proportionality constant equal to $\mathbf{p}^2/3$. In comparison with the latter, the other part of
$\mathbbm{j}_{\mathrm{C}}^{\mathrm{fin}}$ in eq.~\eqref{E-2.2-14} is suppressed by two further powers of $\eta$,
therefore it is pairwise UV-finite and the three-dimensional limit finds a justification. After these
manipulations, $\mathbbm{j}_{\mathrm{C}}^{\mathrm{fin}}$ becomes
\begin{equation}
\mathbbm{j}_{\mathrm{C}}^{\mathrm{fin}}(\mathbf{p}) \equiv \lim_{d \rightarrow 3} \mathbbm{j}_{\mathrm{C}}^{\mathrm{fin}}(d;\mathbf{p})
= -H(\eta)\frac{\alpha M^2}{4\pi}\frac{\mathbf{p}^2}{3}  + \frac{M}{d} \int_{\mathbb{R}^d}\frac{\mathrm{d}^ds}{(2\pi)^{d-1}}
\frac{\eta^3}{e^{2\pi\eta}-1}\frac{\mathbf{p}^2}{\mathbf{s}^2}  \frac{\mathbf{p}^2}{\mathbf{p}^2-\mathbf{s}^2} ~.
\label{E-2.2-15}
\end{equation}
Due to spherical symmetry, the integration over the angular variables in the last term can be immediately done.
By performing again the substitution $s \mapsto 2\pi\eta = \pi\alpha M/s$, the second term on the r.h.s.
of eq.~\eqref{E-2.2-15} can be simplified as
\begin{equation}
\frac{M}{3} \int_{\mathbb{R}^3}\frac{\mathrm{d}^3s}{(2\pi)^{2}} \frac{\eta^3}{e^{2\pi\eta}-1}\frac{\mathbf{p}^4}{\mathbf{s}^2}
\frac{1}{\mathbf{p}^2-\mathbf{s}^2} = - \frac{\mathbf{p}^2}{3}\frac{\alpha M^2}{(2\pi)^3}  \left[ \int_0^{+\infty}
\frac{\mathrm{d}x}{x}\frac{x^2}{e^x-1} - \int_0^{+\infty}\mathrm{d}x \frac{x a^2}{(e^x-1)(x^2+a^2)} \right] ~.
\label{E-2.2-16}
\end{equation}
where $a \equiv i\mathrm{}\pi\alpha M/|\mathbf{p}|$. The first of the two integrals on the r.h.s. of eq.~\eqref{E-2.2-16} can be evaluated
by means of the following identity
\begin{equation}
\zeta(\omega)\Gamma(\omega) = \int_0^{+\infty}\frac{\mathrm{d}t}{t}\frac{t^{\omega}}{e^t-1}~,\label{E-2.2-17}
\end{equation}
connecting Euler's Gamma function with Riemann's Zeta function, while the second one in eq.~\eqref{E-2.2-15}
is analogous to the integral in eq.~(46) of ref.~\cite{KoR00}, modulo a constant factor. Considering the last
two identities, eq.~\eqref{E-2.2-16} can be recast into
\begin{equation}
\frac{\mathbf{p}^2}{3}\frac{\alpha M^2}{(2\pi)^3}\Big\{ \zeta(2)\Gamma(2)  - \frac{a^2}{2}\left[\log
\left(\frac{a}{2\pi}\right)-\frac{\pi}{a}-\psi\left(\frac{a}{2\pi}\right)\right]\Big\} 
= \frac{\mathbf{p}^2}{3}\frac{\alpha M^2}{(2\pi)^3}\left[ \frac{\pi^2}{6}  -
\frac{\pi^2\alpha^2M^2}{2\mathbf{p}^2}H(\eta)\right]~, \label{E-2.2-18}
\end{equation}
where the definition of $H(\eta)$  in eq.~\eqref{E-2.1-27} and the fact that $\zeta(2) = \pi^2/6$
have been exploited. The subsequent addition of the last result to the already calculated contribution
to eq.~\eqref{E-2.2-15} yields the sought closed expression for $\mathbbm{j}_{\mathrm{C}}^{\mathrm{fin}}(\mathbf{p})$,
\begin{equation}
\mathbbm{j}_{\mathrm{C}}^{\mathrm{fin}}(\mathbf{p}) = \frac{M}{3} \int_{\mathbb{R}^3}\frac{\mathrm{d}^3s}{(2\pi)^3}
\frac{2\pi\eta}{e^{2\pi\eta}-1}\frac{\mathbf{p}^2}{\mathbf{s}^2}\frac{\mathbf{p}^2}{\mathbf{p}^2-\mathbf{s}^2}(1+\eta^2) 
=  \frac{\alpha M^2}{48\pi}\frac{\mathbf{p}^2}{3} - \frac{\alpha M^2}{4\pi}\frac{\mathbf{p}^2}{3}(1+\eta^2)H(\eta)~.
\label{E-2.2-19}
\end{equation}
Now we focus on the term in the second row of eq.~\eqref{E-2.2-14}. By comparison with the integrand of eq.~(44)
of ref.~\cite{KoR00}, we expect the integral of interest to display an UV singularity. Splitting the polynomial
within the round brackets on the numerator of the integrand, we recognize, in fact, the already
available $J_{0}^{\mathrm{div}}$ in ref.~\cite{KoR00} whose result in the PDS scheme is given in eq.~(53) of the latter
reference,
\begin{equation}
\mathbbm{j}_{\mathrm{C}}^{\mathrm{div}, 1}(d;\mathbf{p}) \equiv  - \frac{M}{d}\int_{\mathbb{R}^d}\frac{\mathrm{d}^ds}{(2\pi)^d}
\frac{2\pi\eta}{e^{2\pi\eta}-1}\frac{\mathbf{p}^2}{\mathbf{s}^2}(1+\eta^2)  = \frac{\mathbf{p}^2}{d}
J^{\mathrm{div}}_{0}(d;\mathbf{p}) - \frac{M}{d}\int_{\mathbb{R}^d}\frac{\mathrm{d}^ds}{(2\pi)^{d-1}} \frac{\eta^3}{e^{2\pi\eta}-1}
 \frac{\mathbf{p}^2}{\mathbf{s}^2}~.\label{E-2.2-20}
\end{equation}
Again, spherical symmetry permits to integrate over the angular variables of the second integral on the r.h.s.
of eq.~\eqref{E-2.2-20} and the substitution $s \mapsto 2\pi\eta = \pi\alpha M/s$ allows for the exploitation of
the integral relation between the Gamma- and the Riemann Zeta function in eq.~\eqref{E-2.2-17}, obtaining 
\begin{equation}
\begin{split}
 - \frac{M}{d}\int_{\mathbb{R}^d}\frac{\mathrm{d}^ds}{(2\pi)^d}\frac{2\pi\eta^3}{e^{2\pi\eta}-1}\frac{\mathbf{p}^2}{\mathbf{s}^2} 
 = -\left(\frac{\mu}{2}\right)^{3-d} \frac{\alpha^{d-2}M^{d-1}\pi^{d/2-4}}{2^{d+1}\Gamma\left(\frac{d}{2}\right)}
 \frac{\mathbf{p}^2}{d}\int_0^{+\infty} \frac{\mathrm{d}x}{x}\frac{x^{5-d}}{e^x-1} \\ = -\left(\frac{\mu}{2}\right)^{3-d}
 \frac{\alpha^{d-2}M^{d-1}\pi^{d/2-4}}{2^{d+1}\Gamma\left(\frac{d}{2}\right)} \frac{\mathbf{p}^2}{d} & \zeta(5-d)\Gamma(5-d)~.
 \label{E-2.2-21}
\end{split}
\end{equation}
Unlike the first term on the r.h.s. of eq.~\eqref{E-2.2-20}, the present integral proves to be
convergent in  three dimensions, since $\zeta(2) = \pi^2/6$ is finite and the argument of the Gamma functions are
positive integers or half-integers. Additionally, no PDS poles are found in the same expression. Therefore,
the limit $d\rightarrow 3$ can be safely taken, yielding
\begin{equation}
\lim_{d\rightarrow 3} - \frac{M}{d}\int_{\mathbb{R}^d}\frac{\mathrm{d}^ds}{(2\pi)^d}\frac{2\pi\eta^3}{e^{2\pi\eta}-1}
\frac{\mathbf{p}^2}{\mathbf{s}^2} = - \frac{\alpha M^2}{16\pi}\frac{\mathbf{p}^2}{9}~.\label{E-2.2-22}
\end{equation}
Plugging the available result in eq.~(53) of ref.~\cite{KoR00}, we can finally write a closed expression
for $\mathbbm{j}_{\mathrm{SC}}^{\mathrm{div}, 1}(\mathbf{p})$ in the PDS regularization scheme,
\begin{equation}
\mathbbm{j}_{\mathrm{C}}^{\mathrm{div}, 1}(\mathbf{p}) = \frac{\alpha M^2}{4\pi}\frac{\mathbf{p}^2}{3}\left[\frac{1}{3-d}
+\log \frac{\mu \sqrt{\pi}}{\alpha M} + \frac{4}{3} - \frac{3}{2}\gamma_E \right] 
- \frac{\mu M}{8\pi}\mathbf{p}^2 -\frac{\alpha M^2}{16\pi}  \frac{\mathbf{p}^2}{9} ~.  \label{E-2.2-23}
\end{equation}
Finally, we concentrate our attention on the term in the third row of eq.~\eqref{E-2.2-14}. From that equation,
we infer that the only difference with respect to integrand of $\mathbbm{j}_{\mathrm{C}}^{\mathrm{div}, 1}$ consists
in the absence of the factor $1/s^2$, which enhances the divergent behaviour of the integral in the $s\rightarrow
+\infty$ limit. Therefore, we expect also this third contribution to $\mathbbm{j}_{\mathrm{C}}$ to be UV divergent.
After splitting the integral as in eq.~\eqref{E-2.2-20}, we obtain
\begin{equation}
\mathbbm{j}_{\mathrm{SC}}^{\mathrm{div}, 2}(d;\mathbf{p}) = - \frac{M}{d}\int_{\mathbb{R}^d}\frac{\mathrm{d}^ds}{(2\pi)^d}
\frac{2\pi\eta}{e^{2\pi\eta}-1} (1+\eta^2)  = - \frac{M}{d}\int_{\mathbb{R}^d}\frac{\mathrm{d}^ds}{(2\pi)^d}
\frac{2\pi\eta}{e^{2\pi\eta}-1} - \frac{M}{d}\int_{\mathbb{R}^d}\frac{\mathrm{d}^ds}{(2\pi)^d}  \frac{2\pi\eta^3}{e^{2\pi\eta}-1}~.
\label{E-2.2-24} 
\end{equation}
Now we focus on the first term on the r.h.s. of the last equation. Rotational invariance allows again for
the integration over the angular variables in $d$ dimensions. Then, change of variables $s\mapsto x\equiv 2\pi\eta$
permits to exploit again the multiplication identity between the Riemann Zeta and the Euler's Gamma functions
(cf. eq.~\eqref{E-2.2-16}). Additionally, thanks to the fundamental properties of the Gamma function and
the definiton of $\epsilon \equiv 3 -d$ we obtain
\begin{equation}
\begin{gathered}
 -\left(\frac{\mu}{2}\right)^{3-d}\frac{M}{d}\frac{2 \pi^{d/2}}{\Gamma\left(\frac{d}{2}\right)}\int_0^{+\infty}
 \frac{\mathrm{d}s~s^{d-1}}{(2\pi)^d}\frac{2\pi\eta}{e^{2\pi\eta}-1} = -\left(\frac{\mu}{2}\right)^{3-d}\frac{M}{d}
 \frac{2 \pi^{d/2}}{\Gamma\left(\frac{d}{2}\right)} \frac{(\alpha M \pi)^d}{(2\pi)^d}\int_0^{+\infty} \frac{\mathrm{d}x}{x}
 \frac{x^{1-d}}{e^x-1} \\ = -\left(\frac{\mu}{2}\right)^{3-d}\frac{\alpha^{d}M^{1+d}\pi^{d/2}}{2^{d-1} \Gamma
 \left(\frac{d}{2}\right) d} \zeta(1-d)\Gamma(1-d) = - \frac{\alpha^3 M^4 \pi^{3/2} (1-\frac{\epsilon}{3})^{-1}}{24
 (1-\frac{\epsilon}{2})(1-\epsilon)}\left(\frac{\mu}{\alpha M\sqrt{\pi}}\right)^{\epsilon}\frac{\zeta(\epsilon -2)
 \Gamma(\epsilon)}{\Gamma\left(\frac{3-\epsilon}{2}\right)}~,\label{E-2.2-25} 
\end{gathered}
\end{equation}
where, in the last step, the Gamma functions and the physical constants have been rewritten in order to highlight
the dependence on the small quantity $\epsilon$.  From the last row of eq.~\eqref{E-2.2-25}, we can infer that,
while the Gamma function has a simple pole for $d\rightarrow 3$, the Riemann Zeta function analytically
continued to the whole complex plane is zero in that limit, since it is evaluated at a negative even integer,
i.e. $\zeta(-2n)=0$ $n\in \mathbb{N}^+$.  Therefore, the fourth expression in eq.~\eqref{E-2.2-25} 
cannot be immediately evaluated in the three-dimensional limit. 
Performing a Taylor expansion of the Zeta function about $-2$, we obtain
\begin{equation}
\zeta(1-d) \equiv \zeta(\epsilon -2) = \zeta(-2) + \zeta'(-2)\epsilon + \mathcal{O}(\epsilon^2) \approx 0
- \frac{\zeta(3)}{4\pi^2}\epsilon~,\label{E-2.2-26}
\end{equation}
where $\zeta(3) \approx 1.20205$ is an irrational number, known as the Ap\'ery constant. Furthermore, also
the expansion of $\Gamma\left(\frac{3-\epsilon}{2}\right)$ about $3/2$ up to first order in $\epsilon$ has
to be taken into account. Combining eq.~\eqref{E-2.2-26} with the Taylor expansion of the physical constants
with exponent $\epsilon$ in the round bracket and the Laurent expansion of the Gamma function,
eq.~\eqref{E-2.2-25} transforms into
\begin{equation}
- \lim_{d \rightarrow 3}  \frac{M}{d}\int_{\mathbb{R}^d}\frac{\mathrm{d}^ds}{(2\pi)^d} \frac{2\pi\eta}{e^{2\pi\eta}-1} 
=  \frac{\alpha^3 M^4}{16\pi} \frac{\zeta(3)}{3}\lim_{\epsilon\rightarrow 0} \frac{ \epsilon\left({\textstyle
\frac{1}{\epsilon}} -\gamma_E\right) \left[1+{\textstyle \frac{\epsilon}{2}}(2 -2\log 2 -\gamma_E)\right]}{(1-\epsilon)
\left(1-{\textstyle \frac{\epsilon}{2}}\right)\left(1-{\textstyle \frac{\epsilon}{3}}\right)} 
=  \frac{\alpha^3 M^4}{16\pi}\frac{\zeta(3)}{3}~,\label{E-2.2-27}
\end{equation}
where negligible terms in $\epsilon$ have been omitted in the intermediate step. As it can be inferred from
eq.~\eqref{E-2.2-27}, the result of the integration becomes finite in the framework of dimensional regularization,
even if the corresponding integral in the first row of eq.~\eqref{E-2.2-25} is divergent for $d=3$ due to the
singularity at $x=0$. Since the original expression in the end of the second row of eq.~\eqref{E-2.2-25} contains a pole at $d=2$
while $\Gamma(1) = 1$ and $\zeta(-1) = -\frac{1}{12}$ in the two-dimensional limit, the PDS correction should
be taken into account. Therefore, the complete application of the PDS scheme into eq.~\eqref{E-2.2-27} gives
\begin{equation}
- \lim_{d \rightarrow 3}  \frac{M}{d}\int_{\mathbb{R}^d}\frac{\mathrm{d}^ds}{(2\pi)^d} \frac{2\pi\eta}{e^{2\pi\eta}-1}
\Big\lvert^{\mathrm{PDS}} =  \frac{\alpha^3 M^4}{16\pi} \frac{\zeta(3)}{3} -\frac{\alpha^2M^3 \pi}{32}\frac{\mu}{3}~.
\label{E-2.2-28}
\end{equation}
Next, we switch to the evaluation of the last term on the r.h.s. of eq.~\eqref{E-2.2-24}. Proceeding exactly as
in eq.~\eqref{E-2.2-25}, we find
\begin{equation}
\begin{gathered}
-\left(\frac{\mu}{2}\right)^{3-d}\frac{M}{d}\frac{2 \pi^{d/2}}{\Gamma\left(\frac{d}{2}\right)}\int_0^{+\infty}
\frac{\mathrm{d}s~s^{d-1}}{(2\pi)^d}\frac{2\pi\eta^3}{e^{2\pi\eta}-1}  =  -\left(\frac{\mu}{2}\right)^{3-d}\frac{M}{d}
\frac{2 \pi^{d/2}}{\Gamma\left(\frac{d}{2}\right)} \frac{(\alpha M \pi)^d}{(2\pi)^{d+2}}\int_0^{+\infty} \frac{\mathrm{d}x}{x}
\frac{x^{3-d}}{e^x-1} \\ = -\left(\frac{\mu}{2}\right)^{3-d}\frac{M}{d}\frac{2 \pi^{d/2}}{\Gamma\left(\frac{d}{2}\right)}
\frac{(\alpha M \pi)^d}{(2\pi)^{d+2}}\zeta(3-d)\Gamma(3-d)  = -\frac{\alpha^3 M^4}{3~2^4 \sqrt{\pi}}
\frac{\zeta(\epsilon)}{1-\frac{\epsilon}{3}} \left(\frac{\mu}{\alpha M\sqrt{\pi}}\right)^{\epsilon}
\frac{\Gamma(\epsilon)}{\Gamma\left(\frac{3-\epsilon}{2}\right)}~.\label{E-2.2-29} 
\end{gathered}
\end{equation}
Differently from the previous case, the Riemann Zeta function is nonzero in the three-dimensional limit and
the only singularity for $\epsilon=0$ belongs to the Gamma function in the numerator of the last row of
eq.~\eqref{E-2.2-29}. Considering the expansions of all the $\epsilon$-dependent functions about zero,
the asymptotic expression for eq.~\eqref{E-2.2-29} is recovered
\begin{equation}
\lim_{d\rightarrow 3} - \frac{M}{d}\int_{\mathbb{R}^d}\frac{\mathrm{d}^ds}{(2\pi)^d}\frac{2\pi\eta^3}{e^{2\pi\eta}-1} 
=  \frac{\alpha^3 M^4}{16 \pi} \frac{1}{3}  \left[ \frac{1}{3-d} -\frac{3}{2}\gamma_E + \frac{4}{3}
+\log \frac{\mu\sqrt{\pi}}{\alpha M}\right]~.\label{E-2.2-30}
\end{equation}
As it can be inferred from eq.~\eqref{E-2.2-29}, also a PDS singularity at $d \rightarrow 2$ is present, since
the Riemann Zeta function displays a simple pole at unit arguments. In particular, the Laurent expansion of
the Zeta function around 1 yields
\begin{equation}
\zeta(3-d) = \zeta(1+2-d)= \frac{1}{2-d} + \gamma_E + \mathcal{O}({\scriptstyle 2-d})~.\label{E-2.2-31}
\end{equation}  
Applying the PDS regularization scheme and subtracting the correction corresponding to the $d=2$ pole,
the expression in eq.~\eqref{E-2.2-30} becomes
\begin{equation}
- \lim_{d \rightarrow 3}  \frac{M}{d}\int_{\mathbb{R}^d}\frac{\mathrm{d}^ds}{(2\pi)^d} \frac{2\pi\eta^3}{e^{2\pi\eta}-1}
\Big\lvert^{\mathrm{PDS}}  = \frac{\alpha^3 M^4}{16 \pi} \frac{1}{3}   \cdot \left[ \frac{1}{3-d} -\frac{3}{2}\gamma_E
+\frac{4}{3} + \log \frac{\mu\sqrt{\pi}}{\alpha M}\right]  -\frac{\alpha^2M^3}{16\pi}\frac{\mu}{2}~.\label{E-2.2-32}
\end{equation}
Thanks to the last expression and eq.~\eqref{E-2.2-32}, a closed form for the third contribution to
the diagonal elements of the $\mathbb{J}_{\mathrm{C}}$ matrix is found,
\begin{equation}
\mathbbm{j}_{\mathrm{C}}^{\mathrm{div},2}(\mathbf{p}) = \frac{\alpha^3 M^4}{16\pi}\frac{\zeta(3)}{3} -
\frac{\alpha^2M^3 \pi}{32}\frac{\mu}{3}  + \frac{\alpha^3 M^4}{16 \pi} \frac{1}{3} \left[ \frac{1}{3-d}
-\frac{3}{2}\gamma_E + \frac{4}{3} +\log \frac{\mu\sqrt{\pi}}{\alpha M}\right]  -\frac{\alpha^2M^3}{16\pi}\frac{\mu}{2}~.
\label{E-2.2-33} 
\end{equation}
Finally, collecting the three results in eqs.~\eqref{E-2.2-19}, ~\eqref{E-2.2-23} and~\eqref{E-2.2-33}, the latter
matrix elements are obtained
\begin{equation}
\begin{gathered}
\mathbbm{j}_{\mathrm{C}}(\mathbf{p}) = \mathbbm{j}_{\mathrm{C}}^{\mathrm{fin}}(\mathbf{p})+
\mathbbm{j}_{\mathrm{C}}^{\mathrm{div},1}(\mathbf{p})+\mathbbm{j}_{\mathrm{C}}^{\mathrm{div},2}(\mathbf{p}) 
= \frac{\alpha^3M^4}{48 \pi}\left[ \frac{1}{3-d}+\zeta(3) -\frac{3}{2}\gamma_E  + \frac{4}{3}
+ \log\frac{\mu\sqrt{\pi}}{\alpha M} \right] \\
+ \frac{\alpha M^2}{4\pi}\frac{\mathbf{p}^2}{3}\left[\frac{1}{3-d}
+ \frac{4}{3} - \frac{3}{2}\gamma_E + \log \frac{\mu\sqrt{\pi}}{\alpha M}\right] 
- \frac{\alpha^2M^3}{32\pi}\frac{\mu}{3}\left(\pi^2-3\right) -\frac{\mu M}{4\pi}\frac{\mathbf{p}^2}{2}
-  \frac{\alpha M^2}{4\pi} \frac{\mathbf{p}^2}{3}H(\eta) (1+\eta^2)~.\label{E-2.2-34} 
\end{gathered}
\end{equation}
A direct comparison with the $\ell=0$ counterpart of the last expression, eqs.~(47) and (53) in ref.~\cite{KoR00},
shows that the QED contributions to $\mathbbm{j}_{\mathrm{C}}$ include terms of higher order in the fine-structure
constant $\alpha$. Moreover, owing to the elements $\mathbbm{j}_{\mathrm{C}}^{\mathrm{fin}}$ and
$\mathbbm{j}_{\mathrm{C}}^{\mathrm{div},1}$, an explicit dependence on the momenta of the incoming fermions
$\pm \mathbf{p}$ outside $H(\eta)$ appears. Since $\mathbbm{j}_{\mathrm{C}}$ contains quadratic terms in
$\mathbf{p}$, eq.~\eqref{E-2.2-34} gives rise to a non-zero value for the effective range parameter
$r_0^{(1)}$ in the effective range expansion formula in eq.~\eqref{E-2.1-26}. Combining the $\ell=1$
component of the T-matrix expansion in terms of Legendre polynomials in eq.~\eqref{E-2.1-24} with eq.~\eqref{E-2.2-14},
an expression for $|\mathbf{p}|^3 (\cot\delta_1-i)$  can be found,
\begin{equation}
|\mathbf{p}|^3 (\cot\delta_1-\mathrm{i}) = -\frac{12\pi}{M}\frac{1 - D(E^*)\mathbbm{j}_{\mathrm{C}}(\mathbf{p})}{D(E^*)~
C_{\eta}^2(1+\eta^2)}~.\label{E-2.2-35} 
\end{equation}
Plugging the last expression into the $\ell=1$ generalized ERE formula, the term of eq.~\eqref{E-2.2-34}
proportional to $H(\eta)$ cancels out with its counterpart in eq.~\eqref{E-2.1-26}, and all the momentum-independent
contributions can be collected, yielding the expression for the Coulomb-corrected $\ell=1$ scattering length,
\begin{equation}
\frac{1}{a_{\mathrm{C}}^{(1)}} = \frac{12\pi}{M D(E^*)} + \frac{\alpha^2M^2\mu}{8}\left(\pi^2-3\right) 
- \frac{\alpha^3 M^3}{4}\left[ \frac{1}{3-d} +\zeta(3) -\frac{3}{2}\gamma_E 
+ \frac{4}{3} +\log \frac{\mu\sqrt{\pi}}{\alpha M}\right] ~.\label{E-2.2-36}
\end{equation}
which represents the measured P-wave fermion-fermion scattering length. As in the $\ell=0$ case, the ultraviolet
pole is expected to be removed by counterterms which describe short-distance electromagnetic and other
isospin-breaking interactions due to the differences between the quark masses \cite{EpM99}. The subsidiary
terms will transform the coupling constant $D(E^*)$ into a renormalization mass dependent coefficient, $D(E^*,\mu)$,
which allows for a redefinition of the scattering length as in eq.~(55) of ref.~\cite{KoR00},
\begin{equation}
\frac{1}{a_{\mathrm{C}}^{(1)}(\mu)} = \frac{12\pi}{M D(E^*,\mu)} + \frac{\alpha^2 M^2\mu}{8}\left(\pi^2-3\right)~.
\label{E-2.2-37}
\end{equation}
The latter quantity is non-measurable and depends on the renormalization point $\mu$, related to the physical
scattering length through the relation
\begin{equation}
\frac{1}{a_{\mathrm{C}}^{(1)}(\mu)} = \frac{1}{a_{\mathrm{C}}^{(1)}} + \frac{\alpha^3 M^3}{4}\left[\zeta(3)
-\frac{3}{2}\gamma_E +\frac{4}{3} + \log\frac{\mu\sqrt{\pi}}{\alpha M}\right]~,\label{E-2.2-38} 
\end{equation}
which is the $\ell=1$ counterpart of eq.~(56) in ref.~\cite{KoR00}. Besides, grouping the quadratic terms
in the momentum of the fermions arising in the l.h.s. of eq.~\eqref{E-2.1-26}, an expression for the effective range
is recovered, 
\begin{equation}
r_0^{(1)} = \alpha M \left[ \frac{2}{3-d} + \frac{8}{3} -3\gamma_E + 2\log\frac{\mu\sqrt{\pi}}{\alpha M} \right] -3\mu~,
\label{E-2.2-39} 
\end{equation}
As in the case of the inverse of the scattering length in eq.~\eqref{E-2.2-38}, $r_0^{(1)}$ possesses a simple pole
at $d=3$. Now the energy-dependent coefficient of our P-wave interaction $D(E^*)$ is replaced by $D_0$, the
singularity can be removed by means of counterterms coming from the $\mathbf{p}^2$-dependent $\ell=0$ interactions,
proportional to $(\psi \overleftrightarrow{\nabla}^2 \psi)^{\dagger} \psi \overleftrightarrow{\nabla}^2 \psi$ in
momentum space.  These interactions correspond to the term with coefficient $C_2$ of the potential in eq.~\eqref{E-2.0-02}
in momentum space and yield the leading contribution to the effective range in the low-momentum regime
when only zero-angular-momentum interactions are present. Despite the difference in the $\mathrm{SO(3)}$
transformation properties induced by the interaction, both the Lagrangian density with $\ell=0$
(cf. eq.~\eqref{E-2.0-02}) interactions and the one with $\ell=1$ (cf. eq.~\eqref{E-2.0-03}) potentials give
rise to a scattering amplitude $T_{\mathrm{S}}(\mathbf{p},\mathbf{p}')$ whose $|\mathbf{p}|^{2\ell+1} \cdot
(\cot\delta_{\ell}~-~\mathrm{i})$ factor leads to a vanishing effective range. As soon as the Coulomb interaction
is included in the Lagrangian, when the potential couples the fermions to one unit of angular momentum, a purely
electrostatic non-zero effective range emerges, in contrast with the $\ell=0$ case, see sec.~3.3 in ref.~\cite{KoR00}.
Therefore, we shall expect that, for higher angular momentum interactions further coefficients in the
generalized expansion of $|\mathbf{p}|^{2\ell+1}\cot\delta_{\ell}$ in even powers of the momentum of the fermions in the
CoM frame become non-zero when the colliding particles are allowed to exchange Coulomb photons.

%--------------------------------------------------------------------------------------------------------------------------------------------------------------------

\subsection{\textsf{Attractive channel}}\label{S-2.3}

We consider the scattering of two non-relativistic fermions with opposite charges, such as fermion-antifermion
pairs. Concerning  elastic scattering, the continuum eigenstates are again represented by the spherical wave solutions in
eqs.~\eqref{E-2.1-14}-\eqref{E-2.1-15}, with $\eta$ now given by $-\alpha M/|\mathbf{p}|$. Besides, the
phenomenology of the scattering process is now enriched by the presence of bound states. In addition, annihilation
is possible, but this will not be considered here. The Coulomb Green's
function, in fact, is enriched by discrete states, $\phi_{n,\ell,m}(\mathbf{r})$, corresponding to bound states
with principal quantum number $n \ge 1$ and rotation group labels given by $(\ell, m)$,
\begin{equation}
\langle \mathbf{r}' | \hat{G}_{\mathrm{C}}^{(\pm)}|\mathbf{r}\rangle = \sum_{n=1}^{+\infty}\sum_{\ell=0}^{+\infty}
\sum_{m=-\ell}^{\ell} \frac{\phi_{n,\ell,m}(\mathbf{r}') \phi^*_{n,\ell,m}(\mathbf{r}')}{E-E_{n}} 
+ \int_{\mathbb{R}^3}\frac{\mathrm{d}^3\mathbf{s}}{(2\pi)^3}\frac{\psi_{\mathbf{s}}^{(\pm)}(\mathbf{r}')
\psi_{\mathbf{s}}^{(\pm) *}(\mathbf{r})}{E-E_s\pm \mathrm{i}\varepsilon}~,\label{E-2.3-01} 
\end{equation}
where $E_{\mathbf{s}}$ is equal to $\alpha^2 M/4\eta^2$ and the bound state eigenvalues, $E_n$, are given by
Bohr's formula for a system with reduced mass equal to $M/2$,
\begin{equation}
E_n = - \frac{\alpha^2 M}{4n^2}~,\label{E-2.3-02} 
\end{equation}
in natural units. As in the previous case, the Coulomb-corrected strong scattering amplitude of the
elastic scattering process in configuration space takes the form
\begin{equation}
\overbar{T}_{\mathrm{SC}}(\mathbf{p}',\mathbf{p}) = \nabla'\psi_{\mathbf{p}'}^{(-)*}(\mathbf{r}')\Big\lvert_{\mathbf{r}'=0}
\cdot \frac{\overbar{D}(E^*)}{\mathbbm{1}-\overbar{D}(E^*)\overbar{\mathbb{J}}_{\mathrm{C}}} \nabla
\psi_{\mathbf{p}}^{(+)}(\mathbf{r})\Big\lvert_{\mathbf{r}=0}~,\label{E-2.3-03} 
\end{equation}
where $\overbar{D}(E^*)$ is the strong P-wave coupling constant in presence of attractive electrostatic
interaction and the matrix $\overbar{\mathbb{J}}_{\mathrm{C}}$ is, now, given by 
\begin{equation}
\overbar{\mathbb{J}}_{\mathrm{C}} = \overbar{\mathbb{J}}_{\mathrm{C}}^{\mathrm{d}} + \overbar{\mathbb{J}}_{\mathrm{C}}^{\mathrm{c}}~,
\label{E-2.3-04}
\end{equation}
which corresponds to the addition of the contributions from discrete and continuum states,
\begin{equation}
\overbar{\mathbb{J}}_{\mathrm{C}}^{\mathrm{d}} = \sum_{n=1}^{+\infty}\sum_{\ell=0}^{+\infty}\sum_{m=-\ell}^{+\ell}
\int_{\mathrm{R}^3}\mathrm{d}^3r'~\delta(\mathbf{r}') \int_{\mathrm{R}^3}\mathrm{d}^3r~\delta(\mathbf{r}) 
\frac{\nabla'\phi_{n,\ell,m}(\mathbf{r}') \otimes \nabla\phi_{n,\ell,m}^*(\mathbf{r})}{E-E_{n}+\mathrm{i}\varepsilon}~,
\label{E-2.3-05}
\end{equation}
 and
\begin{equation}
\overbar{\mathbb{J}}_{\mathrm{C}}^{\mathrm{c}} =  \int_{\mathrm{R}^3}\mathrm{d}^3r'~\delta(\mathbf{r}') \int_{\mathrm{R}^3}
\mathrm{d}^3r~\delta(\mathbf{r})\int_{\mathrm{R}^3} \frac{\mathrm{d}^3\mathbf{s}}{(2\pi)^3} 
\frac{\nabla'\psi_{\mathbf{s}}^{(+)}(\mathbf{r}') \otimes \nabla\psi_{\mathbf{s}}^{(+)*}(\mathbf{r})}{E-E_{\mathbf{s}}
+\mathrm{i}\varepsilon}~,\label{E-2.3-06}
\end{equation}
respectively. Let us start by evaluating the term $\overbar{\mathbb{J}}_{\mathrm{C}}^{\mathrm{d}}$. With reference
to the expression of the eigenfunctions belonging to the discrete spectrum,
\begin{equation}
\phi_{n,\ell,m}(\mathbf{r}) = \sqrt{\left(\frac{\alpha M}{n}\right)^3 \frac{n-\ell-1!}{n+\ell!~2n}} 
 e^{-\frac{\alpha M}{2n}r} \textstyle{\left(\frac{\alpha M r}{n}\right)^{\ell}} L_{n-\ell-1}^{2\ell+1}
\textstyle{\left(\frac{\alpha M}{n} r \right)}  Y_{\ell}^{m}(\theta,\varphi)~,\label{E-2.3-07}
\end{equation}
where $L_{k}^{n}(x)$ are the associated Laguerre polynomials, we first evaluate the integrals containing the gradient
of the latter in the expression for $\bar{\mathbb{J}}_{\mathrm{C}}^{\mathrm{d}}$ in eq.~\eqref{E-2.3-05}, that can be
performed separately for each of the wavefunctions, since the denominator does not depend on the coordinates.
The application of the gradient on the bound state wavefunctions, $\nabla\phi_{n,\ell,m}(\mathbf{r})
\Big\lvert_{\mathbf{r}=0}$, yields
\begin{equation}
\begin{split}
- \left(\frac{\alpha M}{n}\right)^{\frac{3}{2}} \sqrt{\frac{n-\ell-1!}{n+\ell!~2n}}\int_0^{+\infty}\mathrm{d}r
~\frac{\delta(r)}{4\pi} \Big\{\nabla\left[e^{-\frac{\alpha M}{2n}r}\textstyle{\left(\frac{\alpha M}{n} r
\right)^{\ell}}L_{n-\ell-1}^{2\ell+1}\textstyle{\left(\frac{\alpha M}{n} r \right)}\right] \int_{\partial S^2}
\mathrm{d}\Omega~  Y_{\ell}^{m}(\Omega)  \\ + \frac{e^{-\frac{\alpha M}{2n}r}}{r} \textstyle{\left(\frac{\alpha M}{n}
r\right)^{\ell}} L_{n-\ell-1}^{2\ell+1} \textstyle{\left(\frac{\alpha M}{n} r \right)}  \int_{\partial S^2}
\mathrm{d}\Omega~\nabla & Y_{\ell}^{m}(\Omega) \Big\}~,\label{E-2.3-08}
\end{split}
\end{equation}
where the spherical symmetry of the Dirac delta has been exploited. Of the latter equation, we consider
now the first term on the right hand side. Firstly, expressing the radius vector componentwise as a spherical
tensor of rank 1 (cf. eq.~(5.24) and sec.~5.1 in ref.~\cite{MaG96}), the aforementioned part of
eq.~\eqref{E-2.3-08} becomes 
\begin{equation}
\begin{split}
- \frac{1}{\sqrt{4\pi}} \left(\frac{\alpha M}{n}\right)^{\frac{3}{2}} \sqrt{\frac{n-\ell-1!}{n+\ell!~2n}} \sum_{\mu = -1}^{1}
\int_0^{+\infty}\mathrm{d}r~\delta(r)~\frac{e^{-\frac{\alpha M}{2}r}}{r} \textstyle{\left(\frac{\alpha M}{n}
r\right)^{\ell}} \Big\{\left[\ell - \frac{\alpha M}{2n} r \right]L_{n-\ell-1}^{2\ell+1}\textstyle{\left(\frac{\alpha M}{n}
r \right)} \\  -\frac{\alpha M}{n} r L^{2\ell+2}_{n-\ell-2}\textstyle{\left(\frac{\alpha M}{n} r \right)}\Big\} 
\int_0^{2\pi}\mathrm{d}\varphi \int_0^{\pi}\mathrm{d}\theta \sin\theta~(110|\mu-\mu 0)   Y_1^{\mu}(\theta,\varphi)
Y_{\ell}^{m}(\theta,\varphi) & ~\mathbf{e}_{-\mu}~.\label{E-2.3-09}
\end{split}
\end{equation}
Now, recalling the expression of the constant term of the associated Laguerre polynomials, 
\begin{equation}
L_{m}^{k}(0) = \frac{m + k!}{m!~k!}~,\label{E-2.3-10}
\end{equation}
eq.~\eqref{E-2.3-09} can be concisely recast into
\begin{equation}
- \frac{1}{\sqrt{4\pi}}\frac{\alpha M}{n}\left(\frac{\alpha M}{n}\right)^{\frac{3}{2}} \sqrt{\frac{n-\ell-1!}{n+\ell!~2n}}
\frac{n+1!}{n-2!~3!}  (110|-mm0) (-1)^{m}  \delta_{\ell 1}~\mathbf{e}_m~,\label{E-3.2.3-11}
\end{equation}
where the integration over the angular variables $\Omega$ has been performed. After replacing the Clebsch-Gordan
coefficient $(110|-m m 0)$ with $(-1)^{m+1}/\sqrt{3}$, and performing few manipulations, the sought expression
is recovered, 
\begin{equation}
\frac{1}{12}\left(\frac{\alpha^5 M^5}{6\pi n^5}\right)^{\frac{1}{2}}\sqrt{(n+1)(n-1)}~\delta_{\ell 1}~\mathbf{e}_m~.
\label{E-2.3-12}
\end{equation}
Concerning the second term on the r.h.s. of eq.~\eqref{E-2.3-08}, the rewriting of the gradient of a spherical
harmonic into linear combination of spherical tensors (cf. eqs.~(5.24) and (5.27) in ref.~\cite{MaG96}) gives
\begin{equation}
\begin{gathered}
\frac{1}{\sqrt{4\pi}} \left(\frac{\alpha M}{n}\right)^{3/2} \sqrt{\frac{n-\ell-1!}{n+\ell!~2n}}\int_0^{+\infty}\mathrm{d}r~
\delta(r) \frac{e^{-\frac{\alpha M}{2n}r}}{r}  \textstyle{\left(\frac{\alpha M}{n} r \right)}^{\ell} L_{n-\ell-1}^{2\ell+1}
\textstyle{\left(\frac{\alpha M}{n} r \right)} \\ \cdot \sum_{\mu,\mu'} \int_{0}^{2\pi}\mathrm{d}\varphi \int_{0}^{\pi}\mathrm{d}
\theta~\sin\theta 
\cdot \left[\sqrt{\frac{\ell(\ell + 1)^2}{2\ell+1}} (\ell-1~1~\ell|\mu~\mu'~m) Y_{\ell-1}^{\mu}(\theta,\varphi)
Y_{0}^{0*}(\theta,\varphi)~\mathbf{e}_{\mu'}  \right. \\ \left. + \sqrt{\frac{\ell^2(\ell+1)}{2\ell+1}}
(\ell+1~1~\ell|\mu~\mu'~m)  Y_{\ell+1}^{\mu}(\theta,\varphi)Y_0^{0*}(\theta,\varphi)~\mathbf{e}_{\mu'}\right]~,
\label{E-2.3-13}
\end{gathered}
\end{equation}
thus, allowing again for an immediate integration over the angular variables,
\begin{equation}
\frac{1}{\sqrt{4\pi}} \left(\frac{\alpha M}{n}\right)^{3/2} \sqrt{\frac{n-\ell-1!}{n+\ell!~2n}} \frac{n+1!}{n-2!} 
 \frac{\alpha M}{6 n} \frac{2}{\sqrt{3}}\sum_{\mu} (011|0\mu m)  \delta_{\ell 1}~\mathbf{e}_{\mu}~,\label{E-2.3-14}
\end{equation}
where the eq.~\eqref{E-2.3-10} for the evaluation of the Laguerre polynomials at the origin has been exploited.
Subsequently, the replacement $(011|0 m m) = 1$ gives the desired expression for the second term of eq.~\eqref{E-2.3-08}, 
\begin{equation}
\frac{1}{6}\left(\frac{\alpha^5 M^5}{6\pi n^5}\right)^{\frac{1}{2}} \sqrt{(n+1)(n-1)} \delta_{\ell 1}~\mathbf{e}_m~.
\label{E-2.3-15}
\end{equation}
Equipped with the results in eqs.~\eqref{E-2.3-12} and \eqref{E-2.3-15}, the original integral can be immediately evaluated,  
\begin{equation}
\int_{\mathbb{R}^3} \mathrm{d}^3r~\delta(\mathbf{r}) \nabla \phi_{n,\ell,m}(\mathbf{r}) = \frac{1}{4}\left(
\frac{\alpha^5 M^5}{6\pi n^5}\right)^{\frac{1}{2}}  \sqrt{(n+1)(n-1)}\delta_{\ell 1}~ \mathbf{e}_m~.\label{E-2.3-16}
\end{equation}
Now, taking the tensor product of the latter expression with its complex-conjugate version, as required by
eq.~\eqref{E-2.3-05}, $\overbar{\mathbb{J}}_{\mathrm{C}}^{\mathrm{d}}$ reduces to 
\begin{equation}
\overbar{\mathbb{J}}_{\mathrm{C}}^{\mathrm{d}} = \sum_{n=1}^{+\infty} \sum_{m=-1}^1 \frac{\alpha^5 M^5}{16}
\frac{(n+1)(n-1)}{6\pi n^5} \frac{\mathbf{e}_m \otimes \mathbf{e}_m^*}{E-E_n} = \frac{\alpha^3 M^4}{24\pi}
\sum_{n=1}^{+\infty} \frac{\eta^2 (n+1)(n-1)}{n^3(n^2+\eta^2)}\sum_{m=-1}^1 \mathbf{e}_{m}\otimes \mathbf{e}_{m}^*~.
\label{E-2.3-17}
\end{equation}
Since the diagonal form of the matrix in the spherical complex basis (cf. eq.~(2.141) in ref.~\cite{MaG96}) is
preserved in the Cartesian basis %,
%\begin{equation}
%\sum_{\mu=-1}^1 \mathbf{e}_{\mu}\otimes \mathbf{e}_{\mu}^* = \mathbf{e}_x \otimes  \mathbf{e}_x + \mathbf{e}_y 
%\otimes  \mathbf{e}_y + \mathbf{e}_z \otimes  \mathbf{e}_z~,\label{E-2.3-17bis}
%\end{equation}
and the sum over the principal quantum number can be decomposed and evaluated
in terms of the Digamma function $\psi(z)$, 
\begin{equation}
\sum_{n=1}^{+\infty} \frac{\eta^2}{n(n^2+\eta^2)} = \zeta(1) + \frac{1}{2}\psi(\mathrm{i}\eta) + \frac{1}{2}
\psi(-\mathrm{i}\eta)~,\label{E-2.3-18}
\end{equation}
and
\begin{equation}
\sum_{n=1}^{+\infty} \frac{\eta^2}{n^3(n^2+\eta^2)} = -\frac{\zeta(1)}{\eta^2} + \zeta(3) - \frac{1}{2\eta^2}
\psi(-\mathrm{i}\eta) - \frac{1}{2\eta^2}\psi(\mathrm{i}\eta) ~,\label{E-2.3-19}
\end{equation}
the contribution to the scattering matrix due to the discrete states,
\begin{equation}
\overbar{\mathbb{J}}_{\mathrm{C}}^{\mathrm{d}} \equiv  \overbar{\mathbbm{j}}_{\mathrm{C}}^{\mathrm{d}}(\mathbf{p})~\mathbbm{1}~,
\label{E-2.3-20}
\end{equation}
can be ultimately rewritten as 
\begin{equation}
\overbar{\mathbbm{j}}_{\mathrm{C}}^{\mathrm{d}}(\mathbf{p}) = (1+\eta^2)\frac{\alpha M^2}{3\pi} \frac{\mathbf{p}^2}{2}
\left[\zeta(1) +\frac{1}{2}\psi(-\mathrm{i}\eta)  +\frac{1}{2}\psi(\mathrm{i}\eta)\right] -  \frac{\alpha^3M^4}{24\pi}
\zeta(3)  ~.\label{E-2.3-21}
\end{equation}
As underlined in sec.~3.4 of ref.~\cite{KoR00}, the divergent sum of the harmonic series, $\zeta(1)$, appears in
the last formula. Its presence is only due to the numerable infinity of states in the discrete spectrum,
whose energy depends on the inverse square of $n$, while the modulus square of the gradient of the
eigenfunctions evaluated at the origin yields a factor $ \propto n^2$. The replacement of $\zeta(1)$ in
eq.~\eqref{E-2.3-21} by its Cauchy principal value, equal to $\gamma_E$, allows to assign a finite value
to $\overbar{\mathbbm{j}}_{\mathrm{C}}^{\mathrm{d}}$ and, thus, circumvent the divergence.

At this stage, we switch to the continuous contribution to the auxiliary scattering matrix,
$\overbar{\mathbb{J}}_{\mathrm{C}}^{\mathrm{c}}$. As for the repulsive counterpart in sec.~\ref{S-2.2}, the possible
divergences in the three-dimensional limit require the rewriting of the relevant intergrals in arbitrary complex
dimension $d$. Therefore, the dimensionally regularized version of the second term on the r.h.s. of eq.~\eqref{E-2.3-04}
gives
\begin{equation}
\overbar{\mathbb{J}}_{\mathrm{C}}^{\mathrm{c}}(d) = \mathbbm{1} \frac{M}{d}\int_{\mathbb{R}^d}\frac{\mathrm{d}^ds}{(2\pi)^d}
\frac{-2\pi\eta}{e^{-2\pi\eta}-1}\frac{1+\eta^2}{\mathbf{p}^2-\mathbf{s}^2+\mathrm{i}\varepsilon}\mathbf{s}^2 
+ \mathbbm{1}\frac{M}{d}\int_{\mathbb{R}^d} \frac{d^ds}{(2\pi)^d} \frac{-2\pi\eta}{\mathbf{s}^{-2}}\frac{1+\eta^2}{\mathbf{p}^2
-\mathbf{s}^2+\mathrm{i}\varepsilon}~,\label{E-2.3-22}
\end{equation}
where the initial integral has been split into two parts, making use of the trick in eq.~\eqref{E-2.2-14}.
Due to the sign change in $\eta$, the first term on the r.h.s. of eq.~\eqref{E-2.3-22}, 
\begin{equation}
\overbar{\mathbbm{j}}_{\mathrm{C}}^{\mathrm{old}}(d;\mathbf{p}) = \frac{M}{d}\int_{\mathbb{R}^d}\frac{\mathrm{d}^ds}{(2\pi)^d}
\frac{-2\pi\eta}{e^{-2\pi\eta}-1}\frac{1+\eta^2}{\mathbf{p}^2-\mathbf{s}^2+\mathrm{i}\varepsilon}\mathbf{s}^2~,\label{E-2.3-23}
\end{equation}
can be immediately evaluated, since it coincides with eq.~\eqref{E-2.2-12}. Therefore the result in eq.~\eqref{E-2.2-34}
can be directly exported, rewriting eq.~\eqref{E-2.3-23} as
\begin{equation}
\begin{split}
\overbar{\mathbbm{j}}_{\mathrm{C}}^{\mathrm{old}}(\mathbf{p}) = \frac{\alpha^3M^4}{48\pi}\left[\frac{1}{3-d} + \zeta(3)
- \frac{3}{2}\gamma_E + \frac{4}{3} + \log \frac{\mu\sqrt{\pi}}{\alpha M}\right]  - (1+\eta^2) \frac{\alpha M^2}{4\pi}
\frac{\mathbf{p}^2}{3} \left[\psi(-\mathrm{i}\eta) -\frac{1}{2\mathrm{i}\eta} \right. \\ \left. - \log (-\mathrm{i}\eta) \frac{}{} \right] 
+ \frac{\alpha M^2}{4\pi}\frac{\mathbf{p}^2}{3}\left[\frac{1}{3-d} + \frac{4}{3} - \frac{3}{2}\gamma_E
+ \log\frac{\mu\sqrt{\pi}}{\alpha M}\right]    
- \frac{\alpha^2 M^3}{32\pi}\frac{\mu}{3}(\pi^2-3) - \frac{\mu M}{4\pi} \frac{\mathbf{p}^2}{2} ~,\label{E-2.3-24}
\end{split}
\end{equation}
where the $H(-\eta)$ function has been replaced by its definition in terms of the Digamma function in
eq.~\eqref{E-2.1-27}, in sight of the next developments. Subsequently, we evaluate the second term on the
r.h.s. of eq.~\eqref{E-2.3-22}, the new part of the continuum states contribution. In order to bring $\mathbf{s}^2$
to the denominator, we apply again the trick introduced in eq.~\eqref{E-2.2-14} and split the integral into three parts, 
\begin{equation}
\begin{split}
\overbar{\mathbbm{j}}_{\mathrm{C}}^{\mathrm{new}}(d;\mathbf{p}) = \frac{M}{d}\int_{\mathbb{R}^d} \frac{d^ds}{(2\pi)^d}
\frac{-2\pi\eta}{\mathbf{s}^{-2}}\frac{1+\eta^2}{\mathbf{p}^2-\mathbf{s}^2+\mathrm{i}\varepsilon} 
= -\frac{M}{d}\int_{\mathbb{R}^d}\frac{\mathrm{d}^ds}{(2\pi)^d} (-2\pi\eta) \\ +  \frac{M}{d}\int_{\mathbb{R}^d}
\frac{\mathrm{d}^ds}{(2\pi)^d}\frac{-2\pi\eta\mathbf{p}^2}{ \mathbf{p}^2 - \mathbf{s}^2} 
+ \frac{\alpha^2 M^3}{4 d}  \int_{\mathbb{R}^d}\frac{\mathrm{d}^ds}{(2\pi)^d} & \frac{-2\pi\eta}{\mathbf{p}^2-\mathbf{s}^2}~.
\label{E-2.3-25}
\end{split}
\end{equation}
Concerning the first term on the r.h.s. of the latter equation, it vanishes in dimensional regularization,
see eq.~(4.3.1a) in ref.~\cite{Col84}. Therefore, we can switch to the subsequent term of eq.~\eqref{E-2.3-25}
and apply  Feynman's trick for denominators, finding
\begin{equation}
\frac{M}{d}\int_{\mathbb{R}^d}\frac{\mathrm{d}^ds}{(2\pi)^d}\frac{-2\pi\eta\mathbf{p}^2}{ \mathbf{p}^2 - \mathbf{s}^2}
=  -\frac{\alpha M^2}{d} \frac{\Gamma\textstyle{\left(\frac{3}{2}\right)}}{\Gamma(1)\Gamma\textstyle{\left(
\frac{1}{2}\right)}}  \int_0^1 \mathrm{d}\omega~\omega^{-1/2} \int_{\mathbb{R}^d}\frac{\mathrm{d}^d\mathbf{s}}{(2\pi)^d}
\frac{\pi \mathbf{p}^2}{\left[\mathbf{s}^2 - (1-\omega)\mathbf{p}^2\right]^{3/2}}~,\label{E-2.3-26}
\end{equation}
Defining again the auxiliary variable $\gamma = -\mathrm{i}|\mathbf{p}|$, we perform the momentum integration
in eq.~\eqref{E-2.3-26},
\begin{equation}
-\frac{\alpha M^2}{d} \frac{\mathbf{p}^2 \pi}{\gamma^{3-d}(4\pi)^{d/2}} \frac{\Gamma\left(\frac{3}{2} -
\frac{d}{2}\right)}{\Gamma\textstyle{\left(\frac{1}{2}\right)}} \int_0^1 \mathrm{d}\omega~
\frac{(1-\omega)^{\frac{d}{2}-\frac{3}{2}}}{\sqrt{\omega}}~.\label{E-2.3-27}
\end{equation}
Then, since the remaining integration over $\omega$ turns out to be finite in two dimensions and the rest
of the expression does not display any PDS singularity, we can directly reintroduce $\epsilon \equiv 3-d$ and
consider the three-dimensional limit. In particular, the integral over $\omega$ in eq.~\eqref{E-2.3-27}
can be evaluated in first-order approximation in $\epsilon$, obtaining
\begin{equation}
\int_0^1 \mathrm{d}\omega~\frac{(1-\omega)^{-\frac{\epsilon}{2}}}{\sqrt{\omega}} \approx  2 + 2\epsilon - 2\epsilon \log 2~.
\label{E-2.3-28}
\end{equation}
Second, the terms depending on $\epsilon$ in the exponent can be grouped and expanded to first order in $\epsilon$
as in eq.~\eqref{E-2.2-27}, whereas the Gamma function can be expressed in Laurent series up to order $\epsilon^0$.
Performing few manipulations and taking the $\epsilon \rightarrow 0$ limit, the original expression in
eq.~\eqref{E-2.3-26} becomes
\begin{equation}
\begin{split}
\lim_{d\rightarrow 3} \frac{M}{d}\int_{\mathbb{R}^d}\frac{\mathrm{d}^ds}{(2\pi)^d}\frac{-2\pi\eta\mathbf{p}^2}{ \mathbf{p}^2
- \mathbf{s}^2} 
= - \frac{\alpha M^2}{3\pi}\frac{\mathbf{p}^2}{4} \lim_{\epsilon \rightarrow 0} \frac{\left(1+\epsilon -\epsilon \log 2\right)
\left(\frac{2}{\epsilon}-\gamma_E \right) }{\left(1-\frac{\epsilon}{3}\right) \left(1-\frac{\epsilon}{2}\log
\frac{\pi \mu^2}{\gamma^2}\right)}
\\ = -\frac{\alpha M^2 \mathbf{p}^2}{6\pi} \left[\frac{1}{3-d} + \frac{4}{3} -\frac{\gamma}{2}
+ \log \frac{\mu\sqrt{\pi}}{\alpha M} \right. & \left. + \log(-\mathrm{i}\eta)\right]~.\label{E-2.3-29}
\end{split}
\end{equation}
Subsequently, we compute the last term on the r.h.s. of eq.~\eqref{E-2.3-25}. As it can be inferred, the integral
coincides with the one in of eq.~\eqref{E-2.3-29}, except for an overall factor of $\alpha^2 M^2/4\mathbf{p}^2 = \eta^2$.
Therefore, its evaluation is straightforward and gives
\begin{equation}
\lim_{d\rightarrow 3} \frac{\alpha^2M^3}{4d}\int_{\mathbb{R}^d}\frac{d^ds}{(2\pi)^d}\frac{-2\pi\eta}{\mathbf{p}^2-\mathbf{s}^2}
= -\frac{\alpha M^2}{6\pi}\frac{\alpha^2M^2}{4}  \cdot \left[ \frac{1}{3-d} +\frac{4}{3} - \frac{\gamma_E}{2}
+ \log \frac{\mu\sqrt{\pi}}{\alpha M} + \log(-\mathrm{i}\eta)\right]  ~.\label{E-2.3-30} 
\end{equation}
Collecting both the results in eqs.~\eqref{E-2.3-29} and \eqref{E-2.3-30}, we obtain the sought expression
for $\overbar{\mathbbm{j}}_{\mathrm{C}}^{\mathrm{new}}(d;\mathbf{p})$ in the three-dimensional limit, 
\begin{equation}
\overbar{\mathbbm{j}}_{\mathrm{C}}^{\mathrm{new}}(\mathbf{p}) = - (1+\eta^2) \frac{\alpha M^2 \mathbf{p}^2}{6\pi}
\left[ \frac{1}{3-d} + \frac{4}{3} - \frac{\gamma_E}{2} + \log (-\mathrm{i}\eta)
+ \log \frac{\mu \sqrt{\pi}}{\alpha M} \right]~.\label{E-2.3-31}
\end{equation}
We now collect all the contributions in eqs.~\eqref{E-2.3-21}, \eqref{E-2.3-25} and ~\eqref{E-2.3-31} and
write a closed form for the diagonal matrix elements of $\mathbb{J}_{\mathrm{C}}$,
\begin{equation}
\begin{gathered}
\overbar{\mathbbm{j}}_{\mathrm{C}}^{\mathrm{d}}(\mathbf{p}) + \overbar{\mathbbm{j}}_{\mathrm{C}}^{\mathrm{old}}(\mathbf{p})
+ \overbar{\mathbbm{j}}_{\mathrm{C}}^{\mathrm{new}}(\mathbf{p})  = - \frac{\alpha M^2}{4\pi} \frac{\mathbf{p}^2}{3}
\left[ \frac{1}{3-d} + \frac{4}{3} - \frac{3}{2}\gamma_E + \log \frac{\mu \sqrt{\pi}}{\alpha M} \right]  - \frac{\mu M}{4\pi}\frac{\mathbf{p}^2}{2} \\
- \frac{\alpha^3 M^4}{48\pi}\left[\frac{1}{3-d} + \zeta(3) -\frac{3}{2}\gamma_E + \frac{4}{3} + \log
\frac{\mu\sqrt{\pi}}{\alpha M} \right]  
+ \frac{\alpha M^2\mathbf{p}^2}{12\pi} \overbar{H}(\eta)(1+\eta^2) - \frac{\alpha^2 M^3}{32\pi}\frac{\mu}{3}(\pi^2-3)~,
\label{E-2.3-32}
\end{gathered}
\end{equation}
where the definition of $\overbar{H}(\eta)$ in eq.~\eqref{E-2.1-29} has been exploited and the Cauchy principal
value of $\zeta(1)$ has been taken. A direct comparison with the repulsive counterpart of the last formula in
eq.~\eqref{E-2.2-34} shows that the map between the two expression is provided by the sign reversal in front
of all the terms containing odd powers of the fine-structure constant and the replacement of $H(\eta)$
by $\overbar{H}(\eta)$. This fact is consistent with the conclusions drawn from eq.~(70) in ref.~\cite{KoR00},
where all the PDS-corrective terms remained unaffected by the sign change in the charge of one of the interacting
fermions.

We conclude this section with the derivation of an expression for the scattering length and the effective range,
by making use of the attractive counterpart of the generalized effective-range expansion in eq.~\eqref{E-2.1-26},
obtained by replacing again $H(\eta)$ by $\overbar{H}(\eta)$ with $\eta <0$. As a consequence of the attraction
of the electrostatic interaction, the Coulomb corrections in the strong scattering parameters change sign,
consistent with eq.~\eqref{E-2.3-32}. Concerning the scattering length, we have
\begin{equation}
\frac{1}{\overbar{a}_{\mathrm{C}}^{(1)}} = \frac{12\pi}{M \overbar{D}(E^*)}  + \frac{\alpha^3 M^3}{4}
\left[ \frac{1}{3-d} +\zeta(3)  -\frac{3}{2}\gamma_E + \frac{4}{3} +\log
\frac{\mu\sqrt{\pi}}{\alpha M}\right] + \frac{\alpha^2M^2\mu}{8}  \left(\pi^2-3\right) ~,\label{E-2.3-33}
\end{equation}
where the divergence can be reabsorbed by the P-wave strong coupling constant. Analogously to eq.~(72) of
ref.~\cite{KoR00}, the renormalized version of the scattering length, $\overbar{a}_{\mathrm{C}}^{(1)}(\mu)$, can be
defined in terms of the physical one, $\overbar{a}_{\mathrm{C}}^{(1)}$, 
\begin{equation}
\frac{1}{\overbar{a}_{\mathrm{C}}^{(1)}(\mu)} = \frac{1}{\overbar{a}_{\mathrm{C}}^{(1)}} - \frac{\alpha^3 M^3}{4}
\left[\zeta(3) -\frac{3}{2}\gamma_E +\frac{4}{3} + \log\frac{\mu\sqrt{\pi}}{\alpha M}\right]~.\label{E-2.3-34}
\end{equation}
Finally, the terms proportional to the square of the momentum of the fermions $\mathbf{p}$ give rise to
a nonzero value for the effective range, as in eq.~\eqref{E-2.2-39}, 
\begin{equation}
\overbar{r}_0^{(1)} = - \alpha M \left[ \frac{2}{3-d} + \frac{8}{3} -3\gamma_E + 2\log\frac{\mu\sqrt{\pi}}{\alpha M}
\right] -3\mu~,\label{E-2.3-35}
\end{equation}
whose divergent part, in case the energy-dependent coefficient of the $\ell=1$ interaction $D(E^*)$ is replaced
by $D_0$, can be again reabsorbed by counterterms coming from $\mathbf{p}^2$-dependent $\ell=0$ interactions.

%--------------------------------------------------------------------------------------------------------------------------------------------------------------------

\section{\textsf{The Lattice environment}}\label{S-3.0}

\begin{figure}[h]
\begin{center}
\includegraphics[width=0.50\columnwidth]{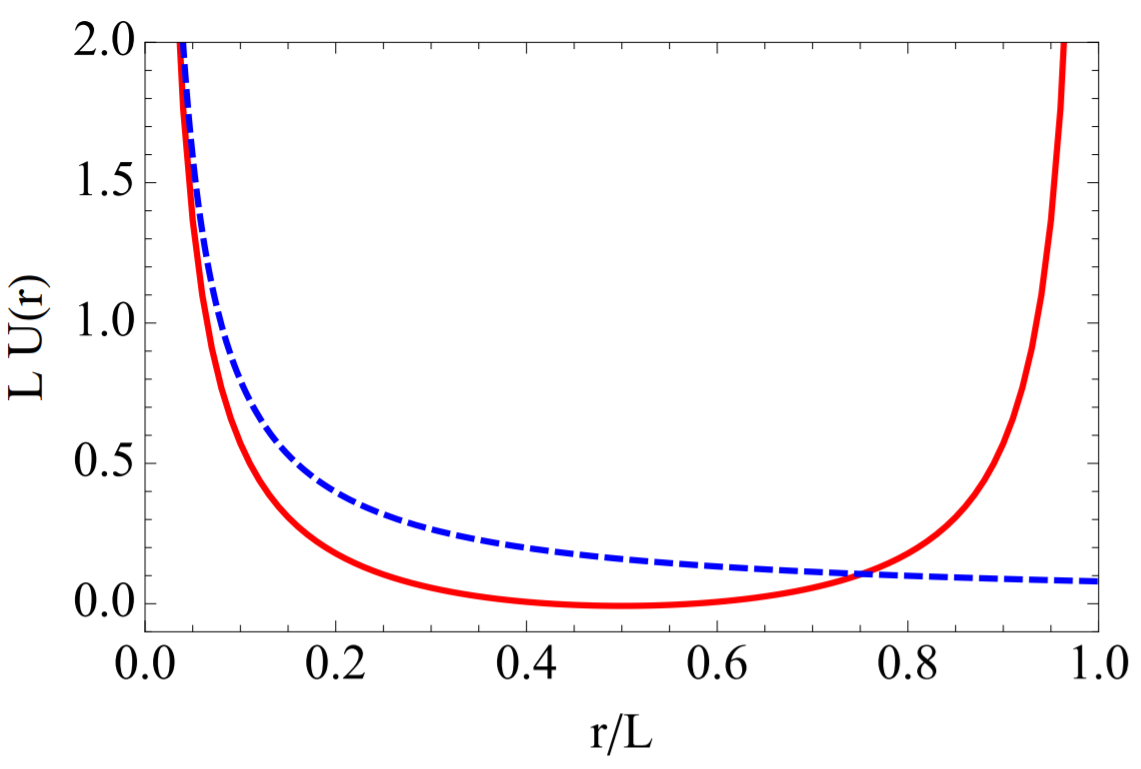}
\end{center}
\caption{Behaviour of the finite volume Coulomb potential energy between unit charges along a lattice axis
(solid red curve) obtained from $U(\mathbf{r},L)$ in eq.~\eqref{E-3.0-01} and the infinite volume
Coulomb potential (dashed blue curve) \cite{DaS14,BeS14}.}
\label{F-3-01}
\end{figure}

At this stage, we transpose the physical system of non-relativistic spinless fermions interacting via Coulomb pho\-tons
onto a cubic lattice with $L$ points per (spatial) dimension and spacing $a$. In this environment, it is customary to
continue analytically the fields and the wavefunctions outside the lattice by means of periodic boundary
conditions (PBCs). It follows that a free particle subject to PBCs carries a momentum $\mathbf{p}=2\pi\mathbf{n}/L$,
where $\mathbf{n}$ is a dimensionless three-vector of integers. Unlike QCD fields, the photon field in QED is
truncated and modified by the boundary of the volume. In particular, when PBCs are implemented, the validity of
Amp\`ere's and Gauss's law is compromised. The problem is circumvented by introducing an uniform background
charge density, a procedure that proves to be equivalent to the removal of the zero modes of the photon
\cite{BeS14}. Once the latter are canceled, the Coulomb potential between two identical charges $e$ becomes
(cf. fig.~\ref{F-3-01})
\begin{equation}
U(\mathbf{r},L) = \frac{\alpha}{\pi L}\sum_{\mathbf{n}\neq 0} \frac{1}{|\mathbf{n}|^2}e^{\mathrm{i}\frac{2\pi}{L}\mathbf{n}\cdot\mathbf{r}}~,
\label{E-3.0-01}
\end{equation}
where the $\mathbf{n} \in \mathbb{Z}^3$ encodes the dimensionless lattice momenta. Discarding the zero modes,
the latter are restricted to $|\mathbf{p}| \geq 2\pi/L$, whereas the viability of a perturbation treatment of
QED is controlled by the parameter $\eta = \alpha M /2|\mathbf{p}|$, which scales as the inverse of the momentum
of the interacting particles. Combining the above constraint with the definition of $\eta$, it follows that
$\eta \sim \alpha M L$ and the photon field insertions can be treated perturbatively if $ML \ll 1/\alpha$.
As $\eta$ grows linearly with the spatial volume, for any value of $M$ exists a critical value of $L$
that regulates the applicability of perturbation theory. Besides the condition $\eta \ll 1$, we assume henceforth
the limit $M \gg 1/L$, since for the current Lattice QCD calculations large volumes are employed \cite{BeS14}. 

Furthermore, the finite volume QED effects are such that the energy eigenvalues of two charged fermions (e.g. hadrons)
are modified in the same way by their self-interactions and by their interactions with each other, and the shifts
take the form of power laws in $L$ \cite{DaS14}. As a consequence, in the presence of Coulomb photons the kinematics
of two-body processes receives power law modifications in the finite volume context \cite{DaS14,UnH08}.
In particular, if the infinite-volume effective range expansion for the P-wave scattering is rewritten in terms
of the center of mass energy,
\begin{equation}
\begin{split}
\mathbf{p}^2 (1+\eta^2)[C_{\eta}^2|\mathbf{p}|(\cot\delta_1-\mathrm{i}) + \alpha M H(\eta)]
= -\frac{1}{{a_{\mathrm{C}}}^{(1)}} \\ +\frac{1}{2}{r_0}^{(1)}M(E^* -2M) & + r_1^{(1)} M^2 (E^*-2M)^2  + \ldots~,
\label{E-3.0-02}
\end{split}
\end{equation}
then $E^* = 2M + T$ in the above expression is replaced by its finite volume counterpart\footnote{Finite
volume physical quantities will be denoted henceforth with an $L$ in the superscript.}.
Eq.~\eqref{E-3.0-02} thus becomes
\begin{equation}
\begin{split}
\mathbf{p}^2 (1+\eta^2)[C_{\eta}^2|\mathbf{p}|(\cot\delta_1-\mathrm{i}) + \alpha M H(\eta)]
= -\frac{1}{{a_{\mathrm{C}}}^{(1)}} \\ +\frac{1}{2}{r_0}^{(1)}M(E^{*L} -2M) & + r_1^{(1)} M^2 (E^{*L}-2M)^2  + \ldots~.
\label{E-3.0-03}
\end{split}
\end{equation}
The original dependence of the r.h.s. of the last equation on the powers of the finite volume kinetic energy
$T^L = E^{*L}-2M^L$ can be exactly restored by exploiting the expression of the finite-volume shift for the
masses of spinless particles with unit charge in eqs.~(6) and (19) of ref.~\cite{DaS14},
\begin{equation}
\Delta M \equiv M^L - M = \frac{\alpha}{2\pi L}\left[\sum_{\mathbf{n}\neq 0}^{\Lambda_n}\frac{1}{|\mathbf{n}|^2}
- 4\pi\Lambda_n\right] + \mathcal{O}\left(\alpha^2; \textstyle{\frac{\alpha}{L^2}}\right)~,\label{E-3.0-04}
\end{equation}
where the sum of the three-dimensional Riemann series regulated by the spherical cutoff $\Lambda_n$ is denoted
with $\mathcal{I}^{(0)} \approx-8.913632$ (cf. app.~\ref{S-A-5.1}). To this purpose, primed scattering
parameters are introduced
\begin{equation}
\frac{1}{{a'_{\mathrm{C}}}^{(1)}} = \frac{1}{a_{\mathrm{C}}^{(1)}} - \frac{\alpha r_0^{(1)} M}{2\pi L}\mathcal{I}^{(0)}
+ \mathcal{O}\left(\alpha^2; \textstyle{\frac{\alpha}{L^2}}\right)~,\label{E-3.0-05}
\end{equation}
\begin{equation}
{r'_0}^{(1)} = r_0^{(1)} + \frac{4\alpha r_1^{(1)} M }{\pi L}\mathcal{I}^{(0)} + \mathcal{O}\left(\alpha^2;
\textstyle{\frac{\alpha}{L^2}}\right)~,\label{E-3.0-06}
\end{equation}
\begin{equation}
{r_1'}^{(1)} = r_1^{(1)} + \frac{3\alpha r_2^{(1)} M}{\pi L} \mathcal{I}^{(0)} + \mathcal{O}\left(\alpha^2;
\textstyle{\frac{\alpha}{L^2}}\right)~,\label{E-3.0-07}
\end{equation}
\begin{equation}
{r_2'}^{(1)} = r_2^{(1)} + \frac{4\alpha r_3^{(1)} M}{\pi L} \mathcal{I}^{(0)} + \mathcal{O}\left(\alpha^2;
\textstyle{\frac{\alpha}{L^2}}\right)~,\label{E-3.0-08}
\end{equation}
and 
\begin{equation}
{r_3'}^{(1)} = r_3^{(1)} + \frac{5\alpha r_4^{(1)} M}{\pi L} \mathcal{I}^{(0)} + \mathcal{O}\left(\alpha^2;
\textstyle{\frac{\alpha}{L^2}}\right)~,\hspace{0.2cm} \ldots~, \label{E-3.0-09}
\end{equation}
differing from the infinite volume counterparts by corrections of order $\alpha$ and scaling as the inverse of
the lattice size. Explicitly, the infinite volume ERE in eq.~\eqref{E-3.0-03} rewritten in terms of the translated
scattering parameters in eqs.~\eqref{E-3.0-05}-\eqref{E-3.0-09} for scattering states $T^L = \mathbf{p}^2/M$ assumes
the form
\begin{equation}
\mathbf{p}^2 (1+\eta^2)[C_{\eta}^2|\mathbf{p}|(\cot\delta_1-\mathrm{i}) + \alpha M H(\eta)] 
= -\frac{1}{{a'_{\mathrm{C}}}^{(1)}}+\frac{1}{2}{r'_0}^{(1)}\mathbf{p}^2  + {r'_1}^{(1)}\mathbf{p}^4 + \ldots~,
\label{E-3.0-10}
\end{equation}
where the changes in the total energy have been incorporated in the primed scattering parameters. Finally,
also the validity region of the last expansion is modified by the lattice environment, due to the changes in
the analytic structure of the scattering amplitude in the complex $|\mathbf{p}|$ plane. The absence of the zero mode
in the Coulomb potential in eq.~\eqref{E-3.0-01}, in fact, yields a shift in the branch cut of the imaginary
$|\mathbf{p}|$ axis from the origin to $\sqrt{2\pi M/L} + \mathcal{O}(1/M)$, which fixes the inelastic threshold
for the two-hadron state (cf. fig.~2 in ref.~\cite{BeS14}).

The last version of the ERE, combined with the quantization conditions discussed below, will turn out to be
the key ingredient for the derivation of the finite volume energy corrections for scattering and bound states
with one unit of angular momentum. 

%--------------------------------------------------------------------------------------------------------------------------------------------------------------------

\subsection{\textsf{Quantization Condition}}\label{S-3.1}

After introducing the finite and discretized configuration space, we derive the conditions that determine the
counterpart of the $\ell=1$ energy eigenvalues on the lattice. These states transform as the three-dimensional
irreducible representation $T_1$ (in Sch\"onflies's notation \cite{Car97}) of the cubic group \cite{BNL14,BNL15,SEM18},
the finite group of the 24 rotations of the cube that replaces the original SO(3) symmetry in the continuum and
infinite volume context \cite{Joh82}. 

As it can be inferred from eq.~\eqref{E-2.1-18}, the eigenvalues of the full Hamiltonian of the system $\hat{H}_0
+\hat{V}_{\mathrm{C}}+\hat{V}_{\mathrm{S}}$ can be identified with the singularities of the two-point correlation
function $G_{\mathrm{SC}}(\mathbf{r},\mathbf{r}')$ and are called  \textit{quantization conditions}
in the literature \cite{Lue86-01,Lue86-02,Lue91,BeS14}.
The Green's functions $G_{\mathrm{SC}}(\mathbf{r}',\mathbf{r})$ in turn can be
computed from the terms in the expansion over the P-wave interaction insertions stemming from eq.~\eqref{E-2.1-20},
with $\hat{V}_{\mathrm{S}}$ in momentum space given in eq.~\eqref{E-2.0-06}. In particular, the three lowest order
contributions in $D(E^*)$ yield, respectively,
\begin{equation}
\langle \mathbf{r}' | \hat{G}_{\mathrm{C}}^{(\pm)}| \mathbf{r}\rangle = G^{(\pm)}_{\mathrm{C}}(\mathbf{r}',\mathbf{r})~,
\label{E-3.1-01}
\end{equation}
\begin{equation}
\langle \mathbf{r}' |  \hat{G}_{\mathrm{C}}^{(\pm)} \hat{V}_{\mathrm{S}} \hat{G}_{\mathrm{C}}^{(\pm)} | \mathbf{r} \rangle 
= D(E^*) \nabla_{\mathbf{r}_1} G_{\mathrm{C}}(\mathbf{r}',\mathbf{r}_1)^{(\pm)} \Big\lvert_{\mathbf{r}_1=\mathbf{0}} \cdot
\nabla_{\mathbf{r}_2} G_{\mathrm{C}}^{(\pm)}(\mathbf{r}_2,\mathbf{r})\Big\lvert_{\mathbf{r}_2=\mathbf{0}}~,\label{E-3.1-02}
\end{equation}
and
\begin{equation}
\begin{split}
\langle \mathbf{r}' |  \hat{G}_{\mathrm{C}}^{(\pm)} \hat{V}_{\mathrm{S}} \hat{G}_{\mathrm{C}}^{(\pm)} \hat{V}_{\mathrm{S}}
\hat{G}_{\mathrm{C}}^{(\pm)} | \mathbf{r} \rangle =   [D(E^*)]^2 \nabla_{\mathbf{r}_1} G_{\mathrm{C}}^{(\pm)}(\mathbf{r},\mathbf{r}_1)
\Big\lvert_{\mathbf{r}_1=\mathbf{0}} \\ \cdot  \nabla_{\mathbf{r}_2} \otimes \nabla_{\mathbf{r}_3} G_{\mathrm{C}}^{(\pm)}(\mathbf{r}_2,
\mathbf{r}_3) \Big\lvert_{\substack{\mathbf{r}_2=\mathbf{0} \\ \mathbf{r}_3=\mathbf{0}}} & \nabla_{\mathbf{r}_4} G_{\mathrm{C}}^{(\pm)}
(\mathbf{r}_4,\mathbf{r})  \Big\lvert_{\mathbf{r}_4=\mathbf{0}}~.\label{E-3.1-03}
\end{split}
\end{equation}
Extending the calculation to higher orders, the expression of $(N+1)^{\mathrm{th}}$ order contribution to the
full two-point correlation function can be derived,
\begin{equation}
\begin{split}
\langle \mathbf{r}' |  \hat{G}_{\mathrm{C}}^{(\pm)} \underbrace{\hat{V}_{\mathrm{S}} \hat{G}_{\mathrm{C}}^{(\pm)} ...}_{\mathrm{N~times}}
\hat{V}_{\mathrm{S}} \hat{G}_{\mathrm{C}}^{(\pm)} | \mathbf{r} \rangle = D(E^*) \nabla_{\mathbf{r}_1} G_{\mathrm{C}}^{(\pm)}(\mathbf{r}',
\mathbf{r}_1) \Big\lvert_{\mathbf{r}_1= \mathbf{0}} \cdot  \\ \prod_{i=2}^{N} \left[ D(E^*) \nabla_{\mathbf{r}_i} \otimes
\nabla_{\mathbf{r}_{i+1}} G_{\mathrm{C}}^{(\pm)}(\mathbf{r}_i,\mathbf{r}_{i+1}) \Big\lvert_{\substack{\mathbf{r}_i=\mathbf{0} \\
\mathbf{r}_{i+1}=\mathbf{0}}} \right] &  \nabla_{\mathbf{r}_{N+2}} G_{\mathrm{C}}^{(\pm)}(\mathbf{r}_{N+2},\mathbf{r})
\Big\lvert_{\mathbf{r}_{N+2}=\mathbf{0}}~,\label{E-3.1-04}
\end{split}
\end{equation}
thus allowing to rewrite the original Green's function in terms of a geometric series of ratio
\begin{equation}
 D(E^*) \nabla_{\mathbf{r}_i}~\otimes~\nabla_{\mathbf{r}_{i+1}} G_{\mathrm{C}}^{(\pm)}(\mathbf{r}_i,\mathbf{r}_{i+1})
 \lvert_{\substack{\mathbf{r}_i =\mathbf{0} \\ \mathbf{r}_{i+1}=\mathbf{0}}}~,\nonumber
\end{equation}
that we identify as $\mathbb{J}_{\mathrm{C}}$ (cf. eq.~\eqref{E-2.2-10}), 
\begin{equation}
G_{\mathrm{SC}}^{(\pm)}(\mathbf{r}',\mathbf{r}) = \langle \mathbf{r}' | \hat{G}_{\mathrm{SC}}^{(\pm)} | \mathbf{r} \rangle
= G^{(\pm)}_{\mathrm{C}}(\mathbf{r}',\mathbf{r})  + \nabla_{\mathbf{r}_1} G_{\mathrm{C}}^{(\pm)}(\mathbf{r}',\mathbf{r}_1)
\Big\lvert_{\mathbf{r}_1=\mathbf{0}} \cdot \frac{D(E^*)}{1 - D(E^*) \mathbb{J}_{\mathrm{C}}} \nabla_{\mathbf{r}_2}
G_{\mathrm{C}}^{(\pm)}(\mathbf{r}_2,\mathbf{r}) \Big\lvert_{\mathbf{r}_2 = \mathbf{0}} ~,\label{E-3.1-05}
\end{equation}
where 
\begin{equation}
\sqrt{D(E^*)}\nabla_{\mathbf{r}_1} G_{\mathrm{C}}^{(\pm)}(\mathbf{r}',\mathbf{0}) \lvert_{\mathbf{r}_1=\mathbf{0}}
 \hspace{0.3cm}\mathrm{and}\hspace{0.3cm} 
\sqrt{D(E^*)}\nabla_{\mathbf{r}_2} G_{\mathrm{C}}^{(\pm)}(\mathbf{r}_2,\mathbf{r}) \lvert_{\mathbf{r}_2=\mathbf{0}}~,
\nonumber
\end{equation}
can be similarly interpreted as a source and a sink coupling the fermions to an P-wave state, respectively.
As in the $\ell=0$ case, the pole in the second term of eq.~\eqref{E-3.1-05} permits to express the infinite
volume quantization conditions,
\begin{equation}
\frac{\mathbbm{1}}{D(E^*)} = \nabla_{\mathbf{r}_i} \otimes \nabla_{\mathbf{r}_{i+1}}
G_{\mathrm{C}}^{(\pm)}(\mathbf{r}_i,\mathbf{r}_{i+1}) \Big\lvert_{\substack{\mathbf{r}_i =\mathbf{0} \\ \mathbf{r}_{i+1}=\mathbf{0}}}~,
\label{E-3.1-06}
\end{equation}
where the identity matrix multiplied by the CoM energy dependent coupling constant is equal to the inverse of
the matrix of the double derivatives of the Coulomb two-point Green's function evaluated at the origin,
$\mathbb{J}_{\mathrm{C}}$. Concentrating again on the retarded two-point correlation function and adopting the notation
of ref.~\cite{BeS14}, the finite-volume counterpart of eq.~\eqref{E-3.1-05} becomes
\begin{equation}
G_{\mathrm{SC}}^{(+),L}(\mathbf{r}',\mathbf{r}) = G^{(+),L}_{\mathrm{C}}(\mathbf{r}',\mathbf{r})  + \nabla_{\mathbf{r}_1}
G_{\mathrm{C}}^{(+),L}(\mathbf{r}',\mathbf{r}_1) \Big\lvert_{\mathbf{r}_1=\mathbf{0}}  \cdot \frac{D^L(E^*)}{1 -
D(E^*) \mathbb{J}_{\mathrm{C}}^L} \nabla_{\mathbf{r}_2} G_{\mathrm{C}}^{(+),L}(\mathbf{r}_2,\mathbf{r})
\Big\lvert_{\mathbf{r}_2 = \mathbf{0}} ~.\label{E-3.1-07}
\end{equation}
Similarly, the pole in the second term on the r.h.s. of the last equation yields the finite volume quantization condition,
 \begin{equation}
\frac{\mathbbm{1}}{D^L(E^*)} = \nabla_{\mathbf{r}_i} \otimes \nabla_{\mathbf{r}_{i+1}} G_{\mathrm{C}}^{(+),L}(\mathbf{r}_i,
\mathbf{r}_{i+1}) \Big\lvert_{\substack{\mathbf{r}_i =\mathbf{0} \\ \mathbf{r}_{i+1}=\mathbf{0}}}~,\label{E-3.1-08}
\end{equation}
that determines the $T_1$ eigenvalues. As in the $\ell=0$ case, we proceed by expanding $G_{\mathrm{C}}^{(+),L}(\mathbf{r}_i,
\mathbf{r}_{i+1})$ in powers of the fine-structure constant and truncate the series to order $\alpha$. Moreover,
we notice that the directional derivatives of the two-point Coulomb Green's function evaluated at the origin
correspond to the pairwise closure of the external legs of the Coulomb ladders in the expansion of $G_{\mathrm{C}}^{(+)}$
in fig.~\ref{F-2-02} to two-fermion vertices, evaluated at the origin in configuration space. As a consequence,
the matrix elements of $\mathbb{J}_{\mathrm{C}}^L$ can be interpreted as bubble diagrams with multiple Coulomb-photon
insertions inside. Analytically, the two lowest order contributions to $\mathbb{J}_{\mathrm{C}}$, that correspond to
bubble diagrams, respectively with and without a Coulomb-photon insertions, read
\begin{equation}
\begin{split}
\mathbb{J}_{\mathrm{C}}(\mathbf{p}) = \nabla_{\mathbf{r}'} \otimes \nabla_{\mathbf{r}} G_{\mathrm{C}}^{(\pm)}(\mathbf{r}',\mathbf{r})
\Big\lvert_{\substack{\mathbf{r} =\mathbf{0} \\ \mathbf{r}' =\mathbf{0}}} = \nabla_{\mathbf{r}'} \otimes \nabla_{\mathbf{r}} \langle
\mathbf{r}' | \hat{G}_{\mathrm{0}}^{(\pm)} | \mathbf{r} \rangle \Big\lvert_{\substack{\mathbf{r} =\mathbf{0} \\ \mathbf{r}' =\mathbf{0} }}
+  \nabla_{\mathbf{r}'} \otimes \nabla_{\mathbf{r}} \langle \mathbf{r}' | \hat{G}_{\mathrm{0}}^{(\pm)} \hat{V}_{\mathrm{C}}
\hat{G}_{\mathrm{0}}^{(\pm)}  | \mathbf{r} \rangle \Big\lvert_{\substack{\mathbf{r} =\mathbf{0} \\ \mathbf{r}' =\mathbf{0}}}  \\ + \ldots 
= - M \int_{\mathbb{R}^3} \frac{\mathrm{d}^3q}{(2\pi)^3}\frac{\mathbf{q}\otimes\mathbf{q}}{\mathbf{q}^2 - \mathbf{p}^2}
+ 4\pi\alpha M^2 \int_{\mathbb{R}^3} \frac{\mathrm{d}^3q}{(2\pi)^3} \int_{\mathbb{R}^3} \frac{\mathrm{d}^3k}{(2\pi)^3}
\frac{1}{\mathbf{q}^2 - \mathbf{p}^2} \frac{1}{\mathbf{k}^2 - \mathbf{p}^2}   \frac{\mathbf{q}\otimes
\mathbf{k}}{|\mathbf{q}-\mathbf{k}|^2} + \ldots &~,\label{E-3.1-09}
\end{split}
\end{equation}
where the Dyson identity between $\hat{G}_{\mathrm{C}}^{(\pm)}$ and $\hat{G}_{0}^{(\pm)}$ (cf. eq.~\eqref{E-2.1-12}) has been
exploited and $\varepsilon$ has been set to zero. Replacing again the integrals over the momenta by sums
over the dimensionless momenta $\mathbf{n},\mathbf{m}\in \mathbb{Z}^3$, the lattice counterpart of eq.~\eqref{E-3.1-09}
is obtained,
\begin{equation}
\mathbb{J}_{\mathrm{C}}^L(\mathbf{p}) = -\frac{M^L}{L^3}\sum_{\mathbf{n}}^{\Lambda_n} \frac{\mathbf{n}\otimes\mathbf{n}}{|\mathbf{n}|^2-\tilde{p}^2} + \frac{\alpha (M^{L})^{2}}{4\pi^3 L^2} \sum_{\mathbf{n}}^{\Lambda_n}\sum_{\mathbf{m}\neq\mathbf{n}}^{\infty}\frac{1}{|\mathbf{n}|^2 - \tilde{p}^2} \frac{1}{|\mathbf{m}|^2 - \tilde{p}^2}\frac{\mathbf{n}\otimes\mathbf{m}}{|\mathbf{n}-\mathbf{m}|^2}  + \ldots~,\label{E-3.1-10}
\end{equation}

where the finite-volume mass $M^L$, the speherical lattice cutoff $\Lambda_n$ and the dimensionless CoM momentum of the
incoming particles $\tilde{p}$ have been reintroduced. With the aim of regulating the sums in eq.~\eqref{E-3.1-10} for
numerical evaluation while maintaining the mass-independent renormalization scheme (cf. sec.~II B of ref.~\cite{BeS14}),
we are allowed to rewrite the finite volume quantization conditions as 
\begin{equation}
\frac{\mathbbm{1}}{D^L(E^*)} - \mathfrak{Re} \mathbb{J}_{\mathrm{C}}^{\{\mathrm{DR}\}} (\mathbf{p})
= \mathbb{J}_{\mathrm{C}}^L(\mathbf{p}) - \mathfrak{Re} \mathbb{J}_{\mathrm{C}}^{\{\Lambda\}} (\mathbf{p})~,\label{E-3.1-11}
\end{equation}
where $\mathbb{J}_{\mathrm{C}}^{\{\Lambda\}} (\mathbf{p})$ and $\mathbb{J}_{\mathrm{C}}^{\{\mathrm{DR}\}} (\mathbf{p})$
denote the $\mathcal{O}(\alpha)$ approximations of $\mathbb{J}_{\mathrm{C}}$ computed in the cutoff- and dimensional
regularization schemes. Starting again from eq.~\eqref{E-3.1-09}, we insert the spherical cutoffs as in its
discrete counterpart (cf. eq.~\eqref{E-3.1-10}), in sight of the evaluation of $\mathbb{J}_{\mathrm{C}}^{\{\Lambda\}}$,
\begin{equation}
\mathbb{J}_{\mathrm{C}}^{\{\Lambda\}} (\mathbf{p}) = - M \int_{S_{\Lambda}^2} \frac{\mathrm{d}^3q}{(2\pi)^3}
\frac{\mathbf{q}\otimes\mathbf{q}}{\mathbf{q}^2 - \mathbf{p}^2} + 4\pi\alpha M^2 
 \int_{S_{\Lambda}^2} \frac{\mathrm{d}^3\mathbf{q}}{(2\pi)^3}  \int_{\mathbb{R}^3} \frac{\mathrm{d}^3\mathbf{k}}{(2\pi)^3}
\frac{1}{\mathbf{q}^2 - \mathbf{p}^2}  \frac{1}{\mathbf{k}^2 - \mathbf{p}^2} \frac{\mathbf{q}\otimes
\mathbf{k}}{|\mathbf{q}-\mathbf{k}|^2}  + \ldots ~,\label{E-3.1-12}
\end{equation}
where $S_{\Lambda}^2$ denotes the three-dimensional sphere with radius $\Lambda$. Isolating the $\mathcal{O}(\alpha)$
contribution, we obtain
\begin{equation}
\begin{split}
M \int_{S^2_{\Lambda}}\frac{\mathrm{d}^3 q}{(2\pi)^3} \frac{\mathbf{q}\otimes\mathbf{q}}{\mathbf{p}^2 -\mathbf{q}^2}
= - \frac{\mathbbm{1}}{3} \frac{M}{2\pi^2} \int_{0}^{\Lambda} \mathrm{d} q~\mathbf{q}^2  - \frac{\mathbbm{1}}{3}
\frac{M}{2\pi^2} \int_{0}^{\Lambda} \mathrm{d} q~\mathbf{p}^2 \\+  \frac{\mathbbm{1}}{3} \frac{M}{2\pi^2} \int_{0}^{\Lambda}
\mathrm{d} q~\frac{\mathbf{p}^4}{\mathbf{p}^2-\mathbf{q}^2} & = - \frac{M\Lambda}{6\pi^2} \left( \frac{\Lambda^2}{3}
+ \mathbf{p}^2\right) \mathbbm{1} + \mathcal{O}(\Lambda^0)~,\label{E-3.1-13}
\end{split}
\end{equation}
where the isotropy of the cutoff has been exploited in the second step and $\mathcal{O}(\Lambda^0)$ denotes
constant or vanishing terms in the $\Lambda \rightarrow +\infty$ limit. Concerning the $\mathcal{O}(\alpha)$ term,
the integral can be simplified as follows
\begin{equation}
4\pi\alpha M^2\int_0^1\mathrm{d}\omega \int_{S_{\Lambda}^2} \frac{\mathrm{d}^3\mathbf{q}}{(2\pi)^3} \frac{1}{\mathbf{q}^2
- \mathbf{p}^2} \int_{\mathbb{R}^3}\frac{\mathrm{d}^3\mathbf{k}}{(2\pi)^3} \frac{\mathbf{q} \otimes
\mathbf{k}}{[\mathbf{k}^2 - 2(1-\omega)\mathbf{k}\cdot\mathbf{q} + \Xi_1]^2}~,\label{E-3.1-14}
\end{equation}
where $\Xi_1 \equiv (1-\omega)\mathbf{q}^2 -\omega\mathbf{p}^2$ and Feynman parametrization for the denominators
has been applied. The subsequent integration over the momentum $\mathbf{k}$ through eq.~(B17) in ref.~\cite{Ram97}
and the exploitation of rotational symmetry in the outcoming integrand (cf. eq.~(4.3.1a) in ref.~\cite{Col84}) gives
\begin{equation}
\frac{\Gamma\left(\frac{1}{2}\right)}{(4\pi)^{\frac{3}{2}}} \frac{\mathbbm{1}}{3} \int_0^1\mathrm{d}\omega \int_{S_{\Lambda}^2}
\frac{\mathrm{d}^3\mathbf{q}}{(2\pi)^3} \frac{4\pi\alpha M^2  }{\mathbf{q}^2 - \mathbf{p}^2}
\frac{(1-\omega)\mathbf{q}^2 }{[\omega(1-\omega)\mathbf{q}^2 -\mathbf{p}^2\omega]^{\frac{1}{2}}} ~.\label{E-3.1-15}
\end{equation}
Then, it is convenient to split the integrand of eq.~\eqref{E-3.1-15} into two parts and to simplify the numerator, 
\begin{equation}
\begin{split}
\frac{\Gamma\left(\frac{1}{2}\right)}{(4\pi)^{\frac{3}{2}}} \frac{\mathbbm{1}}{3} \int_0^1\mathrm{d}\omega
\int_{S_{\Lambda}^2} \frac{\mathrm{d}^3\mathbf{q}}{(2\pi)^3} \frac{4\pi\alpha M^2 (1-\omega)}{[\omega(1-\omega)\mathbf{q}^2
-\mathbf{p}^2\omega]^{\frac{1}{2}}}   \\ +  \frac{\Gamma\left(\frac{1}{2}\right)}{(4\pi)^{\frac{3}{2}}} \frac{\mathbbm{1}}{3}
\int_0^1\mathrm{d}\omega \int_{S_{\Lambda}^2} \frac{\mathrm{d}^3\mathbf{q}}{(2\pi)^3} \frac{\mathbf{p}^2}{\mathbf{q}^2
- \mathbf{p}^2} & \frac{4\pi\alpha M^2 (1-\omega)}{[\omega(1-\omega)\mathbf{q}^2 -\mathbf{p}^2\omega]^{\frac{1}{2}}} 
\equiv \mathfrak{J}_1 + \mathfrak{J}_2~,\label{E-3.1-16}
\end{split}
\end{equation}
where $\mathfrak{J}_1$ ($\mathfrak{J}_2$) corrresponds to the first (second) integral on the l.h.s. of the
last equation and it will generate the leading contributions in $\Lambda$. Beginning with the latter term,
integration over the momentum $\mathbf{l}$ yields
\begin{equation}
\mathfrak{J}_1 = \frac{\alpha M^2}{16 \pi^2} \frac{\mathbbm{1}}{3} \int_0^1\mathrm{d}\omega  
\left[ \frac{2\mathbf{p}^2}{1-\omega}\mathrm{arccoth} \left( \frac{1}{\Xi_2} \right) + 2\Lambda^2\Xi_2 
- \frac{\mathrm{i}\pi\mathbf{p}^2}{1-\omega} \right] ~.\label{E-3.1-17}
\end{equation}
where $\Xi_2$ is an ancillary variable,
\begin{equation}
\Xi_2 \equiv \sqrt{1 - \frac{\mathbf{p}^2}{\Lambda^2 (1-\omega)}}~.\nonumber
\end{equation}
Exploiting the fact that $\mathbf{p}/\Lambda \ll 1$, the integrand in the latter expression can be
considerably simplified. Exploting the results
\begin{equation}
\int_0^1\mathrm{d}\omega~\frac{1}{\sqrt{\omega(1-\omega)}}  = 2\int_0^1\mathrm{d}\omega~\sqrt{\frac{1-\omega}{\omega}}
= \pi~\label{E-3.1-18}
\end{equation}
and
\begin{equation}
\int_0^1\mathrm{d}\omega~\frac{\log\sqrt{1-\omega}}{\sqrt{\omega(1-\omega)}} = -\pi\log 2~,\label{E-3.1-19}
\end{equation}
the remaining integration can be performed, obtaining the desired expression for $\mathfrak{I}_1$
\begin{equation}
\mathfrak{J}_1 = \frac{\alpha M^2}{16\pi} \frac{\mathbbm{1}}{3} \left[ \Lambda^2 - \frac{2\mathrm{i}}{\pi}|
\mathbf{p}|\Lambda + 2\mathbf{p}^2  \log \left(\frac{\Lambda}{|\mathbf{p}|} \right) \right] + \mathcal{O}(\Lambda^0)~.
\label{E-3.1-20}
\end{equation}
Concerning the $\mathfrak{I}_2$ term, its rotational symmetry and integration over the angular variables permits to
split it in turn into two terms,
\begin{equation}
\begin{split}
\mathfrak{J}_2 = M^2 4\pi\alpha \frac{\Gamma\left(\frac{1}{2}\right)}{(4\pi)^{\frac{3}{2}}} \frac{\mathbbm{1}}{3}
\frac{\mathbf{p}^2}{2\pi^2} \int_0^1\mathrm{d}\omega \int_0^{\Lambda} \mathrm{d}q~\frac{1}{\left[ \mathbf{q}^2 -
\frac{\mathbf{p}^2}{1-\omega}\right]^{\frac{1}{2}}}  + \\
M^2 4\pi\alpha \frac{\Gamma\left(\frac{1}{2}\right)}{(4\pi)^{\frac{3}{2}}} \frac{\mathbbm{1}}{3} \frac{\mathbf{p}^2}{2\pi^2}
\int_0^1\mathrm{d}\omega \int_0^{\Lambda} \mathrm{d}q~ & \frac{\mathbf{p}^2}{\mathbf{q}^2-\mathbf{p}^2}
\frac{1}{\left[ \mathbf{q}^2 -\frac{\mathbf{p}^2}{1-\omega}\right]^{\frac{1}{2}}} ~.\label{E-3.1-21}
\end{split}
\end{equation}
Considering the first term on the r.h.s. of eq.~\eqref{E-3.1-21}, integration over the radial momentum
yields again an $\mathrm{arccoth}(x)$ function, which is eventually responsible of a further logarithmic divergence
in the UV region, 
\begin{equation}
\frac{\alpha M^2 \mathbf{p}^2}{4\pi^2} \frac{\mathbbm{1}}{3} \int_0^1\mathrm{d}\omega \int_0^{\Lambda}
\mathrm{d}q~\frac{1}{\left[ \mathbf{q}^2 -\frac{\mathbf{p}^2}{1-\omega}\right]^{\frac{1}{2}}} = 
\frac{\alpha M^2 \mathbf{p}^2}{4\pi^2} \frac{\mathbbm{1}}{3} \int_0^1\mathrm{d}\omega \left[ -\frac{\mathrm{i}\pi}{2}
+ \mathrm{arccoth}\left( \frac{1}{\Xi_2} \right) \right]~.\label{E-3.1-22}
\end{equation}
Approximating the expression again under the assumption $\mathbf{p}/\Lambda \ll 1$ and performing the integration
over $\omega$ (cf. eq.~\eqref{E-3.1-18}), the expression on the r.h.s. of eq.~\eqref{E-3.1-22} becomes
\begin{equation}
\frac{\alpha M^2 \mathbf{p}^2}{4\pi^2} \frac{\mathbbm{1}}{3} \int_0^1 \mathrm{d}\omega \sqrt{\frac{1-\omega}{\omega}}
\left[-\frac{\mathrm{i}\pi}{2} + \log\left(\frac{2\Lambda}{|\mathbf{p}|}\right) + \log\sqrt{1-\omega} \right] 
= \frac{\alpha M^2}{8\pi} \frac{\mathbbm{1}}{3} \mathbf{p}^2\log\left(\frac{\Lambda}{|\mathbf{p}|}\right)
+ \mathcal{O}(\Lambda^0)~,\label{E-3.1-23}  
\end{equation}
i.e. it carries the second logarithmic contribution to $\mathbb{J}_{\mathrm{C}}^{\{\Lambda\}}$ to order $\alpha$ in
the perturbative expansion. Finally, we consider the second term on the r.h.s. of eq.~\eqref{E-3.1-21} and
introduce the auxiliary variables $\gamma^2 \equiv -\mathbf{p}^2$ and $\Xi_3 \equiv \gamma^2/(1-\omega)$.
The integration over the radial momentum $|\mathbf{q}|$ yields
\begin{equation}
\frac{\alpha M^2 \mathbf{p}^2}{4\pi^2} \frac{\mathbbm{1}}{3} \int_0^1\mathrm{d}\omega~\sqrt{\frac{1-\omega}{\omega}} 
\left[ \frac{\log\left( 1 + \Lambda \sqrt{\frac{\gamma^2 - \Xi_3}{\gamma^2(\Xi_3 + \Lambda^2)}}
\right)}{2\sqrt{\gamma^2(\gamma^2 - \Xi_3)}} - \frac{\log\left( 1 + \frac{\Lambda}{\gamma^2}
\sqrt{\frac{\gamma^2(\gamma^2 - \Xi_3)}{\gamma^2 + \Lambda^2}} \right)}{2\sqrt{\gamma^2(\gamma^2 - \Xi_3)}} \right]~,
\label{E-3.1-24}  
\end{equation}
an expression that can be simplified in the large coutoff limit, $\mathbf{p}/\Lambda \ll 1$, obtaining
\begin{equation}
\frac{\alpha M^2}{4\pi^2} \frac{\mathbbm{1}}{3} \mathbf{p}^2 \int_0^1 \mathrm{d}\omega \frac{1-\omega}{\omega}
\mathrm{arccot}\left(\sqrt{\frac{\omega}{1-\omega}}\right) + \mathcal{O}(\Lambda^0)~.\label{E-3.1-25} 
\end{equation}
Although the remaining integral is unbound, the overall expression is independent of the cutoff $\Lambda$,
therefore it can be neglected as the whole $\mathcal{O}(\Lambda^0)$ contributions. This divergence is analogous
to the one found in the $\ell=0$ case, and turns out to disappear if a translation in the momenta such as $\mathbf{k}
\mapsto \mathbf{k} - \mathbf{q}$ in the original expression of the $\mathcal{O}(\alpha)$ term of
$\mathbb{J}_{\mathrm{C}}^{\{\Lambda\}}$ in eq.~\eqref{E-3.1-12} is performed. Now, collecting the results in
eqs.~\eqref{E-3.1-13} and \eqref{E-3.1-20}, the cutoff-regularized version of $\mathbb{J}_{\mathrm{C}}$ is obtained
to the desired order in the fine-structure constant,
\begin{equation}
\begin{split}
\mathbb{J}_{\mathrm{C}}^{\{\Lambda\}}(\mathbf{p}) = - \frac{M\Lambda}{2\pi^2} \frac{\mathbbm{1}}{3}\left( \frac{\Lambda^2}{3}
+ \mathbf{p}^2\right)   + \frac{\alpha M^2}{16\pi} \frac{\mathbbm{1}}{3}\left[\Lambda^2 - \frac{2\mathrm{i}}{\pi}|
\mathbf{p}|\Lambda + 4\mathbf{p}^2\log\frac{\Lambda}{|\mathbf{p}|}  \right] +\mathcal{O}(\alpha^2)~,
\label{E-3.1-26}
\end{split}
\end{equation}
where the $\mathcal{O}(\Lambda^0)$ contributions have been discarded. Now we proceed with the calculation in
dimensional regularization of $\mathbb{J}_{\mathrm{C}}$. To this purpose, it is convenient to start from the
exact expression of $\mathbb{J}_{\mathrm{C}}$ to all orders in $\alpha$ in arbitrary $d$ dimensions
(cf. eq.~\eqref{E-2.2-11}),
\begin{equation}
\begin{split}
\mathbb{J}_{\mathrm{C}}^{\{\mathrm{DR}\}} (\mathbf{p}) = \int_{\mathbb{R}^d}\frac{\mathrm{d}^d \mathbf{q}}{(2\pi)^d} \frac{2\pi
M \eta(\mathbf{q})}{e^{2\pi\eta(\mathbf{q})}-1}\frac{1 +\eta^2(\mathbf{q})}{\mathbf{p}^2 -\mathbf{q}^2 +\mathrm{i}\varepsilon}
\mathbf{q} \otimes \mathbf{q} \\ =  M\int_{\mathbb{R}^d}\frac{\mathrm{d}^d \mathbf{q}}{(2\pi)^d}\frac{\mathbf{q}\otimes
\mathbf{q}}{\mathbf{p}^2 -\mathbf{q}^2 +\mathrm{i}\varepsilon}  - \alpha M^2 & \int_{\mathbb{R}^d}
\frac{\mathrm{d}^d\mathbf{q}}{(2\pi)^d} \frac{\pi}{2|\mathbf{q}|} \frac{\mathbf{q} \otimes\mathbf{q}}{\mathbf{p}^2
-\mathbf{q}^2 +\mathrm{i}\varepsilon} + \mathcal{O}(\alpha^2)~,\label{E-3.1-27}
\end{split}
\end{equation}
where the integrand has been expanded up to the first order in $\alpha$. In particular, the $\alpha$-independent
contribution in eq.~\eqref{E-3.1-27} gives
\begin{equation}
M \int_{\mathbb{R}^d} \frac{\mathrm{d}^d\mathbf{q}}{(2\pi)^d} \frac{\mathbf{q}\otimes \mathbf{q}}{\mathbf{p}^2-\mathbf{q}^2}
=  - M \frac{\mathbbm{1}}{d}  \int_{\mathbb{R}^d} \frac{\mathrm{d}^dq}{(2\pi)^d}  - M \frac{\mathbbm{1}}{d}
\int_{\mathbb{R}^d} \frac{\mathrm{d}^dq}{(2\pi)^d}  \frac{\mathbf{p}^2}{\mathbf{p}^2-\mathbf{q}^2}~,\label{E-3.1-28}
\end{equation}
where the rotational invariance of the integrand has been exploited and $\varepsilon$ has been set to zero. Since the
integrand is a polynomial in the momentum, the first contribution on the r.h.s. of the last equation vanishes 
in dimensional regularization, whereas the remaining term turns out to coincide
with the purely strong counterpart of $\mathbb{J}_{\mathrm{C}}(d;\mathbf{p})$ in eq.~\eqref{E-2.0-18}, 
\begin{equation}
\mathbb{J}_{0}(d;\mathbf{p}) = - M \frac{\mathbbm{1}}{d} \int_{\mathbb{R}^d} \frac{\mathrm{d}^dq}{(2\pi)^d}
\frac{\mathbf{p}^2}{\mathbf{p}^2-\mathbf{q}^2} = M\mathbf{p}^2 \frac{\mathbbm{1}}{3}
\frac{(-\mathbf{p}^2)^{\frac{d}{2}-1}}{(4\pi)^{\frac{d}{2}}} \Gamma\left(\textstyle{1-\frac{d}{2}}\right)~,\label{E-3.1-29}
\end{equation}
thus is finite and the limit $d \rightarrow 3$ can be safely taken, obtaining
\begin{equation}
\mathbb{T}_{\mathrm{S}}^{\{\mathrm{DR}\}}(\mathbf{p}) = \frac{\mathbbm{1}}{3} \frac{\mathrm{i} M|\mathbf{p}|^3}{4\pi}~.
\label{E-3.1-30}
\end{equation}
Due to the presence of the imaginary unit in the r.h.s. of the last equation, it turns out that the
$\alpha$-independent component of $\mathbb{J}_{\mathrm{C}}^{\{\mathrm{DR}\}}$ does not contribute in eq.~\eqref{E-3.1-11},
since only real parts are retained. Regarding the $\mathcal{O}(\alpha)$ term of $\mathbb{J}_{\mathrm{C}}$ in
arbitrary dimension, the integral can be recast as
\begin{equation}
\begin{split}
M \int_{\mathbb{R}^d}\frac{\mathrm{d}^d\mathbf{q}}{(2\pi)^d} \frac{\pi \eta(\mathbf{q})~\mathbf{q}\otimes
\mathbf{q}}{\mathbf{q}^2-\mathbf{p}^2}  = M \frac{\mathbbm{1}}{d} \int_{\mathbb{R}^d}
\frac{\mathrm{d}^d\mathbf{q}}{(2\pi)^d} \frac{\pi \eta(\mathbf{q})~\mathbf{q}^2}{\mathbf{q}^2-\mathbf{p}^2} \\
= \frac{\alpha M^2 \pi}{2}\frac{\mathbbm{1}}{d} \int_{\mathbb{R}^d}
\frac{\mathrm{d}^d\mathbf{q}}{(2\pi)^d} \frac{1}{|\mathbf{q}|}  + \frac{\alpha M^2 \pi}{2} \frac{\mathbbm{1}}{d}
\frac{\Gamma(\frac{3}{2})}{\Gamma(\frac{1}{2})} & \int_0^1 \frac{\mathrm{d}\omega}{\sqrt{\omega}} \int_{\mathbb{R}^d }
\frac{\mathrm{d}^d\mathbf{q}}{(2\pi)^d}\frac{\mathbf{p}^2}{\left[\mathbf{q}^2+\gamma^2(1-\omega)\right]^{\frac{3}{2}}}~,
\label{E-3.1-31}
\end{split}
\end{equation}
where $\gamma \equiv -\mathrm{i}|\mathbf{p}|$ and the Feynman parametrization for the denominators has been adopted.
The subsequent momentum integration in the latter yields
\begin{equation}
\frac{\mathbbm{1}}{d} \frac{\alpha M^2 \pi }{2 (4\pi)^{d/2}} \frac{\mathbf{p}^2 \Gamma(\frac{3}{2}-
\frac{d}{2})}{\Gamma(\frac{1}{2})} \left(\frac{\mu}{2}\right)^{3-d}\int_0^1 \frac{\mathrm{d}\omega}{\sqrt{\omega}}
[\gamma^2(1-\omega)]^{\frac{d}{2}-\frac{3}{2}}~,\label{E-3.1-32}
\end{equation}
where the renormalization scale $\mu$ has been introduced. Conversely, the first term in the second row
of eq.~\eqref{E-3.1-31} vanishes like the first integral on the r.h.s. of eq.~\eqref{E-3.1-28}. Introducing the
small quantity $\epsilon = 3-d$, the integral over $\omega$ can be computed to first order in $\epsilon$, obtaining
\begin{equation}
\int_0^1\frac{\mathrm{d}\omega}{\sqrt{\omega}} (1-\omega)^{-\frac{\epsilon}{2}} = 2 + 2\epsilon - \epsilon\log 2
+ \mathcal{O}(\epsilon^2) ~.\label{E-3.1-33}
\end{equation}
Exploiting the result in the last formula, eq.~\eqref{E-3.1-31} partially expanded to order $\epsilon$ becomes
\begin{equation}
\frac{\mathbbm{1}}{3}\frac{\alpha M^2\mathbf{p}^2}{8\pi}\left(\frac{\mu \sqrt{\pi}}{\gamma}\right)^{\epsilon}
\Gamma\textstyle{\left(\frac{\epsilon}{2}\right)}\frac{1+\epsilon -\epsilon\log 2+\mathcal{O}(\epsilon^2)}{1-
\frac{\epsilon}{3} + \mathcal{O}(\epsilon^2)}~.\label{E-3.1-34}
\end{equation}
Expanding in turn the Gamma function and the power term $\mu \sqrt{\pi}/\gamma$ in Laurent and Taylor series, respectively,
and truncating the expansion to order $\epsilon^0$, the desired expression for $\mathbb{T}_{\mathrm{SC}}(\mathbf{p})$
in dimensional regularization is recovered
\begin{equation}
\mathbb{J}_{\mathrm{C}}^{\{\mathrm{DR}\}}(\mathbf{p}) = \mathbbm{1} \frac{\alpha M^2}{4\pi} \frac{\mathbf{p}^2}{3}
\left[\frac{1}{\epsilon} - \frac{\gamma_E}{2} + \frac{4}{3} +\mathrm{i}\pi + \log\left(\frac{\mu\sqrt{\pi}}{2|\mathbf{p}|}
\right) \right]~.\label{E-3.1-35}
\end{equation}
Then, we bring further simplification to the finite volume quantization condition by taking the trace of eq.~\eqref{E-3.1-11}, 
\begin{equation}
\frac{3}{D^L(E^*)} - \mathrm{tr}\left[\mathfrak{Re} \mathbb{J}_{\mathrm{C}}^{\{\mathrm{DR}\}}(\mathbf{p})\right]
= \mathrm{tr}\left[\mathbb{J}_{\mathrm{C}}^L(\mathbf{p})\right] - \mathrm{tr}
\left[\mathfrak{Re}\mathbb{J}_{\mathrm{C}}^{\{\Lambda\}}(\mathbf{p})\right]~,\label{E-3.1-36}
\end{equation}
thus transforming a matrix identity into a scalar one as the one for S-waves. Finally, we take the real
part of the expressions in eqs.~\eqref{E-3.1-26} and \eqref{E-3.1-34} and replace the fermion mass by its
lattice couterpart, obtaining the regulated version of the finite volume quantization condition
(cf. eq.~\eqref{E-3.1-08}) in explicit form,
\begin{equation}
\begin{split}
\frac{1}{D^L(E^*)}  = - \frac{M^L}{3L^3}\left[\sum_{\mathbf{n}}^{\Lambda_n}1 - \frac{4\pi}{3}\Lambda_n^3\right] -\frac{M^L}{4\pi^2 L}\frac{\mathbf{p}^2}{3}\left[\sum_{\mathbf{n}}^{\Lambda_n}\frac{1}{\mathbf{n}^2-\tilde{\mathbf{p}}^2} - 4\pi\Lambda_n \right]  +  \frac{\alpha (M^L)^2}{4\pi} \frac{\mathbf{p}^2}{3}\left[\frac{1}{\epsilon} - \frac{\gamma_E}{2} \right. \\ \left. + \frac{4}{3} + \log\left(\frac{\mu L}{4\sqrt{\pi}}\right) \right]
 + \frac{\alpha (M^L)^2}{4\pi^3 L^2} \frac{1}{3} \left[\sum_{\mathbf{n}}^{\Lambda_n}\sum_{\mathbf{m}\neq \mathbf{n}}^{\infty} \frac{1}{\mathbf{n}^2-\tilde{\mathbf{p}}^2} \frac{1}{\mathbf{m}^2-\tilde{\mathbf{p}}^2}\frac{\mathbf{m}\cdot \mathbf{n}}{|\mathbf{n}-\mathbf{m}|^2} - 4\pi^4\tilde{\mathbf{p}}^2\log\Lambda_n - \pi^4 \Lambda_n^2 \right]~.\label{E-3.1-37}
\end{split}
\end{equation}
Thanks to the last identity, the lattice version of the effective range expansion in eq.~\eqref{E-3.0-10} can be
rewritten in an explicit form. Besides, we drop in the next developments of the derivation the pole $1/\epsilon$ arising from
dimensional regularization, since it does not deliver information on the energy eigenvalues of our two-body system.

%--------------------------------------------------------------------------------------------------------------------------------------------------------------------

\subsection{\textsf{Finite Volume Effective Range Expansion}}\label{S-3.2}

First of all, we concentrate on the infinite volume version of the effective range expansion. By inserting
the expression of $T_{\mathrm{SC}}$ given in eq.~\eqref{E-2.2-12} into eq.~\eqref{E-2.1-26} and exploiting the
closed form of $\mathbbm{j}_{\mathrm{C}}$ in dimensional regularization to all orders in $\alpha$ given in
eq.~\eqref{E-2.2-34} we obtain a relation between the P-wave phase shift and the strong coupling constant $D(E^*)$
in the presence of Coulomb photon exchanges, 
\begin{equation}
\mathbf{p}^2(1+\eta^2)[C_{\eta}^2|\mathbf{p}|(\cot\delta_1-\mathrm{i}) + \alpha M H(\eta)]  = -\frac{12\pi}{M D(E^*)}
+ \alpha M \mathbf{p}^2\left[\frac{4}{3} - \frac{3}{2}\gamma_E + \log\left(\frac{\mu\sqrt{\pi}}
{\alpha M}\right) \right]~.\label{E-3.2-01}
\end{equation}
Since the asymptotic behaviour of the momentum integrals in the ultraviolet region is left invariant by
discretization, eq.~\eqref{E-3.2-01} can be straightforwardly adapted to the cubic lattice case, provided the
infinte volume parameters are replaced by finite volume ones, 
\begin{equation}
\mathbf{p}^2(1+\eta^2)[C_{\eta}^2|\mathbf{p}|(\cot\delta_1^L-\mathrm{i}) + \alpha M^L H(\eta)] 
= -\frac{12\pi}{M^L D^L(E^*)} + \alpha M^L \mathbf{p}^2\left[\frac{4}{3} - \frac{3}{2}\gamma_E
+ \log\left(\frac{\mu\sqrt{\pi}}{\alpha M^L}\right) \right]~,\label{E-3.2-02}
\end{equation}
which is valid to all orders in the fine-structure constant. Now, the quantization conditions derived
in the previous section can be exploited by replacing the inverse of the finite-volume strong-coupling constant
with the expression in eq.~\eqref{E-3.1-36}. Thus, the following equation is obtained,
\begin{equation}
\begin{gathered}
\mathbf{p}^2(1+\eta^2)[C_{\eta}^2|\mathbf{p}|(\cot\delta_1^L-\mathrm{i}) + \alpha M^L H(\eta)] 
=  \frac{4\pi}{L^3}\left[\sum_{\mathbf{n}}^{\Lambda_n}1 - \frac{4\pi}{3}\Lambda_n^3\right] +\frac{\mathbf{p}^2}{\pi L}
\left[\sum_{\mathbf{n}}^{\Lambda_n}\frac{1}{\mathbf{n}^2-\tilde{\mathbf{p}}^2} - 4\pi\Lambda_n \right] \\
- \frac{\alpha M^L}{\pi^2 L^2} \left[\sum_{\mathbf{n}}^{\Lambda_n}\sum_{\mathbf{m}\neq \mathbf{n}}^{\infty} \frac{1}{\mathbf{n}^2
-\tilde{\mathbf{p}}^2} \frac{1}{\mathbf{m}^2-\tilde{\mathbf{p}}^2}\frac{\mathbf{m}\cdot \mathbf{n}}
{|\mathbf{n}-\mathbf{m}|^2} - \pi^4 \Lambda_n^2 - 4\pi^4\tilde{\mathbf{p}}^2\log\Lambda_n
\right]  \\ +  \alpha M^L\mathbf{p}^2\left[ \log\left(\frac{4\pi}{\alpha M^L L}\right) - \gamma_E\right]~.\label{E-3.2-03}
\end{gathered}
\end{equation}

Similarly to the $\ell=0$ case, we notice that the finite-volume mass of the fermions is
multiplied by the fine-structure constant in all the occurrences. It follows that, in the $\mathcal{O}(\alpha)$
approximation of the Coulomb Green's functions and their derivatives, we can consistently ignore the leading-order
corrections to $M^L$ and replace the latter with $M$, at least in eq.~\eqref{E-3.2-03}. Nevertheless, we
account for the QED power law modifications of the masses \cite{DaS14} in the finite-volume version of the
effective-range expansion in eq.~\eqref{E-2.1-26} by means of the shifted scattering parameters
(cf. eqs.~\eqref{E-3.0-05}-\eqref{E-3.0-09}), 
\begin{equation}
\mathbf{p}^2(1+\eta^2)[C_{\eta}^2|\mathbf{p}|(\cot\delta_1^L-\mathrm{i}) + \alpha M H(\eta)] 
= -\frac{1}{{a'_{\mathrm{C}}}^{(1)}}+\frac{1}{2}{r'_0}^{(1)}\mathbf{p}^2  + {r'_1}^{(1)} \mathbf{p}^4+\ldots~,\label{E-3.2-04}
\end{equation}
where the ellipsis stands for higher-order scattering parameters. Combining the latter expression with eq.~\eqref{E-3.0-10}
and isolating the regulated sums, we obtain the desired explicit version of the effective-range expansion to order $\alpha$,

\begin{equation}
\begin{gathered}
-\frac{1}{{a'_C}^{(1)}} + \frac{1}{2}{r'_0}^{(1)} \mathbf{p}^2 + {r'_1}^{(1)} \mathbf{p}^4 +  {r'_2}^{(1)} \mathbf{p}^6 +  {r'_3}^{(1)} \mathbf{p}^8 +  ... \\
 =  \frac{4\pi}{L^3}\mathcal{S}_0(\tilde{\mathbf{p}}) + \frac{\mathbf{p}^2}{\pi L}\mathcal{S}_1(\tilde{\mathbf{p}})  
- \frac{\alpha M \mathbf{p}^2}{4\pi^4} \mathcal{S}_2(\tilde{\mathbf{p}})  -\frac{\alpha M}{\pi^2 L^2}\mathcal{S}_3(\tilde{\mathbf{p}}) 
+ ... + \alpha M\mathbf{p}^2 \left[ \log\left(\frac{4\pi}{\alpha M L}\right)-\gamma_E \right]~,\label{E-3.2-05}
\end{gathered}
\end{equation}

in terms of the \textit{L\"uscher functions},
\begin{equation}
\mathcal{S}_0(\tilde{\mathbf{p}}) = \sum_{\mathbf{n}}^{\Lambda_n}1 - \frac{4\pi}{3}\Lambda_n^3~, \hspace{0.5cm}
\mathcal{S}_1(\tilde{\mathbf{p}}) = \sum_{\mathbf{n}}^{\Lambda_n}\frac{1}{\mathbf{n}^2-\tilde{\mathbf{p}}^2}
- 4\pi\Lambda_n~,\label{E-3.2-06}
\end{equation}
\begin{equation}
\mathcal{S}_2(\tilde{\mathbf{p}}) = \sum_{\mathbf{n}}^{\Lambda_n}\sum_{\mathbf{m}\neq \mathbf{n}}^{\infty} \frac{1}{\mathbf{n}^2
-\tilde{\mathbf{p}}^2} \frac{1}{\mathbf{m}^2-\tilde{\mathbf{p}}^2}\frac{1}{|\mathbf{n}-\mathbf{m}|^2} - 4\pi^4 \log
\Lambda_n~,\label{E-3.2-07}
\end{equation}
and
\begin{equation}
\mathcal{S}_3(\tilde{\mathbf{p}}) = \sum_{\mathbf{n}}^{\Lambda_n}\sum_{\mathbf{m}\neq \mathbf{n}}^{\infty} \frac{1}{\mathbf{n}^2
-\tilde{\mathbf{p}}^2} \frac{1}{\mathbf{m}^2-\tilde{\mathbf{p}}^2}\frac{\mathbf{m}\cdot \mathbf{n}
- \tilde{\mathbf{p}}^2}{|\mathbf{n}-\mathbf{m}|^2} - \pi^4 \Lambda_n^2~.\label{E-3.2-08}
\end{equation}
In particular, the $\tilde{p}$-independent L\"uscher sum in eq.~\eqref{E-3.2-06} vanishes, as shown in
app.~\ref{S-A-5.1}, whereas $\mathcal{S}_3(\tilde{\mathbf{p}})$ generates the new series of double sums to be
computed in the low-$\mathbf{p}$ limit.

%--------------------------------------------------------------------------------------------------------------------------------------------------------------------

\subsection{\textsf{Approximate Energy Eigenvalues}}\label{S-3.3}

Since the Sommerfeld factor is not a rational function of the momentum of the colliding particles in the CoM frame,
a non-perturbative counterpart in $\alpha$ of the eq.~\eqref{E-3.2-05} in the low-momentum limit would allow
only numerical solutions for $\mathbf{p}^2$, which lie beyond our purpose.
Nevertheless, under the hypothesis that the expansions are perturbative in $1/L$ times the length scale characterizing
the strength of the interaction, governed by the scattering parameters, and assuming that $ML \ll 1/\alpha$,
the Coulomb photon insertions in the diagrams can be treated perturbatively.  Under these conditions, the
approximate expression of the ERE presented in eq.~\eqref{E-3.0-10} can be exploited for an analytical
derivation of the finite volume corrections to the energy eigenvalues.  

%--------------------------------------------------------------------------------------------------------------------------------------------------------------------

\subsubsection{\textsf{The Lowest Unbound State}}\label{S-3.3.1}

Differently from the S-wave case in sec.~III~D of ref.~\cite{BeS14}, the perturbative expansion of the arguments
of the summations in the L\"uscher functions around zero lattice momentum $\tilde{p}$, corresponding to a total energy
equal to $2M$ now looses significance, due to symmetry reasons. Given that $\ell=1$ states in the continuum and
infinite volume are mapped to $T_1$ states in a finite cubic lattice, the latter are expected to be
three-fold degenerate. While the multiplicity of the zero energy scattering state is one,
its nearest neighbour with $|\tilde{\mathbf{p}}| = 1$ is six-fold degenerate with total energy equal to
$2M + 4\pi^2/ML^2$, thus making it a suitable candidate for both $T_1$ and $T_2$ eigenstates. Moreover, if the lattice
size is large enough, the momentum $|\mathbf{p}| = 2\pi/L$ is small, the effective-range expansion in the l.h.s.
of eq.~\eqref{E-3.2-05} remains valid and can be truncated at any power of $\mathbf{p}^2$. Otherwise, the
non-perturbative approach in sec.~2.5 of ref.~\cite{Lue86-01} needs to be considered, and the energy eigenvalues
would be expressed in terms of the phase shift $\delta_1$. Following the small-momentum approach, we expand the
L\"uscher functions in eqs.~\eqref{E-3.2-06}-\eqref{E-3.2-08} in Taylor series about $\tilde{p}^2 = 1$ and
retain only small or negative powers of the quantity $\delta\tilde{p}^2\equiv\tilde{p}^2 -1$. Concerning the L\"uscher
function $\mathcal{S}_1(\tilde{p})$, the approximation yields
\begin{equation}
\begin{split}
\mathcal{S}_1(\tilde{\mathbf{p}}) = -\frac{6}{\delta\tilde{p}^2} + \sum_{|\mathbf{n}|\neq 1}^{\Lambda_n}
\frac{1}{|\mathbf{n}|^2-1} - 4\pi \Lambda_n + \delta\tilde{p}^2\sum_{|\mathbf{n}|\neq 1}^{\infty}
\frac{1}{(|\mathbf{n}|^2-1)^2}  \\ + (\delta\tilde{p}^2)^2 \sum_{|\mathbf{n}| \neq 1}^{\infty} \frac{1}{(|\mathbf{n}|^2-1)^3} 
+  (\delta\tilde{p}^2)^3 \sum_{|\mathbf{n}|\neq 1}^{\infty} \frac{1}{(|\mathbf{n}|^2-1)^4} & + (\delta\tilde{p}^2)^4
\sum_{|\mathbf{n}| \neq 1}^{\infty} \frac{1}{(|\mathbf{n}|^2-1)^{5}} + \ldots~,\label{E-3.3.1-01}
\end{split}
\end{equation}
where the dots denote terms of order $(\delta\tilde{p}^2)^6$ and the large $\Lambda_n$  limit is understood.
In the notation of app.~\ref{S-A-5.0} and ref.~\cite{BeS14}, eq.~\eqref{E-3.3.1-01} is concisely recast into
\begin{equation}
\mathcal{S}_1(\tilde{\mathbf{p}}) = -\frac{6}{\delta\tilde{p}^2} + \mathcal{I}^{(1)} + \delta\tilde{p}^2\mathcal{J}^{(1)} 
+ (\delta\tilde{p}^2)^2 \mathcal{K}^{(1)} +  (\delta\tilde{p}^2)^3 \mathcal{L}^{(1)}  + (\delta\tilde{p}^2)^4 \mathcal{O}^{(1)}
+ \ldots~,\label{E-3.3.1-02}
\end{equation}
where the sums of the implied three-dimensional Riemann series are reported in app.~\ref{S-A-5.1}. 
Regarding the function $\mathcal{S}_2(\tilde{p})$, we proceed by isolating and expanding the double sums with
$|\mathbf{n}|$ or $|\mathbf{m}|$ equal to one,
\begin{equation}
\sum_{|\mathbf{n}|=1}\sum_{|\mathbf{m}|\neq 1}^{\infty} \frac{1}{1 - \tilde{p}^2} \frac{1}{\mathbf{m}^2 - \tilde{p}^2}
\frac{1}{|\mathbf{n}-\mathbf{m}|^2} + \sum_{|\mathbf{m}|=1}\sum_{|\mathbf{n}|\neq 1}^{\Lambda_n} \frac{1}{1 - \tilde{p}^2}
\frac{1}{\mathbf{n}^2 - \tilde{p}^2}  \frac{1}{|\mathbf{n}-\mathbf{m}|^2}~.\label{E-3.3.1-03}
\end{equation}
The subsequent expansion of the argument of the two series in the same fashion of eq.~\eqref{E-3.3.1-01} shows that
all the resulting sums converge in the infinite-$\Lambda_n$ limit. Therefore, we are allowed to remove the
spherical cutoff and the two double sums in eq.~\eqref{E-3.3.1-03} merge, yielding
\begin{equation}
2\sum_{|\mathbf{n}|=1}\sum_{|\mathbf{m}|\neq 1}^{\infty} \frac{1}{1 - \tilde{p}^2} \frac{1}{\mathbf{m}^2 - \tilde{p}^2}
\frac{1}{|\mathbf{n}-\mathbf{m}|^2}  = - \frac{2}{\delta\tilde{p}^2}\chi_1 - 2\chi_2  -2\delta\tilde{p}^2\chi_3
- 2(\delta\tilde{p}^2)^2\chi_4 + \ldots~.\label{E-3.3.1-04}
\end{equation}
where, differently from the ones in the appendix of ref.~\cite{BeS14}, the series $\chi_i$ with $i \geq 1$ 
include $\mathbf{n} = \mathbf{0}$ and are defined as
\begin{equation}
\chi_i = \sum_{|\mathrm{n}| = 1}\sum_{|\mathbf{m}| \neq 1}^{\infty} \frac{1}{(|\mathbf{m}|^2-1)^i}\frac{1}{|\mathbf{n}
-\mathbf{m}|^2}~.\label{E-3.3.1-05}
\end{equation}
Furthermore, the expansion of the remaining term of the L\"uscher function $\mathcal{S}_2$ leads to results in
analogous to the ones of the S-wave case, 
\begin{equation}
\begin{split}
\sum_{|\mathbf{n}|\neq 1}^{\Lambda_n}\sum_{\substack{|\mathbf{m}|\neq 1 \\ \mathbf{m} \neq \mathbf{n}}}^{\infty} \frac{1}{|\mathbf{n}|^2
-\tilde{p}^2}\frac{1}{|\mathbf{m}|^2-\tilde{p}^2}\frac{1}{|\mathbf{n}-\mathbf{m}|^2} - 4\pi\log\Lambda_n \\
= \mathcal{R}^{(1)}  + \delta\tilde{p}^2 \left(\mathcal{R}_{24}^{(1)} + \mathcal{R}_{42}^{(1)}\right) &
+ (\delta\tilde{p}^2)^2  \left(\mathcal{R}_{44}^{(1)} + \mathcal{R}_{26}^{(1)} + \mathcal{R}_{62}^{(1)} \right)
+  \ldots~,\label{E-3.3.1-06}
\end{split}
\end{equation}
where the ellipsis stands for terms of order $(\delta\tilde{p}^2)^3$. The $\mathcal{R}^{(1)}$ and $\mathcal{R}_{ij}^{(1)}$
sums in the last equation coincide with the P-wave counterparts of the sums in eqs.~(A6) and (A9) in ref.~\cite{BeS14},
namely
\begin{equation}
\mathcal{R}^{(1)} \equiv \sum_{|\mathbf{n}|\neq 1}^{\Lambda_n}\sum_{\substack{|\mathbf{m}|\neq 1 \\ \mathbf{m} \neq \mathbf{n}}}^{\infty}
\frac{1}{|\mathbf{n}|^2-1}\frac{1}{|\mathbf{m}|^2-1}\frac{1}{|\mathbf{n}-\mathbf{m}|^2} - 2\pi^4\log\Lambda_n~,
\label{E-3.3.1-07}
\end{equation}
and 
\begin{equation}
\mathcal{R}_{2i~2j}^{(1)} \equiv \sum_{|\mathbf{n}|\neq 1}^{\infty}\sum_{\substack{|\mathbf{m}|\neq 1 \\ \mathbf{m} \neq \mathbf{n}}}^{\infty}
\frac{1}{(|\mathbf{n}|^2-1)^{i}}\frac{1}{(|\mathbf{m}|^2-1)^{j}}\frac{1}{|\mathbf{n}-\mathbf{m}|^2} ~,\label{E-3.3.1-08}
\end{equation}
with $i,j \geq 2$ and $i+j \geq 6$, which are invariant under permutation of the lower indices, $\mathcal{R}_{ij}^{(1)}
= \mathcal{R}_{ji}^{(1)}$, and convergent for $i,j \geq 1$ and $i+j > 2$. Exploiting the symmetry property
under index exchange and combining the results in eqs.~\eqref{E-3.3.1-07} and \eqref{E-3.3.1-08},
we find the desired result
\begin{equation}
\mathcal{S}_2(\tilde{p}) = - \frac{2}{\delta\tilde{p}^2}\chi_1 + \mathcal{R}^{(1)} - 2\chi_2 + 2\delta\tilde{p}^2
\left(\mathcal{R}_{24}^{(1)} -\chi_3\right)  + (\delta\tilde{p}^2)^2  \left(\mathcal{R}_{44}^{(1)} + 2\mathcal{R}_{26}^{(1)}
-2\chi_4 \right) + \ldots~.\label{E-3.3.1-09}
\end{equation}
Subsequently, we treat the genuinely new L\"uscher function, $\mathcal{S}_3(\tilde{p})$. To this purpose, we decompose
the initial double series into three pieces, 
\begin{equation}
\begin{split}
\mathcal{S}_3(\tilde{p}) = \lim_{\Lambda_m \rightarrow + \infty} \Big\{ \frac{1}{2} \sum_{\mathbf{n}}^{\Lambda_n}
\sum_{\mathbf{m} \neq \mathbf{n}}^{\Lambda_m} \frac{1}{|\mathbf{m}|^2-\tilde{p}^2}\frac{1}{|\mathbf{n}-\mathbf{m}|^2} 
+ \frac{1}{2} \sum_{\mathbf{n}}^{\Lambda_n}\sum_{\mathbf{m} \neq \mathbf{n}}^{\Lambda_m} \frac{1}{|\mathbf{n}|^2-\tilde{p}^2}
\frac{1}{|\mathbf{n}-\mathbf{m}|^2}  \\  - \frac{1}{2} \sum_{\mathbf{n}}^{\Lambda_n}\sum_{\mathbf{m} \neq \mathbf{n}}^{\Lambda_m}
\frac{1}{|\mathbf{m}|^2-\tilde{p}^2} \frac{1}{|\mathbf{n}|^2-\tilde{p}^2} \Big\} & - \pi^4\Lambda_n^2~,\label{E-3.3.1-10}
\end{split}
\end{equation}
where a subsidiary spherical cutoff $\Lambda_m$ has been introduced in order to highlight the divergent nature of
the three terms. In particular, the first and the second contribution to $\mathcal{S}_3(\tilde{p})$ on the r.h.s.
of the last equation, can be recast as
\begin{equation}
\begin{split}
 - \frac{1}{\delta\tilde{p}^2}\mathcal{P}^{(1)} + \chi_1  + \frac{1}{2}\left(\mathcal{P}_{022}^{(1)} +
 \mathcal{P}_{202}^{(1)} \right) + \frac{\delta\tilde{p}^2}{2} \left(\mathcal{P}_{042}^{(1)} + \mathcal{P}_{402}^{(1)}
 + 2\chi_2 \right) \\ + \frac{(\delta\tilde{p}^2)^2}{2} \left(\mathcal{P}_{062}^{(1)} + \mathcal{P}_{602}^{(1)}
 \right. & \left. + 2\chi_3 \right) + \ldots~,\label{E-3.3.1-11}
\end{split}
\end{equation}
where the dots denote terms of order $(\delta\tilde{p}^2)^3$, while the non-symmetric and divergent generalizations
of $\chi_0$ in eq.~\eqref{E-A-5.1-01} and of $\mathcal{R}_{ij}^{(1)}$ in  eq.~\eqref{E-3.3.1-08} have been introduced,
\begin{equation}
\mathcal{P}^{(1)} = \sum_{|\mathbf{m}|=1} \sum_{|\mathbf{n}|\neq 1}^{\Lambda_n} \frac{1}{|\mathbf{m}-\mathbf{n}|^2},\label{E-3.3.1-12}
\end{equation}
and 
\begin{equation}
\mathcal{P}_{2i~2j~2k}^{(1)} = \sum_{|\mathbf{n}|\neq 1}^{\Lambda_n}\sum_{\substack{|\mathbf{m}|\neq 1 \\ \mathbf{m} \neq \mathbf{n}}}^{\Lambda_m}
\frac{1}{(|\mathbf{n}|^2-1)^{i}}\frac{1}{(|\mathbf{m}|^2-1)^{j}}\frac{1}{|\mathbf{n}-\mathbf{m}|^{2k}}~,\label{E-3.3.1-13}
\end{equation}
respectively. Quite similarly, the third contribution to $\mathcal{S}_{3}(\tilde{p})$ in eq.~\eqref{E-3.3.1-10} can be
subdivided and expanded as follows
\begin{equation}
 - \frac{1}{2} \sum_{|\mathbf{n}|\neq 1}^{\Lambda_n}\sum_{\substack{|\mathbf{m}| \neq 1\\ \mathbf{n} \neq \mathbf{m} }}^{\Lambda_m}
 \frac{1}{|\mathbf{m}|^2-\tilde{p}^2}  \frac{1}{|\mathbf{n}|^2-\tilde{p}^2}   +  \frac{3}{\delta\tilde{p}^2}
 \sum_{|\mathbf{n}|\neq 1}^{\Lambda_n}\frac{1}{|\mathbf{n}|^2-\tilde{p}^2} +  \frac{3}{\delta\tilde{p}^2} 
 \sum_{|\mathbf{m}|\neq 1}^{\Lambda_m} \frac{1}{|\mathbf{m}|^2-\tilde{p}^2} ~,\label{E-3.3.1-14} 
\end{equation}
where the two terms involving single sums can be, in turn, expanded in pairs, obtaining
\begin{equation}
\frac{3}{\delta\tilde{p}^2}\left( \mathcal{I}_{\Lambda_n}^{(1)} + \mathcal{I}_{\Lambda_m}^{(1)}\right)
+ 6 \mathcal{J}^{(1)} + 6 \delta\tilde{p}^2 \mathcal{K}^{(1)}   + 6(\delta\tilde{p}^2)^2 \mathcal{L}^{(1)}
+ \ldots ~,\label{E-3.3.1-15} 
\end{equation}
where the neglected terms are again of order $(\delta\tilde{p}^2)^3$ and the for the resulting convergent series
the limit $\Lambda_n, \Lambda_m\rightarrow+\infty$ is understood. On the other hand, this limit can not be
taken for the non-regularized counterpart of  $\mathcal{I}^{(1)}$,
\begin{equation}
\mathcal{I}_{\Lambda_s}^{(1)} = \sum_{\mathbf{s}\neq\mathbf{0}}^{\Lambda_s} \frac{1}{|\mathbf{s}|^2}~,\label{E-3.3.1-16} 
\end{equation}
whose divergence will cancel with the one from $\mathcal{P}^{(1)}$ in eq.~\eqref{E-3.3.1-12}. Secondly,
the expansion of the double sum in eq.~\eqref{E-3.3.1-14} yields the appearance of further $\mathcal{P}_{2i~2j~2k}^{(1)}$ terms, 
\begin{equation}
 -\frac{1}{2}\mathcal{P}_{220}^{(1)} - \frac{1}{2} \delta\tilde{p}^2 \left( \mathcal{P}_{240}^{(1)} + \mathcal{P}_{420}^{(1)}
 \right)  - \frac{1}{2} (\delta\tilde{p}^2)^2  \left( \mathcal{P}_{440}^{(1)} + \mathcal{P}_{260}^{(1)}
 + \mathcal{P}_{620}^{(1)} \right) ~.\label{E-3.3.1-17} 
\end{equation}
Collecting the expansions of the three contributions in eqs.~\eqref{E-3.3.1-15} and \eqref{E-3.3.1-17} we can finally write
\begin{equation}
\begin{split}
- \frac{1}{2} \sum_{\mathbf{n}}^{\Lambda_n}\sum_{\mathbf{m} \neq \mathbf{n}}^{\infty}\frac{1}{|\mathbf{m}|^2-\tilde{p}^2}
\frac{1}{|\mathbf{n}|^2-\tilde{p}^2}  = \frac{3}{\delta\tilde{p}^2}\left(\mathcal{I}_{\Lambda_n}^{(1)}  +
\mathcal{I}_{\Lambda_m}^{(1)}\right)   + 6 \mathcal{J}^{(1)}
 -\frac{1}{2}\mathcal{P}_{220}^{(1)}  \\ +  \delta\tilde{p}^2 \left[6\mathcal{K}^{(1)} - \frac{1}{2}\left(\mathcal{P}_{240}^{(1)}
 + \mathcal{P}_{420}^{(1)} \right) \right]  + (\delta\tilde{p}^2)^2 \left[ 6 \mathcal{L}^{(1)}  -
 \frac{1}{2} \right. & \left. \left(\mathcal{P}_{440}^{(1)} + \mathcal{P}_{260}^{(1)} + \mathcal{P}_{620}^{(1)} \right)  \right]
 + \ldots~.\label{E-3.3.1-18} 
\end{split}
\end{equation}
Now, the two partial results in eqs.~\eqref{E-3.3.1-11} and \eqref{E-3.3.1-18} can be summed together, obtaining
the sought expression of $\mathcal{S}_3(\tilde{p})$ as a power series in $\delta\tilde{p}^2$. In particular,
we notice that the sum of all the series appearing at each order in $\delta\tilde{p}^2$ has to be finite in
the limit $\Lambda_n,\Lambda_m\rightarrow+\infty$, irrespective of the convergent or divergent behaviour of each
individual sum. In particular, we observe that the latter limit can be directly taken to all orders in the small
quantity, with the only exception of the sums in the $\delta\tilde{p}^2$-independent contribution, that are
regularized quadratically in the cutoff $\Lambda_n$. As a consequence, we can simplify our $\delta\tilde{p}^2$-expansion
for $\mathcal{S}_3(\tilde{p})$ by grouping the divergent sums order by order and defining the finite coefficients
\begin{equation}
\Qoppa_0 \equiv \lim_{\substack{\Lambda_n\rightarrow +\infty \\ \Lambda_m\rightarrow +\infty}}-\frac{1}{2}\left(3\mathcal{I}_{\Lambda_n}^{(1)}
+ 3\mathcal{I}_{\Lambda_m}^{(1)} - \mathcal{P}^{(1)} \right) = \sum_{|\mathbf{n}|=1}\sum_{|\mathbf{m}| \neq 1}^{\infty}
\frac{ \mathbf{m}\cdot\mathbf{n}-1}{(|\mathbf{m}|^2 - 1)|\mathbf{m} - \mathbf{n}|^2}~,\label{E-3.3.1-19}
\end{equation}
\begin{equation}
\Qoppa_1 \equiv \lim_{\Lambda_m\rightarrow +\infty} \frac{1}{2}\left(\mathcal{P}_{022}^{(1)} + \mathcal{P}_{202}^{(1)} -
\mathcal{P}_{220}^{(1)}\right) - \pi^4\Lambda_n^2 = \sum_{|\mathbf{n}|\neq 1}^{\Lambda_n} \sum_{\substack{\mathbf{m}\neq\mathbf{n} \\
|\mathbf{m}| \neq 1}}^{\infty} \frac{1}{|\mathbf{n}|^2 - 1}\frac{1}{|\mathbf{m}|^2 - 1} \frac{\mathbf{n}\cdot
\mathbf{m} - 1}{|\mathbf{n}-\mathbf{m}|^2} - \pi^4\Lambda_n^2 ~,\label{E-3.3.1-20}
\end{equation}
\begin{equation}
\begin{split}
\Qoppa_2 \equiv \lim_{\substack{\Lambda_n \rightarrow +\infty \\ \Lambda_m\rightarrow +\infty}} \frac{1}{2}\left( \mathcal{P}_{042}^{(1)}
+ \mathcal{P}_{402}^{(1)} - \mathcal{P}_{240}^{(1)} - \mathcal{P}_{420}^{(1)}\right) \\ =  \sum_{|\mathbf{n}|\neq 1}^{\infty}
\sum_{\substack{\mathbf{m}\neq\mathbf{n} \\ |\mathbf{m}| \neq 1}}^{\infty}  & \frac{1 - 2\mathbf{m}\cdot\mathbf{n} - |\mathbf{m}|^2
|\mathbf{n}|^2 + \mathbf{n}\cdot\mathbf{m}(|\mathbf{m}|^2 + |\mathbf{n}|^2)}{(|\mathbf{m}|^2-1)^2(|\mathbf{n}|^2-1)^2
|\mathbf{m}-\mathbf{n}|^2}  ~,\label{E-3.3.1-21}
\end{split}
\end{equation}
and 
\begin{equation}
\Qoppa_3 \equiv \lim_{\substack{ \Lambda_n \rightarrow +\infty \\ \Lambda_m\rightarrow +\infty}} \frac{1}{2}\left( \mathcal{P}_{062}^{(1)}
+ \mathcal{P}_{602}^{(1)} - \mathcal{P}_{440}^{(1)} - \mathcal{P}_{620}^{(1)}  - \mathcal{P}_{620}^{(1)}\right) 
=  \sum_{|\mathbf{n}|\neq 1}^{\infty} \sum_{\substack{\mathbf{m}\neq\mathbf{n} \\ |\mathbf{m}| \neq 1}}^{\infty}
\frac{q_S(\mathbf{n},\mathbf{m}) + \mathbf{m}\cdot\mathbf{n}~q_X(\mathbf{n},\mathbf{m})}{(|\mathbf{m}|^2-1)^3
(|\mathbf{n}|^2-1)^3 |\mathbf{m}-\mathbf{n}|^2} ~,\label{E-3.3.1-22}
\end{equation}
where the polynomials $q_S(\mathbf{n},\mathbf{m})$ and $q_X(\mathbf{n},\mathbf{m})$ are defined as
\begin{equation}
q_S(\mathbf{n},\mathbf{m}) = -1 + 3 |\mathbf{n}|^2 |\mathbf{m}|^2 - |\mathbf{m}|^2 |\mathbf{n}|^2
(|\mathbf{n}|^2+|\mathbf{m}|^2)\label{E-3.3.1-23}
\end{equation}
and
\begin{equation}
q_X(\mathbf{n},\mathbf{m}) = 3 - 3(|\mathbf{n}|^2 + |\mathbf{m}|^2) +  |\mathbf{m}|^2|\mathbf{n}|^2
+ |\mathbf{m}|^4 + |\mathbf{n}|^4~.\label{E-3.3.1-24}
\end{equation}
Equipped with the definitions in eqs.~\eqref{E-3.3.1-19}-\eqref{E-3.3.1-22}, we can write compactly the
final expression for the expansion of $\mathcal{S}_3(\tilde{p})$  in terms of convergent sums up to the
quadratic order in $\delta\tilde{p}^2$,
\begin{equation}
\mathcal{S}_3(\tilde{p}) = -\frac{2}{\delta\tilde{p}^2}\Qoppa_0 + \chi_1  + 6 \mathcal{J}^{(1)} + \Qoppa_1  
+  \delta\tilde{p}^2 \left[\chi_2 + 6\mathcal{K}^{(1)} + \Qoppa_2  \right]  + (\delta\tilde{p}^2)^2 \left[\chi_3
 +  6 \mathcal{L}^{(1)} + \Qoppa_3 \right] + \ldots~.\label{E-3.3.1-25}
\end{equation}
After redefining the regularized sum in eq.~\eqref{E-3.3.1-09} through the addition of the last term on the r.h.s.
of eq.~\eqref{E-3.2-05},
\begin{equation}
\tilde{\mathcal{R}}^{(1)} \equiv \mathcal{R}^{(1)} - 4\pi^4 \left[ \log\left(\frac{4\pi}{\alpha M L}\right)-
\gamma_E \right]~,\label{E-3.3.1-26}
\end{equation}
we plug the expressions in eqs.~\eqref{E-3.3.1-02}, \eqref{E-3.3.1-11}, \eqref{E-3.3.1-12} and \eqref{E-3.3.1-25}
in the finite volume effective range expansion and obtain
\begin{equation}
\begin{gathered}
-\frac{1}{{a'_{\mathrm{C}}}^{(1)}} + \frac{1}{2} \frac{4\pi^2 {r_{0}'}^{(1)}}{L^2} \tilde{p}^2 + \frac{16
\pi^4 {r_{1}'}^{(1)}}{L^4} \tilde{p}^4   + \frac{64 \pi^6 {r_{2}'}^{(1)}}{L^6} \tilde{p}^6 +
\frac{256 \pi^6 {r_{3}'}^{(1)}}{L^8} \tilde{p}^8 +  \ldots \\ = -\frac{\alpha M}{\pi^2 L^2} \Big\{ -
\frac{2}{\delta\tilde{p}^2}\Qoppa_0  + \chi_1  + 6 \mathcal{J}^{(1)} + \Qoppa_1  +  \delta\tilde{p}^2 \left[\chi_2
+ 6\mathcal{K}^{(1)} + \Qoppa_2  \right]   + (\delta\tilde{p}^2)^2 \left[\chi_3 +  6 \mathcal{L}^{(1)} + \Qoppa_3 \right]
+ \ldots \Big\}  \\ - \frac{\alpha M \mathbf{p}^2}{4\pi^4} \left[ - \frac{2}{\delta\tilde{p}^2}\chi_1 + \mathcal{R}^{(1)}
- 2\chi_2 + 2 \delta\tilde{p}^2 \left(\mathcal{R}_{24}^{(1)} -\chi_3\right) + (\delta\tilde{p}^2)^2
\left(\mathcal{R}_{44}^{(1)} + 2\mathcal{R}_{26}^{(1)} - 2\chi_4 \right) +  \ldots \right]  \\ + \frac{\mathbf{p}^2}{\pi L}
\left[ -\frac{6}{\delta\tilde{p}^2} + \mathcal{I}^{(1)} + \delta\tilde{p}^2\mathcal{J}^{(1)} + (\delta\tilde{p}^2)^2
\mathcal{K}^{(1)} + \ldots \right] +\ldots ~. \label{E-3.3.1-27}
\end{gathered}
\end{equation}
where the shifted higher-order scattering parameters ${r_{2}'}^{(1)}$ and ${r_{3}'}^{(1)}$ on the l.h.s. of the last
equation have been included, see eqs.~\eqref{E-3.0-05} and \eqref{E-3.0-09}. The subsequent multiplication of the
last equation by the scattering length permits to introduce coefficients identical to the ones in eq.~(37) of
ref.~\cite{BeS14}, 
\begin{equation}
d_0 =  \xi \frac{{a'_{\mathrm{C}}}^{(1)}}{\pi L} \hspace{4mm} d_1 = \xi \frac{\alpha  M}{4\pi^4}{a'_{\mathrm{C}}}^{(1)}
\hspace{4mm} d_2 = \xi^2 {a'_{\mathrm{C}}}^{(1)} {r_{0}'}^{(1)}~,\nonumber
\end{equation}
\begin{equation}
d_3 = \xi^3 {a'_{\mathrm{C}}}^{(1)} {r_{1}'}^{(1)} \hspace{4mm} d_4 = \xi^4 {a'_{\mathrm{C}}}^{(1)} {r_{2}'}^{(1)} \hspace{4mm}
d_5 = \xi^5 {a'_{\mathrm{C}}}^{(1)} {r_{3}'}^{(1)}~,\label{E-3.3.1-28}
\end{equation}
modulo an overall damping factor $\xi \equiv 4\pi^2/L^2$. If the scattering parameters are small and of the
same magnitude of $1/L$ as in the S-wave case, the importance of the auxiliary parameters just introduced can
be quantitatively assessed. In particular, by assigning one unit of 'weight' for each scattering parameter in the
effective-range expansion and one unit for $1/L$, we find that, neglecting the fine-structure constant, the
largest parameter is $d_1$ (order three), followed by $d_0$ and $d_2$ (order four), whereas the constants
$d_3$, $d_4$ and $d_5$ are of order six, eight and ten respectively. With the aim of finding a perturbative
formula exact to third order in $d_0$, we observe that in the final expression for the squared momentum
shift $\delta\tilde{p}^2$, only terms of order smaller or equal to twelve in $1/L$ and in the scattering
parameters should be retained. This fact justifies the inclusion of the above higher order scattering
parameters in the effective range expansion in eq.~\eqref{E-3.3.1-27}. Furthermore, rewriting $\tilde{p}^2$ in
the latter equation as $\delta\tilde{p}^2+1$, the approximated effective range expansion can be rewritten as a
power series of the squared momentum shift, 
\begin{equation}
\begin{gathered}
0 =  \frac{2}{\delta\tilde{p}^2}(\chi_1 d_1 + \Qoppa_0 d_1 - 3d_0) + 1 + d_0(\mathcal{I}^{(1)} - 6) 
+ d_1(\chi_1+2\chi_2 - 6\mathcal{J}^{(1)} -\Qoppa_1 -\tilde{\mathcal{R}}^{(1)}) \\ - \frac{d_2}{2}-d_3-d_4-d_5 
+ \delta\tilde{p}^2 \left[d_0(\mathcal{I}^{(1)} + \mathcal{J}^{(1)}) 
+ d_1\left( \chi_2 + 2\chi_3 -6\mathcal{K}^{(1)} - \Qoppa_2 - 2\mathcal{R}_{24}^{(1)} - \tilde{\mathcal{R}}^{(1)}\right)
 \right. \\ \left. -\frac{d_2}{2} -2d_3 -3d_4 - 4d_5 \right]  + (\delta\tilde{p}^2)^2 \left[ d_0(\mathcal{K}^{(1)} 
+ \mathcal{J}^{(1)}) - d_3  - 3d_4 - 6d_5 \right. \\ \left. + d_1\left(\chi_3 + 2\chi_4 -6\mathcal{L}^{(1)} - \Qoppa_3 
- \mathcal{R}_{44}^{(1)} - 2\mathcal{R}_{26}^{(1)} - 2\mathcal{R}_{24}^{(1)}\right) \right]~.\label{E-3.3.1-29}
\end{gathered}
\end{equation}
Due to the smallness of $\delta\tilde{p}^2$, contributions multiplied by higher positive powers of the lattice
momentum are increasingly suppressed. It follows that the dominant finite volume corrections are expected
to be found by solving the truncated version of eq.~\eqref{E-3.3.1-29} to order zero in $\delta\tilde{p}^2$,
\begin{equation}
\begin{split}
\frac{2}{\delta\tilde{p}^2}(\chi_1 d_1 + \Qoppa_0 d_1 - 3d_0) + 1 + d_0(\mathcal{I}^{(1)} - 6) - \frac{d_2}{2}-d_3 \\
+ d_1\left(\chi_1+2\chi_2 - 6\mathcal{J}^{(1)} -\Qoppa_1 -\tilde{\mathcal{R}}^{(1)}\right) &  -d_4-d_5 = 0~.
\label{E-3.3.1-30}
\end{split}
\end{equation}
Solving the last equation for $\delta\tilde{p}^2$ and expanding the denominator up to order twelve in the
small constants, we find
\begin{equation}
\begin{gathered}
\delta\tilde{p}^2 = d_1 \Big\{ -2(\chi_1+\Qoppa_0)\left[1+ \frac{d_2}{2} + \frac{d_2^2}{4} + d_3 + d_4 \right] 
+ d_0(1+d_2)\left[ 2(\mathcal{I}^{(1)} - 6)(\Qoppa_0 + \chi_1) \right. \\ \left.+ 6(\Qoppa_1-\chi_1 -2\chi_2 
+\tilde{\mathcal{R}}^{(1)} + 6 \mathcal{J}^{(1)}) \right] + d_0^2 (\mathcal{I}^{(1)}-6)\left[ \frac{}{} 12(\Qoppa_1 -\chi_1
-2\chi_2 \right.  \\ \left. + \tilde{\mathcal{R}} + 6\mathcal{J}^{(1)}) + (\mathcal{I}^{(1)} - 6)(\Qoppa_0 + 2\chi_1)
\right]\Big\}  + 6 d_0\left[1 + \frac{d_2}{2} - d_0(1+d_2)(\mathcal{I}^{(1)} - 6) \right. \\ \left.
+ d_0^2(\mathcal{I}^{(1)} - 6)^2 + \frac{d_2^2}{4}  + d_3 + d_4   \right]~.\label{E-3.3.1-31}
\end{gathered}
\end{equation}
Denoting the outcome of the first iteration as $\delta\tilde{p}^2_0$, we proceed with an improvement of the last
result by plugging the latter into the term proportional to $\delta\tilde{p}^2$ of the expansion in
eq.~\eqref{E-3.3.1-29} truncated to the terms quadratic in the squared momentum shift, 
\begin{equation}
\begin{gathered}
\frac{1}{\delta\tilde{p}^2}(2\chi_1 d_1 + 2\Qoppa_0 d_1 - 6d_0) + 1 + d_0(\mathcal{I}^{(1)} - 6) -d_3-d_4 
+ d_1(\chi_1+2\chi_2 - 6\mathcal{J}^{(1)} -\Qoppa_1 -\tilde{\mathcal{R}}^{(1)}) \\ - \frac{d_2}{2}-d_5 
+ \delta\tilde{p}_0^2 \left[d_0(\mathcal{I}^{(1)} + \mathcal{J}^{(1)}) -\frac{d_2}{2} -2d_3 -3d_4 - 4d_5 \right. \\
\left. + d_1\left( \chi_2 + 2\chi_3 -6\mathcal{K}^{(1)} - \Qoppa_2 - 2\mathcal{R}_{24}^{(1)} - \tilde{\mathcal{R}}^{(1)}\right)
\right] = 0~.\label{E-3.3.1-32}
\end{gathered}
\end{equation}
Solving the last equation and retaining only terms up to order twelve in $1/L$ and the scattering parameters,
we obtain a refined version of the momentum shift, 
\begin{equation}
\begin{gathered}
\delta\tilde{p}_2^2 \equiv \delta\tilde{p}^2 = d_1 \Big\{ -2(\chi_1+\Qoppa_0)\left[1+ d_2 + d_2^2 + d_3 + d_4 \right] 
+ d_0(1+2d_2)\left[ 2(\mathcal{I}^{(1)} - 6)(\Qoppa_0 + \chi_2) \right. \\ \left.  + 6(\Qoppa_1-\chi_1 -2\chi_2
+\tilde{\mathcal{R}}^{(1)} + 6 \mathcal{J}^{(1)}) \right]  + 12d_0 d_2(\Qoppa_0 + \chi_1)  - d_0^2 (\mathcal{I}^{(1)}-6)
\left[12(\Qoppa_1-\chi_1  -2\chi_2 \right. \\ \left. + \tilde{\mathcal{R}} + 6\mathcal{J}^{(1)}) + 2(\mathcal{I}^{(1)} - 6)
(\Qoppa_0 + \chi_1)  \right] + 12d_0^2\left[ 3(\Qoppa_2 + \tilde{\mathcal{R}}^{(1)}  + 6\mathcal{K}^{(1)}  +2\mathcal{R}_{24}) \right. \\
\left.- 3(\chi_2+2\chi_3) + 2(\mathcal{I}^{(1)}+\mathcal{J}^{(1)})(\Qoppa_0 + \chi_1)\right]\Big\} 
+ 6 d_0\left[1 + d_2 + d_2^2 + d_3 + d_4 \right. \\ \left. - d_0(1+2d_2)(\mathcal{I}^{(1)} - 6) +3d_0d_2 +
d_0^2(\mathcal{I}^{(1)} - 6)^2 - 6d_0^2(\mathcal{I}^{(1)}+\mathcal{J}^{(1)})\right]~,  
\end{gathered}
\label{E-3.3.1-33}
\end{equation}
in which the new contributions appearing at each order in the auxiliary constants (cf. eq~\eqref{E-3.3.1-31}) have
been isolated. Restoring the dimensional constants in the momenta, the energy of the lowest $T_1$ scattering state can be now obtained in few steps,
\begin{equation}
\begin{gathered}
E_{\mathrm{S}}^{(1,T_1)}  =  \frac{4 \pi^2}{M L^2}  +  \frac{4 \pi^2\delta\tilde{p}_2^2}{M L^2}  = \frac{4 \pi^2}{M L^2} 
+ 6 \xi \frac{4\pi {a'_{\mathrm{C}}}^{(1)}}{ML^3}\left[1 + \xi^2 {a'_{\mathrm{C}}}^{(1)} {r_{0}'}^{(1)}
+ \xi^3 {a'_{\mathrm{C}}}^{(1)} {r_{1}'}^{(1)}  \right. \\ \left. + \xi^4 {a'_{\mathrm{C}}}^{(1)}\left({a'_{\mathrm{C}}}^{(1)}
({r_{0}'}^{(1)})^2+  {r_{2}'}^{(1)} \right)     - \xi \left(\frac{{a'_{\mathrm{C}}}^{(1)}}{\pi L}\right)\left(1
+2\xi^2 {a'_{\mathrm{C}}}^{(1)} {r_{0}'}^{(1)}\right)(\mathcal{I}^{(1)} - 6)  \right. \\ \left.  +  3  \xi^3
\left(\frac{{a'_{\mathrm{C}}}^{(1)}}{\pi L}\right){a'_{\mathrm{C}}}^{(1)}{r_{0}'}^{(1)} + \xi^2 \left(
\frac{{a'_{\mathrm{C}}}^{(1)}}{\pi L}\right)^2(\mathcal{I}^{(1)} - 6)^2      - 6 \xi^2 \left(\frac{{a'_{\mathrm{C}}}^{(1)}}{\pi L}
\right)^2(\mathcal{I}^{(1)}+\mathcal{J}^{(1)}) + \ldots  \right] \\ + \xi \frac{\alpha~{a'_{\mathrm{C}}}^{(1)}}{L^2\pi^2}
\Big\{ -(2\chi_1+2\Qoppa_0)\left[1+ \xi^2 {a'_{\mathrm{C}}}^{(1)} {r_{0}'}^{(1)} + \xi^3 {a'_{\mathrm{C}}}^{(1)} {r_{1}'}^{(1)}
+ \xi^4 {a'_{\mathrm{C}}}^{(1)}\left({a'_{\mathrm{C}}}^{(1)} ({r_{0}'}^{(1)})^2+  {r_{2}'}^{(1)} \right) \right] \\
  + \xi\left(\frac{{a'_{\mathrm{C}}}^{(1)}}{\pi L}\right)\left(1+2\xi^2 {a'_{\mathrm{C}}}^{(1)} {r_{0}'}^{(1)}\right)\left[
(\mathcal{I}^{(1)} - 6)(2\Qoppa_0 + 2\chi_1) + 6(\Qoppa_1-\chi_1 -2\chi_2+\tilde{\mathcal{R}}^{(1)} + 6 \mathcal{J}^{(1)})
\right] \\  - \xi^2 \left(\frac{{a'_{\mathrm{C}}}^{(1)}}{\pi L}\right)^2 (\mathcal{I}^{(1)}-6)\left[12(\Qoppa_1-\chi_1 -2\chi_2
+ \tilde{\mathcal{R}}  + 6\mathcal{J}^{(1)}) + (\mathcal{I}^{(1)} - 6)(2\Qoppa_0 + 2\chi_1)  \right] \\
+ 12\xi^2\left(\frac{{a'_{\mathrm{C}}}^{(1)}}{\pi L}\right)^2\left[ 3(\Qoppa_2 + \tilde{\mathcal{R}}^{(1)}
+ 6\mathcal{K}^{(1)} +2\mathcal{R}_{24})  - 3(\chi_2+2\chi_3) + (\mathcal{I}^{(1)}+\mathcal{J}^{(1)})(2\Qoppa_0 + 2\chi_1)
\right] \\ + 6\xi^3 \left(\frac{{a'_{\mathrm{C}}}^{(1)}}{\pi L}\right){a'_{\mathrm{C}}}^{(1)}{r_{0}'}^{(1)}(2\Qoppa_0 + 2\chi_1)
+ \ldots \Big\}  + \ldots ~,\label{E-3.3.1-34}
\end{gathered}
\end{equation}
where the ellipsis stands for terms of higher order in the scattering and $1/L$ parameters and in the fine-structure
constant. Besides, all the terms on the r.h.s of the last equation with the only exception of the first represent
the modifications of a free $T_1$ lattice state with energy $\xi/M$ induced by the strong and the electromagnetic
interactions. Analogously to the $\ell=0$ case, only the interplay between strong and electromagnetic forces
generates the linear corrections in $\alpha$, and the leading QED corrections are of the same order of the
modifications due to the QCD forces alone. By comparison with the S-wave counterpart of eq.~\eqref{E-3.3.1-34}
(cf. eq.~(37) in ref.~\cite{BeS14}), we observe that contributions from higher order scattering parameters
such as  ${r_{2}'}^{(1)}$ and ${r_{2}'}^{(1)}$ begin to appear, whereas all the terms arising from ${r_{3}'}^{(1)}$,
included in the original version of the ERE, vanish in the order twelve expansion in the scattering parameters.
Furthermore, by explicit computation it can be proven that the subsequent iteration step for the improvement
of the squared momentum shift, $\delta\tilde{p}_4^2$, does not lead to the appearance of further addend on the
r.h.s. of eq.~\eqref{E-3.3.1-34} in the chosen approximation scheme. Finally, we conclude the treatment by
isolating the corrections in the last equation and restoring the infinite volume scattering parameters, 

\begin{equation}
\begin{gathered}
\Delta E_{\mathrm{S}}^{(1,T_1)}  = 6 \xi \frac{4\pi a_{\mathrm{C}}^{(1)}}{ML^3}\Big\{1 + \xi^2 a_{\mathrm{C}}^{(1)} r_{0}^{(1)}
+ \xi^3 a_{\mathrm{C}}^{(1)} r_{1}^{(1)} + \xi^4 a_{\mathrm{C}}^{(1)}\left(a_{\mathrm{C}}^{(1)} {r_{0}^{(1)}}^2+  r_{2}^{(1)} \right) \\
+ \xi \left(\frac{a_{\mathrm{C}}^{(1)}}{\pi L}\right)\left[ 3\xi^2 a_{\mathrm{C}}^{(1)}r_{0}^{(1)} - \left(1+2\xi^2 a_{\mathrm{C}}^{(1)}
r_{0}^{(1)}\right)(\mathcal{I}^{(1)} - 6) \right] + \xi^2 \left(\frac{a_{\mathrm{C}}^{(1)}}{\pi L}\right)^2\left[ (\mathcal{I}^{(1)}
- 6)^2 - 6 (\mathcal{I}^{(1)}+\mathcal{J}^{(1)}) \right] \\ + \ldots  \Big\} + \xi \frac{\alpha~a_{\mathrm{C}}^{(1)}}{L^2\pi^2}
\Big\{ 3a_{\mathrm{C}}^{(1)}\mathcal{I}^{(0)}\left(r_0^{(1)} + 8\xi^2r_1^{(1)} + 6\xi^4 r_2^{(1)}\right) + 3\xi {a_{\mathrm{C}}^{(1)}}^2
r_0^{(1)}\mathcal{I}^{(0)}(1+\xi)\left(r_0^{(1)} + \xi^2 r_1^{(1)}\right)  \\ -(2\chi_1+2\Qoppa_0)\left[1+ \xi^2 a_{\mathrm{C}}^{(1)}
{r_{0}'}^{(1)} + \xi^3 a_{\mathrm{C}}^{(1)} r_{1}^{(1)} + \xi^4 a_{\mathrm{C}}^{(1)}\left(a_{\mathrm{C}}^{(1)} {r_{0}^{(1)}}^2+  r_{2}^{(1)}
\right) \right] + 6\xi^3 \left(\frac{a_{\mathrm{C}}^{(1)}}{\pi L}\right)a_{\mathrm{C}}^{(1)}r_{0}^{(1)}(2\Qoppa_0 + 2\chi_1) \\
+ \xi \left(\frac{a_{\mathrm{C}}^{(1)}}{\pi L}\right)\left(1+2\xi^2 a_{\mathrm{C}}^{(1)} r_{0}^{(1)}\right)\left[ (\mathcal{I}^{(1)}
- 6)(2\Qoppa_0 + 2\chi_1) + 6(\Qoppa_1-\chi_1 -2\chi_2+\tilde{\mathcal{R}}^{(1)} + 6 \mathcal{J}^{(1)}) \right] \\
- \xi^2 \left(\frac{a_{\mathrm{C}}^{(1)}}{\pi L}\right)^2 (\mathcal{I}^{(1)}-6)\left[12(\Qoppa_1-\chi_1 -2\chi_2
+ \tilde{\mathcal{R}} + 6\mathcal{J}^{(1)}) + (\mathcal{I}^{(1)} - 6)(2\Qoppa_0 + 2\chi_1)  \right] \\
+ 12\xi^2\left(\frac{a_{\mathrm{C}}^{(1)}}{\pi L}\right)^2\left[ 3(\Qoppa_2 + \tilde{\mathcal{R}}^{(1)} +
6\mathcal{K}^{(1)} +2\mathcal{R}_{24}) - 3(\chi_2+2\chi_3) + (\mathcal{I}^{(1)}+\mathcal{J}^{(1)})(2\Qoppa_0 + 2\chi_1)\right] \\
- 3 \xi \left(\frac{a_{\mathrm{C}}^{(1)}}{\pi L}\right)a_{\mathrm{C}}^{(1)}r_0^{(1)} (1+\xi) \mathcal{I}^{(0)}
\left(\mathcal{I}^{(1)}-6\right)  + \ldots \Big\}  + \ldots ~,\label{E-3.3.1-35}
\end{gathered}
\end{equation}

where the differences with respect to eq.~\eqref{E-3.3.1-34} involve only the linear term in the fine-structure
constant and are proportional to $\mathcal{I}^{(0)}$. 

\subsubsection{\textsf{The Lowest Bound State}}\label{S-3.3.2} 

Although bound states between two hadrons of the same charge have not been observed in nature, at unphysical values
of the quark masses in Lattice QCD such states do appear \cite{BCD12,YIK12,BCC13-01,BCC13-02}. Moreover,
two-boson bound states originated by strong forces are expected to explain certain features of heavy quark compounds.
In particular, the interpretation of observed lines Y(4626), Y(4630) and Y(4660) of the hadron sprectrum in terms of
P-wave $[cs][\bar{c} \bar{s}]$ tetraquark states with $1^{--}$ seems promising \cite{DCP20}.\\
Additionally, loosely bound binary compounds of hadrons appearing in the vicinity of a P-wave strong decay threshold
are not forbidden by the theory of hadronic molecules \cite{GHM08}. Possible candidates of such two-body systems are
represented by the hidden charm pentaquark states $P_c^+(4380)$ and $P_c^+(4450)$, located slightly below the
$\bar{D}\Sigma_c^*$ and $\bar{D}^*\Sigma_c$ energy thresholds at $4385.3$~MeV and $4462.2$~MeV, respectively.
Although a wide variety of different studies on the two states have been conducted \cite{MeO15,YaS17,HeJ17},
a very recent one advances the molecular hypotesis \cite{SYH19} with orbital angular momentum equal to one in
the framework of heavy quark spin symmetry (HQSS).\\
Therefore, it remains instructive to study the lowest two-fermion $T_1$ bound state, by switching to
imaginary momenta $\mathbf{p}=\mathrm{i}\boldsymbol{\kappa}$, where $\kappa = |\boldsymbol{\kappa}|$
represents the imaginary part of the momentum. To this purpose, we rewrite the FV effective range expansion
in eq.~\eqref{E-3.2-05}, truncated on the l.h.s. to the sextic term in the binding momentum,
\begin{equation}
\begin{split}
-\frac{1}{{a'_C}^{(1)}} - \frac{1}{2}{r'_0}^{(1)} \kappa^2 + {r'_1}^{(1)} \kappa^4 -  {r'_2}^{(1)} \kappa^6  
=  \frac{4\pi}{L^3}\mathcal{S}_0(\mathrm{i}\tilde{\kappa}) + \frac{\tilde{\kappa}^2}{\pi L}\mathcal{S}_1(\mathrm{i}
\tilde{\kappa}) \\ - \frac{\alpha M \tilde{\kappa}^2}{4\pi^4} \mathcal{S}_2(\mathrm{i}\tilde{\kappa})   
-\frac{\alpha M}{\pi^2 L^2}\mathcal{S}_3(\mathrm{i}\tilde{\kappa}) + \alpha M\tilde{\kappa}^2 & \left[ \log
\left(\frac{4\pi}{\alpha M L}\right)-\gamma_E \right]~.\label{E-3.3.2-01}
\end{split}
\end{equation}
First, we consider the limit of large lattice binding momentum, $\tilde{\kappa} = |\tilde{\boldsymbol{\kappa}}| \gg 1$,
which corresponds to a tightly-bound state. Thus, an approximation for the L\"uscher functions in this regime
becomes necessary. In particular, we observe that the asymptotic behaviour of $\mathcal{S}_1(\mathrm{i}\tilde{\kappa})$
and $\mathcal{S}_2(\mathrm{i}\tilde{\kappa})$ is already available in literature and gives, 
\begin{equation}
\mathcal{S}_1(\mathrm{i}\tilde{\kappa}) = \sum_{\mathbf{n}}^{\Lambda_n} \frac{1}{|\mathbf{n}|^2
+ \tilde{\kappa}^2} -4\pi\Lambda_n \rightarrow - 2\pi^2\tilde{\kappa} \label{E-3.3.2-02}
\end{equation}
and 
 \begin{equation}
\mathcal{S}_2(\mathrm{i}\tilde{\kappa})  = \sum_{\mathbf{n}}^{\Lambda_n}\sum_{\mathbf{m} \neq \mathbf{n}}^{\infty} \frac{1}{|\mathbf{n}|^2
+ \tilde{\kappa}^2} \frac{1}{|\mathbf{m}|^2 + \tilde{\kappa}^2} \frac{1}{|\mathbf{n}-\mathbf{m}|^2} 
- 4\pi^4\log\Lambda_n \rightarrow -4\pi^4 \log(2\tilde{\kappa})+ \frac{\pi^2}{\tilde{\kappa}} \mathcal{I}^{(0)}~,
\label{E-3.3.2-03}
\end{equation}
see eqs.~(43,44) in ref.~\cite{BeS14}, respectively. On the other hand, the large binding energy limit of
$\mathcal{S}_3(\mathrm{i}\tilde{\kappa})$ requires a new derivation, presented in detail in app.~\eqref{S-A-6.0},
\begin{equation}
\begin{split}
\mathcal{S}_3(\mathrm{i}\tilde{\kappa})=  \sum_{\mathbf{n}}^{\Lambda_n}\sum_{\mathbf{m} \neq \mathbf{n}}^{\infty} \frac{1}{|\mathbf{n}|^2
+ \tilde{\kappa}^2} \frac{1}{|\mathbf{m}|^2 + \tilde{\kappa}^2} \frac{\mathbf{n}\cdot\mathbf{m} + \tilde{\kappa}^2}
{|\mathbf{n}-\mathbf{m}|^2} - \pi^4 \Lambda_n^2 \rightarrow \frac{\pi^2}{2\tilde{\kappa}} - 2\pi^2\tilde{\kappa}
\mathcal{I}^{(0)} - 2\pi^4 \tilde{\kappa}^2 ~.\label{E-3.3.2-04}
\end{split}
\end{equation}
Recalling the fact that $\mathcal{S}_0(\tilde{\kappa}) ) = \Upsilon = 0$ (cf. eq.~\eqref{E-A-5.1-01})
and collecting the results in eqs.~\eqref{E-3.3.2-02}-\eqref{E-3.3.2-04}, the above finite volume effective
range expansion becomes,
\begin{equation}
-\frac{1}{{a'_C}^{(1)}} - \frac{1}{2}{r'_0}^{(1)} \kappa^2 + {r'_1}^{(1)} \kappa^4 -  {r'_2}^{(1)} \kappa^6 =  \kappa^3
+ \frac{\alpha M}{2} \kappa^2 + \frac{3}{2}\frac{\alpha M}{\pi L}\mathcal{I}^{(0)}\kappa - \frac{\alpha M \pi}{L^3 \kappa}
- \alpha M \kappa^2\left[ \log\left(\frac{4\kappa}{\alpha M}\right)   - \gamma_E\right]~.\label{E-3.3.2-05}
\end{equation}
Second, we highlight the dependence on the fine-structure constant in the last equation by rewriting the
binding momentum in a power series,
\begin{equation}
\kappa = \kappa_0+  \kappa_1 +  \kappa_2 + \ldots~,\label{E-3.3.2-06}
\end{equation}
where $\kappa_0$ results from strong interactions alone and the subscript corresponds to the power of $\alpha$ on
which each term in the expansion depends. Replacing in eq.~\eqref{E-3.3.2-05} the shifted scattering
parameters with the infinite volume ones in eqs.~\eqref{E-3.0-05}-\eqref{E-3.0-09} and discarding all the terms of
order $\alpha^2$ or higher in the expansions, eq.~\eqref{E-3.3.2-05} transforms into
\begin{equation}
\begin{gathered}
-\frac{1}{a_{\mathrm{C}}^{(1)}} + \frac{\alpha  M \mathcal{I}^{(0)}}{2\pi L} r_0^{(1)} - \frac{1}{2} r_0^{(1)} \kappa_0^2
- r_0^{(1)} \kappa_0 \kappa_1 + r_1^{(1)} \kappa_0^4   - \frac{2\alpha M \mathcal{I}^{(0)}}{\pi L} r_1^{(1)} \kappa_0^2
+ 4 r_1^{(1)} \kappa_0^3\kappa_1 \\ + \frac{3\alpha M \mathcal{I}^{(0)}}{\pi L}r_2^{(1)} \kappa_0^4   - r_2^{(1)} \kappa_0^6 
- 6 r_2^{(1)} \kappa_0^5 \kappa_1 - \frac{4\alpha M \mathcal{I}^{(0)}}{\pi L} r_3^{(1)} \kappa_0^6 = \kappa_0^3
+ \frac{3}{2}\frac{\alpha M}{2\pi L} \mathcal{I}^{(0)} \kappa_0 \\ - 3\kappa_0^2 \kappa_1 - \frac{\alpha M \pi}{L^3 \kappa_0}
+ \frac{\alpha M}{2} \kappa_0^2 + \alpha M \kappa_0^2 \left[ \log\left(\frac{4\kappa_0}{\alpha M}\right)
- \gamma_E \right]~.\label{E-3.3.2-07}
\end{gathered}
\end{equation}
Grouping all the terms independent on $\alpha$, we observe that the following equality
\begin{equation}
-\frac{1}{a_{\mathrm{C}}^{(1)}}  - \frac{1}{2} r_0^{(1)} \kappa_0^2   + r_1^{(1)} \kappa_0^4  - r_2^{(1)} \kappa_0^6
= \kappa_0^3~,\label{E-3.3.2-08}
\end{equation}
holds, since the remaining terms in eq.~\eqref{E-3.3.2-07} depend linearly on $\alpha$. Therefore, an expression
for $\kappa_1$ can be drawn from the original eq.~\eqref{E-3.3.2-07} by dropping the terms listed in eq.~\eqref{E-3.3.2-08},
\begin{equation}
\begin{split}
\kappa_1 \left(3\kappa_0^2 - r_0^{(1)} \kappa_0 + 4 r_1^{(1)} \kappa_0^3 - 6 \kappa_0^5 r_2^{(1)} \right) 
= \frac{\alpha M \mathcal{I}^{(0)}}{2\pi L} \left(3 \kappa_0  - r_0^{(1)} + 2 r_1^{(1)} \kappa_0^2  \right. \\ \left. 
- 6 r_2^{(1)} \kappa_0^4 + 8 r_3^{(1)} \kappa_0^6 \right)  -\frac{\alpha M\pi}{L^3 k_0} + \alpha M \kappa_0^2 \left[
\log\left( \frac{4\kappa_0}{\alpha M} \right) \right. & \left. - \gamma_E + \frac{1}{2}\right]~.\label{E-3.3.2-09}
\end{split}
\end{equation}
In particular, retaining only the terms depending on the two lowest order scattering parameters as in the zero
angular momentum case (cf. eqs.~(45)-(46) in ref.~\cite{BeS14}), a more approximated expression for $\kappa_1$
in terms of $\kappa_0$, $a_{\mathrm{C}}^{(1)}$ and $r_0^{(1)}$ can be obtained,
\begin{equation}
\kappa_1  \approx \frac{\alpha M }{2\pi L} \frac{\mathcal{I}^{(0)}}{\kappa_0} - \frac{\alpha M}{\pi^3 L^3}
\frac{\pi^4}{3\kappa_0^3 - r_0^{(1)} \kappa_0^2}   + \frac{\alpha M \kappa_0^2}{3\kappa_0^2 - r_0^{(1)} \kappa_0} \left[
\log\left( \frac{4\kappa_0}{\alpha M} \right)  - \gamma_E+\frac{1}{2} \right]  ~.\label{E-3.3.2-10}
\end{equation}
From the latter equation the first two terms generate finite-volume corrections, whereas the third one introduces
QED modifications to the unperturbed binding momentum $\kappa_0$ which do not vanish in the infinite volume limit.
Now, considering the binding energy of the lowest $T_1$ state in the linear approximation in $\alpha$, 
\begin{equation}
E_{B}^{(1,T_1)} = \frac{\kappa^2}{M} = \frac{\kappa_0^2}{M} + 2\frac{\kappa_0\kappa_1}{M} + \ldots~,\label{E-3.3.2-11}
\end{equation}
and substituting the simplified expression of $\kappa_1$ in eq.~\eqref{E-3.3.2-10}, we find an approximate expression
for the energy of the lowest $T_1$ bound state,
\begin{equation}
E_{B}^{(1,T_1)}(L) = \frac{\kappa_0^2}{M} + \frac{2 \alpha \kappa_0^3}{3\kappa_0^2 - r_0^{(1)} \kappa_0} \left[  \log
\left( \frac{4\kappa_0}{\alpha M} \right) - \gamma_E + \frac{1}{2} \right]  + \frac{\alpha\mathcal{I}^{(0)}}{\pi L}
- \frac{\alpha}{\pi^3 L^3} \frac{2\pi^4}{\kappa_0^2} \frac{1}{3k_0 - r_0^{(1)}}~.\label{E-3.3.2-12}
\end{equation}
where the first two terms represent the infinite volume binding energy up to $\mathcal{O}(\alpha)$,
while the third contribution vanishes for $L \rightarrow +\infty$ together with the fourth one.
Furthermore, if $r_0^{(0)}$ is small enough, the denominator in the second and in the fourth term of the
last equation can be expanded in powers of $r_0^{(0)}/\kappa_0 \ll 1$ and the leading order mass shift for
tightly-bound $T_1$ two-body states eventually becomes,
\begin{equation}
\Delta E_{B}^{(1,T_1)} \equiv E_{B}^{(1,T_1)}(\infty) -  E_{B}^{(1,T_1)}(L)  = - \frac{\alpha\mathcal{I}^{(0)}}{\pi L}  
+ \frac{\alpha}{\pi^3 L^3} \frac{2\pi^4}{3k_0^3 - r_0^{(1)}k_0^2} \approx - \frac{\alpha\mathcal{I}^{(0)}}{\pi L}
+ \frac{\alpha}{\pi^3 L^3}  \frac{2\pi^4}{3\kappa_0^3}~.\label{E-3.3.2-13}
\end{equation}
From the r.h.s. of the last expression we can infer that the leading QED corrections to $E_{B}^{(1,T_1)}(L)$ presented
in eq.~\eqref{E-3.3.2-13} are positive ($\mathcal{I}^{(0)} <0$, see eq.~\eqref{E-A-5.1-16}), analogously to the
LO mass shifts for $\ell=1$ states of two-body systems with strong interactions alone in eq.~(53) of
refs.~\cite{KLH12,KLH11}. Also noteworthy is the fact that, when compared to the S-wave case in eq.~(46) in
ref.~\cite{BeS14}, the sign of the P-wave shift is reversed while the magnitude remains unchanged. As discussed
in refs.~\cite{KLH12,KLH11}, the significance of this behaviour can be traced back to the spatial profile of
the $\ell=0$ and $\ell = 1$ two-body wavefunctions associated to the considered bound eigenstates. Qualitatively,
the relationship found between the two finite volume energy corrections means that zero angular momentum states
are more deeply bound when embedded in a finite volume, while the counterpart with one unit of angular momentum are
less bound. In conclusion, together with the derivation of $\Delta E_{B}^{(1,T_1)}$, we have simultaneously proven
that the addition of a long-range force on top of strong forces in two-fermion systems within a cubic lattice
produces changes of the same magnitude on S- and P-wave bound energy eigenvalues. 

%--------------------------------------------------------------------------------------------------------------------------------------------------------------------

\section{\textsf{Outlook}}\label{4.0}

Before drawing the conclusions of our work, we qualitatively outline the possible improvements to the above analysis.
There are two main directions for the generalization of the present study. These consist in the inclusion of
transverse photons in the Lagrangian density of the system and in the treatment of strong interactions
with higher angular momentum couplings.

\begin{figure}[ht]
\includegraphics[width=1.0\columnwidth]{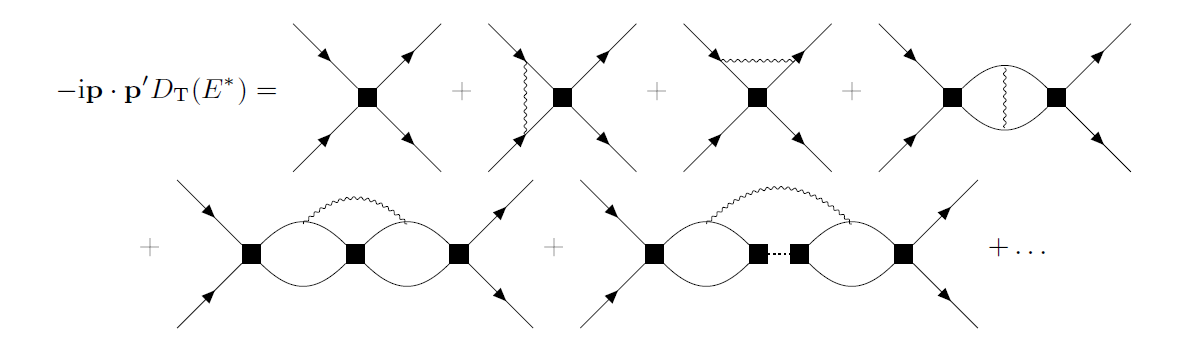}
\caption{The \textit{dressed} strong vertex coupling to one unit of angular momentum as a superposition of
the original fermion-fermion interaction vertex plus a numerable infinity of tree-level and bubble diagrams
containing one transverse-photon exchange. The ellipses represent other diagrams featuring one or
more strong vertices and single transverse-photon insertions.}\label{F-4-01}
\end{figure}

Let us start from the former.  Thanks to their vector nature, transverse photons can couple to the fermionic
fields in several different ways, see eq.~\eqref{E-2.1-01}. Among these, here we retain only the Dipole
vertex (cf. app.~\ref{S-A-1.0}), which is expected to yield the leading-order contributions to the
T-matrix and the full Green's function. As we previously hinted, these photons, denoted by wavy lines,
can propagate also between different bubbles. Consequently, at each order in $D(E^*)$ a numerable infinity
of new diagrams with different topologies and transverse-photon exchanges inside and outside the bubbles appears.
Unfortunately, part of the amplitudes associated to these diagrams can not be written as powers of the
loop (bubble) integrals, since the transverse photons propagating outside the fermion loops introduce a
correlation between the bubbles (cf. fig.~(4) in ref.~\cite{BeS14} and fig.~\ref{F-4-01}). It follows
that an expression for the T-matrix element of the two-body scattering process, $T_{\mathrm{SCT}}$, written in
terms of a geometric series of ratio proportional to the interaction strength, $D(E^*)$, can not be found.
More formally, a self-consistent rewriting of the full Green's function operator $\hat{G}_{\mathrm{SCT}}$ in
the form of a self-consistent equation \textit{\`a la Dyson} separating the QED Green's function operator,
$\hat{G}_{\mathrm{CT}}$, from the strong interaction operator $\mathcal{V}^{(1)}$ (cf. eq.~\eqref{E-2.1-20})
does not exist. These facts prevent the exact determination of $T_{\mathrm{SCT}}$ to all orders in the
fine-structure constant.

Nevertheless, the gapped nature of the momentum in the lattice environment allows for a perturbative
treatment of the whole non-relativistic QED. Therefore, approximate expressions for $T_{\mathrm{SCT}}$ that
incorporate the effects of the transverse photons up to the desired order in $\alpha$ can be derived.

One of these approaches consists in writing the infinite-volume T-matrix element $T_{\mathrm{SCT}}$ exactly as
$T_{\mathrm{SC}}$ in eq.~\eqref{E-2.2-14} with $D(E^*)$ at the denominator replaced by a \textit{dressed}
strong P-wave coupling constant $D_{\mathrm{T}}(E^*)$, that includes the effects of transverse photons up to
first order in the fine-structure constant. Analytically, this energy-depen\-dent constant can be
derived by evaluating the contributions of all the possible bubble diagrams with one tranverse-photon
exchange in fig.~\ref{F-4-01}. Moreover, considering the fact that diagrams with one transverse photon
across $n$-bubbles are suppressed by a factor $(\sqrt{|\mathbf{p}|})^n$, the numerable inifnity of contributions
on the r.h.s. of fig.~\ref{F-4-01} can be reduced to a finite set. The amplitudes corresponding to these
diagrams can be evaluated via dimensional regularization as the ones containing radiation pions
in refs.~\cite{KSW98-02,KSW99,MeS00} or via the cutoff approach and, in finite volume, they can be
constructed by replacing the relevant integrals with three-dimensional sums, eventually regularized
by a spherical cutoff.

Furthermore, the finite-volume quantization conditions can be derived as in sec.~\ref{S-3.1}, keeping
track of the trans\-verse-photon contributions via the aforementioned redefinition of the strong
coupling constant. In the end of the process, an expression for $1/D(E^*)$ analogous to eq.~\eqref{E-3.1-06}
is found, provided the powers of the original strong coupling constant in the QED contributions to
$D_{\mathrm{T}}(E^*)$ in fig.~\ref{F-4-01} are replaced by scattering parameters (cf. sec.~III~B of ref.~\cite{BeS14})
using the expression of $D(E*)$ in eq.~\eqref{E-2.0-25}.

In the present linear approximation in the fine-structure constant, the latter operation is, in fact, justified.
Due to their dependence on the scattering parameters, the new terms arising from the transverse-photon interactions
are not expected to bring the dominant $\mathcal{O}(\alpha)$ contibution to the finite-volume corrections of the
lowest $T_1$ scattering state. Additionally, such terms may have no effect in the final expression for the
corrected version of $\Delta E_{B}^{(1,T_1)}$ in eq.~\eqref{E-3.3.2-13}, due to the approximations done in the
derivation of the latter in sec.~\ref{S-3.3.2}.

As we previously noted, the second main generalization of our work consists of the adoption of strong
interactions coupled to more units of angular momentum. The most significant of these extensions is
represented by the D-wave case, where the strong part of the EFT Lagrangian density becomes
\begin{equation}
\mathcal{L} = \psi^{\dagger}\left[\mathrm{i}\hbar\partial_t - \frac{\hbar^2\nabla^2}{2M}\right]\psi
+ \frac{F(E^*)}{6} (\psi \psi)^{\dagger}(\psi \psi) - \frac{F(E^*)}{32}(\psi \overleftrightarrow{\partial}_i
\overleftrightarrow{\partial}_j\psi)^{\dagger}(\psi \overleftrightarrow{\partial}_i
\overleftrightarrow{\partial}_j \psi) ~,\label{E-4.0-01}
\end{equation} 
where $F(E^*)$ is a new suitable energy-dependent coupling constant (cf. sec.~\ref{S-2.0}).
Due to the presence of higher-order differential operators acting on the fermionic fields, the computation
of the strong scattering amplitude, $T_{\mathrm{S}}$, via a geometric series on the loop integrals, as in
eq.~\eqref{E-2.0-15}, involves a rank-four tensor as a ratio. The elements of the D-wave counterpart of
$\mathbb{J}_0$ correspond to double mixed derivatives of the free Green's function, except for the diagonal
terms, in which an additional contribution proportional to $G_0(\mathbf{0},\mathbf{0})$ is expected to be present.
Also the full Green's function $G_{\mathrm{SC}}(\mathbf{0},\mathbf{0})$ is likely to undergo similar changes,
which lead to the onset of more rapidly UV-divergent integrals in the D-wave counterpart of
$\mathbb{J}_{\mathrm{SC}}$. Besides, some novelties are expected to arise from the quantization condition
stemming from the full two-point correlation function $G_{\mathrm{SC}}$. The D-wave counterpart 
of $\mathbb{J}_{\mathrm{C}}$ in finite volume generates the constraints on $1/F^L(E^*)$ for energy states 
transforming as two distinct representations of the cubic group, $\mathcal{O}$. The $\ell=2$ irreducible 
representation of $SO(3)$, in fact, decomposes into the $E \oplus T_2$ irreps \cite{Joh82,Alt57} of $\mathcal{O}$. 
As a consequence, the one-to-one correspondence between the transformation properties of the selected multiplet 
of states under the operations of SO(3) and $\mathcal{O}$ is no longer valid. It is, thus, possible that a derivation 
of the finite-volume corrections that makes use of the effective range expansion for D-waves in ref.~\cite{KMB82},
\begin{equation}
\left(\eta^4+5\eta^2+4\right)\frac{\mathbf{p}^4}{4}\left[C_{\eta}^2|\mathbf{p}|(\cot\delta_2-\mathrm{i})
+ \alpha M H(\eta)\right]  = -\frac{1}{a_{\mathrm{C}}^{(2)}}+\frac{1}{2}r_0^{(2)}\mathbf{p}^2+ r_1^{(2)}\mathbf{p}^4
+ \ldots~,\label{E-4.0-02}
\end{equation}
has to follow two separate paths for the $E$ and the $T_2$ states. New challenges arise also in the evaluation
of the cutoff-regularized double sums stemming from the $\delta\tilde{p}^2$ expansions (cf. sec.~\ref{S-3.3.1})
of the L\"uscher functions. Nevertheless, the reference lattice state for the lowest unbound energy level
is expected to correspond again to the one with $|\tilde{\mathbf{p}}| = 1$ and energy $4\pi^2/L^2 M$. Therefore,
the final formulae for the finite-volume energy corrections to the lowest $E$ and $T_2$ scattering states are
likely to conserve some resemblance with the one presented in eq.~\eqref{E-3.3.1-35}. Concerning the lowest
bound state, it would be certainly of interest to compare the magnitude of the outcoming corrections with the
existing ones for the energy of the most bound S- and P-wave states.

%--------------------------------------------------------------------------------------------------------------------------------------------------------------------

\section{\textsf{Conclusion}}\label{5.0}

In this paper, we have extended the investigation of QED for low-energy scattering to the
case of spinless fermions strongly coupled to one unit of angular momentum in the framework of pionless EFT.
Among the forces of electromagnetic nature, the electrostatic interaction represents the dominant contribution
to T-matrix elements in the low-energy sector and the Coulomb ladders have to be resummed to all orders in
the fine-structure constant.

A pivotal role in the procedure has been covered by the non-perturbative formalism based on the full
Coulomb propagator developed comprehensively in ref.~\cite{KoR00} in the context of $pp$ scattering.
Differently from the transverse photons, the Coulomb ones do not propagate between the fermionic bubbles
in the diagrams, a crucial feature that permits to rewrite the full two-body Green function operator in terms
of the Coulomb one and operator representing the P-wave strong interaction. It is exactly the same property
that led us to write the T-matrix element of the scattering processes and the full Green's function in closed
form, thus paving the way to the derivation of the quantization conditions. Moreover, though the analysis of
the attractive Coulomb case, the observations on the scattering parameters pointed out in sec.~3.4
of ref.~\cite{KoR00} have found here another confirmation.

Second, the infinite-volume analysis of fermion-fermion scattering in secs.~\ref{S-2.0}-\ref{S-2.3} allowed
us to attain our main goal, the derivation of finite-volume energy corrections for two-body P-wave bound and
scattering states, by providing an extension of the analysis on S-wave states in ref.~\cite{BeS14}. Motivated
by the growing interest for lattice EFTs and, above all, LQCD, we have transposed our system of charged
particles in a cubic box with periodic boudary conditions.

Having regards to the prescriptions from the literature \cite{DET96,HiP83}, we have removed the zero
momentum modes from the relevant three-dimensional sums and considered the QED corrections to the mass of the
spinless particles \cite{DaS14,UnH08}, in sight of the application of our results to realistic baryon-baryon
systems on the lattice. Furthermore, the characteristic size of our cubic box has been chosen to fulfill the
constraint $ML \ll 1/\alpha$, which is required for the viability of the perturbative approach of QED on the
lattice. Under this hypothesis, the non-relativistic relation between the finite-volume energy of two
composite fermions in the $T_1$ representation of the cubic group and its P-wave scattering parameters
receives QED corrections obtainable in closed form.

Although more cumbersome than the S-wave counterpart, the expression we have presented in secs.~3.3.1
for the energy shift of the lowest unbound state resembles the features of the one in sec.~III~D~1 of
ref.~\cite{BeS14}, except for the appearance of higher-order scattering parameters. Besides, the
finite-volume corrections for the P-wave bound state prove to have the opposite sign and the same magnitude of
the ones for the S-wave state in sec.~III~D~3 of ref.~\cite{BeS14}, up to contributions of order $1/L^3$.
This fact confirms the long-standing observations on the $A_1$ and $T_1$ lattice energy eigenvalues in the
analysis of refs.~\cite{KLH12,KLH11}, drawn in the context of a two-body system governed by finite-range
interactions in the non-relativistic regime.

In the latter work, the interplay of parity and angular momentum quantum numbers in the wavefunctions was
found to be responsible of the relation between the leading-order S- and P-wave energy shifts. Only the
generalization of our analysis will tell if the existing relationships between the finite-volume shifts
in tab.~I of ref.~\cite{KLH12} for two-body states with higher angular momentum remain at least
approximately valid in presence of QED.

\section*{\textsf{Acknowledgements}}

First of all, we express our gratitude to Akaki Rusetski, Andria Agadjanov and Wael Elkamhawy for the precious
advice and their expertise in effective field theories and Filippo Stellin for the useful suggestions in the
drafting of the manuscript. 
Secondly, we thank Serdar Elhatisari for his introduction to cuda C++ programming for the GPUs, a crucial tool
for the precise evaluation of the double sums. Then, we recognize the contributions by Timo A. L\"ahde, Andreas Nogga,
Paul Kapinos and 
Dieter an Mey in the technical assistance. We also thank Hans-Werner Hammer for a useful communication.
Besides, we acknowledge financial support from the Deutsche Forschungsgemeinschaft (Sino-German CRC 110, grant No. TRR~110)
and the VolkswagenStiftung (grant No. 93562). The work of UGM was also supported by the Chinese Academy of Sciences
(CAS) President's International Fellowship Initiative (PIFI) (Grant No. 2018DM0034). Finally, we acknowlegde
computational resources provided by RWTH Aachen (JARA 0015 project).

%--------------------------------------------------------------------------------------------------------------------------------------------------------------------

\begin{appendices}

\section{\textsf{Feynman rules}}\label{S-A-1.0}

\begin{figure}[h]
\begin{center}
\begin{minipage}{0.48\columnwidth}
\includegraphics[width=1.0\columnwidth]{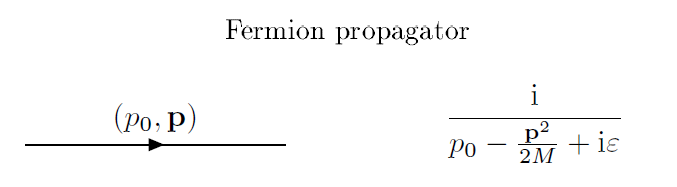}
\end{minipage}
\begin{minipage}{0.48\columnwidth}
\includegraphics[width=1.0\columnwidth]{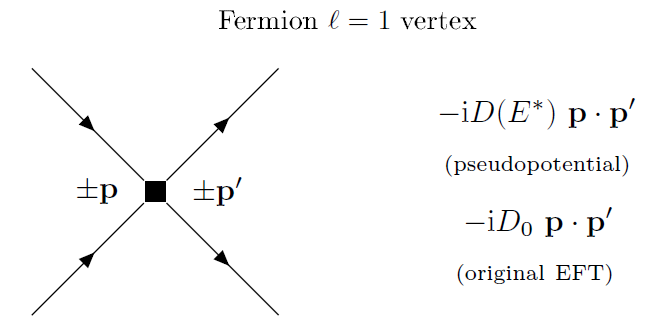}
\end{minipage}\\
\begin{minipage}{0.48\columnwidth}
\includegraphics[width=1.0\columnwidth]{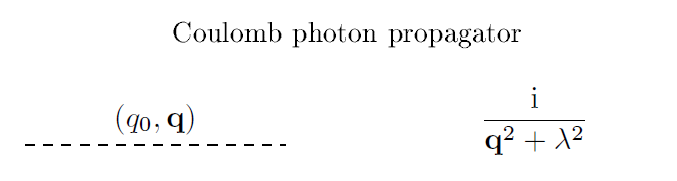}
\end{minipage}
\begin{minipage}{0.48\columnwidth}
\includegraphics[width=1.0\columnwidth]{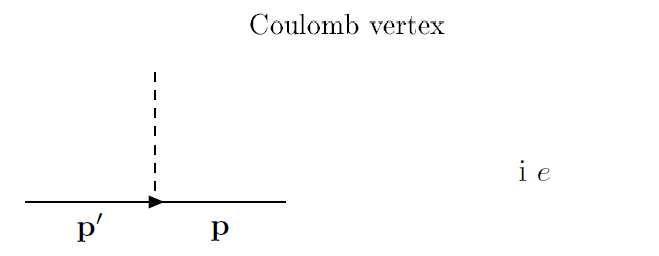}
\end{minipage}\\
\begin{minipage}{0.48\columnwidth}
\includegraphics[width=1.0\columnwidth]{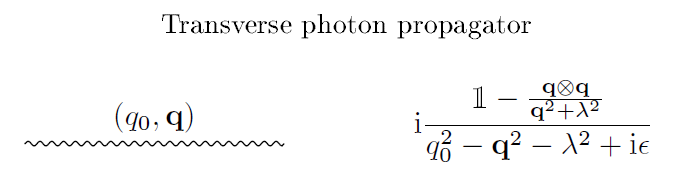}
\end{minipage}
\begin{minipage}{0.48\columnwidth}
\includegraphics[width=1.0\columnwidth]{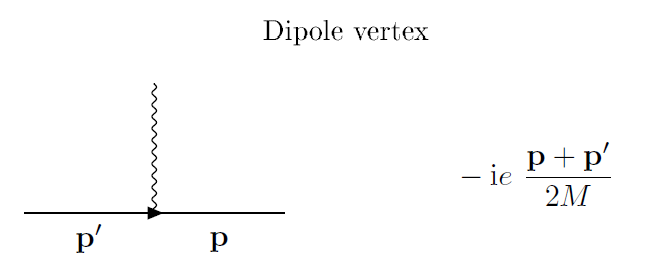}
\end{minipage}
\end{center}
\caption{Feynman rules for spinless fermions and scalar and vector photons in our EFT with NRQED.}\label{F-A-01}
\end{figure}

For the computation of the amplitude associated to each Feynman diagram in the framework of a
non-relativistic effective field theory for spinless fermions with NRQED corrections, the
rules in fig.~\ref{F-A-01} are understood. The imaginary part in the denominator of the retarded fermion propagators $\epsilon$  and the
photon mass $\lambda$ are arbitrarily small quantities.

\section{\textsf{Integrals in Dimensional Regularization}}\label{S-A-2.0}

In this section the evaluation of the second term on the r.h.s. of eq.~\eqref{E-2.1-05} for
$T_{\mathrm{SC}}^{\mathrm{tree}}(\mathbf{p},\mathbf{p}')$ is explicitly shown, that is the
derivation of the new contribution to eq.~\eqref{E-2.1-06}. In $d$ dimensions, the last integral
in eq.~\eqref{E-2.1-05} becomes
\begin{equation}
\mathfrak{I} \equiv D(E^*) M\int_{\mathrm{R}^3}\frac{\mathrm{d}^dl}{(2\pi)^d}\frac{e^2}{\mathbf{l}^2
+\lambda^2}\frac{\mathbf{p}'\cdot\mathbf{l}}{\mathbf{l}^2-2\mathbf{p}\cdot\mathbf{l} +\mathrm{i}\varepsilon}~.
\label{E-A-2.0-01}
\end{equation}
Since the loop integral is not going to be performed in the complex plane, the parameter $\epsilon$ can be set
to zero. Applying the Feynman parametrization for the denominators, we obtain
\begin{equation}
\mathfrak{I} = D(E^*) M e^2 \frac{\Gamma(2)}{\Gamma(1)\Gamma(1)} \int_0^1 \mathrm{d}\omega \int_{\mathrm{R}^d}
\frac{\mathrm{d}^dl}{(2\pi)^d}  \frac{\mathbf{p}'\cdot\mathbf{l}}{[\omega(\mathbf{l}^2+\lambda^2)
+(\mathbf{l}^2-2\mathbf{p}\cdot\mathbf{l})(1-\omega)]^2}  ~.\label{E-A-2.0-02}
\end{equation}
Then, we rewrite the polynomial in the denominator as $\mathbf{l}^2 - 2\mathbf{p}\cdot\mathbf{l}(1-\omega)
+ \omega\lambda^2$ and consider the application of the dimensional regularization formula in eq.~(B.17)
in ref.~\cite{Ram97} for the carrying-out of the momentum integral,
\begin{equation}
\mathfrak{I} = M e^2\frac{\Gamma\left(2-\frac{d}{2}\right)}{(4\pi)^{d/2}\Gamma(2)} \int_0^1 \mathrm{d}\omega
\frac{(1-\omega)~\mathbf{p}\cdot\mathbf{p}'~D(E^*)}{[\omega\lambda^2-\mathbf{p}^2(1-\omega)^2]^{2-d/2}}~.
\label{E-A-2.0-03}
\end{equation}
As the r.h.s. of the last equation is non-singular in three dimensions, the limit $d \rightarrow 3$ can
be safely taken. Furthermore, we define the auxiliary quantities 
\begin{equation}
\beta = 1 + \frac{\lambda^2}{2\mathbf{p}^2} \hspace{0.3cm}\mathrm{and}\hspace{0.3cm} \gamma = \frac{\lambda}{\sqrt{2}|\mathbf{p}|}\sqrt{2+\frac{\lambda^2}{2\mathbf{p}^2}}~\label{E-A-2.0-04}.
\end{equation}
Performing the change of variables $\omega\mapsto\omega'\equiv (\omega-\beta)/\gamma$ in the integrand of
eq.~\eqref{E-A-2.0-03}, we obtain
\begin{equation}
\mathfrak{K} \equiv \int_0^1 \mathrm{d}\omega \frac{(1-\omega)}{[\omega\lambda^2-\mathbf{p}^2(1-\omega)^2]^{1/2}} 
= \int_{-\frac{\beta}{\gamma}}^{\frac{1-\beta}{\gamma}}\mathrm{d}\omega'  \frac{1-\gamma\omega'-\beta}{|\mathbf{p}|
\sqrt{1-(\omega')^{2}}}~.\label{E-A-2.0-05}
\end{equation} 
The last expression can be integrated,
\begin{equation}
- \int_{-\frac{\beta}{\gamma}}^{\frac{1-\beta}{\gamma}}\frac{\mathrm{d}\omega' \omega'\gamma}{|\mathbf{p}|
\sqrt{1-(\omega')^2}} + \int_{-\frac{\beta}{\gamma}}^{\frac{1-\beta}{\gamma}} \frac{(1-\beta)\mathrm{d}\omega'}
{|\mathbf{p}|\sqrt{1-(\omega')^2}} = \frac{\gamma}{|\mathbf{p}|}\sqrt{1-(\omega')^2}\Big
\lvert_{-\frac{\beta}{\gamma}}^{\frac{1-\beta}{\gamma}} + \frac{1-\beta}{|\mathbf{p}|} \mathrm{arcsin}(\omega')
\Big\lvert_{-\frac{\beta}{\gamma}}^{\frac{1-\beta}{\gamma}} ~.\label{E-A-2.0-06}
\end{equation}
After evaluating the results of the integration over $\omega'$ in terms of the original variables, $\mathbf{p}$ and $\lambda$, a closed form for $\mathfrak{K}$ is found,
\begin{equation}
\mathfrak{K} = \frac{\mathrm{i}}{|\mathbf{p}|} + \frac{\lambda}{\mathbf{p}^2} -\frac{\pi\lambda^2}{4|\mathbf{p}|^3}
+  \frac{\mathrm{i}\lambda^2}{2|\mathbf{p}|^3}\log\frac{2|\mathbf{p}|}{\lambda}+  \frac{\mathrm{i}\lambda^2}
{2|\mathbf{p}|^3}\log\left(1 + \frac{\lambda\mathrm{i}}{2|\mathbf{p}|}\right) ~.\label{E-A-2.0-07}
\end{equation}
where the conventions $\log(-\mathrm{i}) = -\mathrm{i}\pi/2$ and $\sqrt{-1} = -\mathrm{i}$ are understood.
Plugging eq.~\eqref{E-A-2.0-07} into eq.~\eqref{E-A-2.0-03}, the desired result is obtained,
\begin{equation}
\mathfrak{I} = \mathbf{p}\cdot\mathbf{p}' D(E^*)\frac{e^2}{4\pi}\frac{M}{2}\Bigg\{ \frac{\mathrm{i}}{|\mathbf{p}|}
+ \frac{\lambda}{\mathbf{p}^2}  +\mathrm{i}\frac{\lambda^2}{2|\mathbf{p}|^3}\left[-\mathrm{i}\frac{\pi}{2}
+ \log\frac{2|\mathbf{p}|}{\lambda} + \log\left(1 + \frac{\lambda\mathrm{i}}{2|\mathbf{p}|}
\right)\right] \Bigg\}~.\label{E-A-2.0-08}
\end{equation}

In the second part of this appendix, we evaluate the momentum integrals appearing in the 1-loop diagram
with one Coulomb photon exchange inside the loop (cf. right part of the fig.~\ref{F-2-02} and eq.~\eqref{E-2.1-09}).
To this aim, we first introduce the notation $\gamma^2 \equiv -\mathbf{p}^2$ for the physical momenta,
we set $\varepsilon$ to zero and rewrite eq.~\eqref{E-2.1-08} in arbitrary dimensions $d$ as
\begin{equation}
\mathfrak{L}\equiv T_{\mathrm{SC}}^{\mathrm{1-loop}}(\mathbf{p},\mathbf{p}';d) = - [D(E^*)]^2 M^2 
 \int_{\mathbb{R}^d}\frac{\mathrm{d}^d \mathbf{k}}{(2\pi)^d}  \int_{\mathbb{R}^d}\frac{\mathrm{d}^d
\mathbf{q}}{(2\pi)^d}\frac{\mathrm{i}\mathbf{p}'\cdot\mathbf{k}}{(\mathbf{q}+\mathbf{k})^2+\gamma^2}
\frac{e^2}{\mathbf{q}^2}  \frac{\mathrm{i}(\mathbf{q}+\mathbf{k})\cdot\mathbf{p}}{\mathbf{k}^2+\gamma^2}~.
\label{E-A-2.0-09}
\end{equation}  
where the fictitious photon mass has been set to zero, since no infrared divergences  occur in the integration
(cf. Appendix A of ref.~\cite{KoR00}). Secondly, we decide to carry out the integration over $\mathbf{k}$
first and merge the relevant denominators again by means of Feynman's trick, obtaining
\begin{equation}
\mathfrak{L} = p_i p_j' M^2 [D(E^*)]^2 \frac{\Gamma(2)}{\Gamma(1)\Gamma(1)} \int_0^1 \mathrm{d}\omega
\int_{\mathbb{R}^d}\frac{\mathrm{d}^d q}{(2\pi)^d} \frac{e^2}{\mathbf{q}^2} 
\int_{\mathbb{R}^d}\frac{\mathrm{d}^d k}{(2\pi)^d} \frac{q_ik_j + k_ik_j}{[\mathbf{k}^2+2(1-\omega)
\mathbf{q}\cdot\mathbf{k} + \Delta^2]^2} ~, \label{E-A-2.0-10}
\end{equation}
where $\Delta^2 \equiv \gamma^2 + \mathbf{q}^2(1-\omega)$ and Einstein's convention for repeated
indices is understood. Now, we can proceed by evaluating the two integrals over $\mathbf{k}$ generated by
$q_ik_j$ and $k_ik_j$ separately. The former integral turns out to be an application of eq.~(B.18)
in ref.~\cite{Ram97} and gives 
\begin{equation}
\int_{\mathbb{R}^d}\frac{\mathrm{d}^dk}{(2\pi)^d} \frac{q_ik_j}{[\mathbf{k}^2+2(1-\omega)\mathbf{q}\cdot\mathbf{k}
+\Delta^2]^2}  = -\frac{q_iq_j(1-\omega)}{[\Delta^2-(1-\omega)^2\mathbf{q}^2]^{2-d/2}} \frac{\Gamma\left(2
-\frac{d}{2}\right)}{(4\pi)^{d/2}\Gamma(2)}~.\label{E-A-2.0-11}
\end{equation}
The second term of eq.~\eqref{E-A-2.0-10}, instead, yields two contributions, one of the two being
proportional to the r.h.s. of eq.~\eqref{E-A-2.0-11}. The application of eq.~(B.18) in ref.~\cite{Ram97}
indeed leads to
\begin{equation}
\begin{split}
\int_{\mathbb{R}^d}\frac{\mathrm{d}^dk}{(2\pi)^d} \frac{k_ik_j}{[\mathbf{k}^2+2(1-\omega)\mathbf{q}\cdot\mathbf{k}
+\Delta^2]^2}  = \frac{1}{(4\pi)^{d/2}}\frac{1}{\Gamma(2)} \\ \cdot \left[\frac{q_iq_j (1-\omega)^2
\Gamma\left(2-\frac{d}{2}\right)}{[\Delta^2-(1-\omega)^2\mathbf{q}^2]^{2-d/2}} \right. & \left. + \frac{1}{2}
\frac{\delta_{ij}\Gamma\left(1-\frac{d}{2}\right)}{[\Delta^2-(1-\omega)^2\mathbf{q}^2]^{1-d/2}}\right]~.
\label{E-A-2.0-12}
\end{split}
\end{equation}
By comparison of the last equation with eq.~\eqref{E-A-2.0-10}, we observe that the $\mathbf{q}$ integrals
involving $q_i q_j$ can be performed together. Therefore, we group the two terms together with the necessary
constants and, after few manipulations, we define
\begin{equation}
\mathfrak{L}_1 \equiv - p_ip_j' [D(E^*)]^2 M^2 \frac{\Gamma\left(2-\frac{d}{2}\right)}{(4\pi)^{d/2}}\int_0^1
\mathrm{d}\omega  \int_{\mathbb{R}^d}\frac{\mathrm{d}^dq}{(2\pi)^d} \frac{q_iq_j(1-\omega)\omega}
{[\gamma^2+\mathbf{q}^2(1-\omega)\omega]^{2-d/2}}  \frac{e^2}{\mathbf{q}^2}\label{E-A-2.0-13}~.
\end{equation}
Then, with the help of the auxiliary parameter $\Xi_4^2 \equiv \gamma^2/[\omega(1-\omega)]$ we apply
again the Feynman parametrization for the two denominators and rewrite the last equation as
\begin{equation}
\mathfrak{L}_1 = - p_ip_j' e^2 D_0^2 M^2 \frac{\Gamma\left(3-\frac{d}{2}\right)}{(4\pi)^{d/2}}\int_0^1\mathrm{d}\omega
[(1-\omega)\omega]^{d/2-1}  \int_0^1\mathrm{d}\phi \phi^{1-d/2} \int_{\mathbb{R}^d}\frac{\mathrm{d}^dq}{(2\pi)^d}
\frac{q_iq_j}{[\mathbf{q}^2+\phi\Xi_4^2]^{3-d/2}} ~.\label{E-A-2.0-14}
\end{equation}
In this form, the application of eq.~(B.18) in ref.~\cite{Ram97} with $\mathbf{q}=0$ suffices for the carrying
out of the momentum integral in $\mathfrak{L}_1$,
\begin{equation}
\int_{\mathbb{R}^d}\frac{\mathrm{d}^dq}{(2\pi)^d} \frac{q_iq_j}{[\mathbf{q}^2+\phi\Qoppa^2]^{3-d/2}} 
= \frac{\delta_{ij}}{2}\frac{\Gamma(2-d)}{\Gamma\left(3-\frac{d}{2}\right)} \frac{\phi^{d-2}\gamma^{2d-4}}
{(4\pi)^{d/2}}[\omega(1-\omega)]^{2-d}~,\label{E-A-2.0-15}
\end{equation}
and eq.~\eqref{E-A-2.0-14} becomes 
\begin{equation}
\mathfrak{L}_1 = -\mathbf{p}\cdot\mathbf{p}' e^2[D(E^*)]^2\frac{M^2}{2} \frac{\gamma^{2d-4}}{(4\pi)^d}
\frac{\Gamma(3-d)}{(2-d)}  \int_0^1\mathrm{d}\omega ~[\omega(1-\omega)]^{1-d/2} \int_0^1\mathrm{d}\phi
\phi^{d/2-1}~,\label{E-A-2.0-16}
\end{equation}
where the Gamma functions have been simplified. While the integration over $\phi$ is straightforward and
gives $2/d$, the remaining one can be performed in an analogous fashion as eq.~(A.6) of ref.~\cite{KoR00}.
By considering $\epsilon \equiv 3-d$, in fact, the integrand can be expanded in a power series in $\epsilon$,
giving
\begin{equation}
\int_0^1\mathrm{d}\omega~[\omega(1-\omega)]^{1-d/2} =  \int_0^1\mathrm{d}\omega e^{\epsilon\log\sqrt{\omega-\omega^2}} 
 [\omega(1-\omega)]^{-1/2}  = \pi - 2\pi\epsilon\log 2  +\mathcal{O}(\epsilon^2)~.\label{E-A-2.0-17}
\end{equation}
Introducing also the renormalization scale $\mu$, the part of the scattering amplitude of interest can be recast as
\begin{equation}
\mathfrak{L}_1 = -  \cos\theta [D(E^*)]^2M^2 e^2 \frac{\mathbf{p}^2}{d}\left(\frac{\mu}{2}\right)^{3-d} 
\cdot \frac{\gamma^{2d-4} \Gamma(\epsilon)}{(4\pi)^{d}(2-d)}  \left[\pi-2\pi\epsilon\log 2 
+\mathcal{O}(\epsilon^2) \right]~.\label{E-A-2.0-18}
\end{equation}
It displays a pole singularity in the limit $d\to 3$ and a simple PDS pole at $d=2$. By exploiting the Laurent
series expansion of the Gamma function for small arguments and truncating it at NLO, $\mathfrak{L}_1$ finally
becomes
\begin{equation}
\mathfrak{L}_1 = -  \cos\theta [D(E^*)]^2 \frac{\alpha M^2}{16\pi} \frac{\mathbf{p}^4}{3}\left[\frac{1}{3-d}
-\gamma_E +\frac{4}{3}+ \mathrm{i}\pi +\log\left(\frac{\pi\mu}{2\mathbf{p}^2}\right) \right]~,\label{E-A-2.0-19}
\end{equation}
where the regular parts of the amplitude have been evaluated in the three-dimensional limit. Now, due to the
presence of a $d=2$ singularity, the PDS correction to $\mathfrak{L}_1$ is non-zero. Introducing again the
renormalization scale $\mu$ of the MS scheme and noticing that the integrand in eq.~\eqref{E-A-2.0-18}
in the $d\rightarrow 2$ limit coincides with 1, the correction turns out to be
\begin{equation}
\delta\mathfrak{L}_1 = \cos\theta [D(E^*)]^2 \frac{\alpha M^2}{4\pi}\frac{\mu\mathbf{p}^2}{4}~.\label{E-A-2.0-20}
\end{equation}
Subtracting the latter equation to eq.~\eqref{E-A-2.0-19}, the PDS corrected part of the amplitude can be obtained,
\begin{equation}
\mathfrak{L}_1^{\mathrm{PDS}} = \mathfrak{L}_1-\delta\mathfrak{L}_1 = -\cos\theta [D(E^*)]^2 \frac{\alpha M^2}{4\pi}
\frac{\mathbf{p}^2}{4} \cdot\left[\frac{\mathbf{p}^2}{3}\left(\frac{1}{3-d} -\gamma_E +\frac{4}{3}
- \log\frac{2\mathbf{p}^2}{\pi\mu} + \mathrm{i}\pi\right) + \mu\right]~.\label{E-A-2.0-21}
\end{equation}
Next, we concentrate on the evaluation of the second term of eq.~\eqref{E-A-2.0-10}.
Considering the original constant factors and the integral over the $\mathbf{q}$, we define 
\begin{equation}
\mathfrak{L}_2 \equiv  \mathbf{p}\cdot\mathbf{p}' [D(E^*)]^2 \frac{M^2}{2} \frac{\Gamma\left(1
-\frac{d}{2}\right)}{(4\pi)^{d/2}} \int_0^1 \mathrm{d}\omega \int_{\mathbb{R}^d}\frac{\mathrm{d}^dq}{(2\pi)^d}
\frac{e^2}{\mathbf{q}^2}\frac{1}{[\gamma^2 + \omega(1-\omega)\mathbf{q}^2]^{1-d/2}}\label{E-A-2.0-22}
\end{equation}
and the constant $\Xi_5^2 \equiv \gamma^2/[\omega(1-\omega)]$, so that we can exploit again the Feynman
parametrization for the denominators, obtaining
\begin{equation}
\mathfrak{L}_2 = \mathbf{p}\cdot\mathbf{p}' e^2 [D(E^*)]^2 \frac{M^2}{2} \frac{\Gamma\left(2-\frac{d}{2}\right)}
{(4\pi)^{d/2}} \int_0^1 \mathrm{d}\omega [\omega(1-\omega)]^{d/2-1}\int_0^1 \mathrm{d}\phi \int_{\mathbb{R}^d}
\frac{\mathrm{d}^dq}{(2\pi)^d}\frac{\phi^{-d/2}}{[\phi\Xi_5^2 + \mathbf{q}^2]^{2-d/2}}~.\label{E-A-2.0-23}
\end{equation}
The integral over the $\mathbf{q}$ can be now carried out as a straightforward application of eq.~(B.16)
in ref.~\cite{Ram97}, yielding
\begin{equation}
\mathfrak{L}_2 = \mathbf{p}\cdot\mathbf{p}' e^2 [D(E^*)]^2 \frac{M^2}{2} \frac{\Gamma\left(2-\frac{d}{2}\right)}
{(4\pi)^{d/2}} \int_0^1 \mathrm{d}\omega [\omega(1-\omega)]^{d/2-1}  \int_0^1 \mathrm{d}\phi
\frac{\phi^{-d/2}(\phi\Qoppa^2)^{d-2}}{(4\pi)^{d/2}}\frac{\Gamma(2-d)}{\Gamma\left(2-\frac{d}{2}\right)}~.
\label{E-A-2.0-24}
\end{equation}
In the last rewriting, the integral over the $\phi$ can be immediately performed, while the Gammas can
again be simplified and reduced, so that eq.~\eqref{E-A-2.0-24} transforms into
\begin{equation}
\mathfrak{L}_2 = -\mathbf{p}\cdot\mathbf{p}' e^2 [D(E^*)]^2 \gamma^{2d-4} \frac{M^2}{4} \left(\frac{\mu}{2}
\right)^{3-d} \frac{\Gamma(3-d)}{(4\pi)^{d}\left(1-\frac{d}{2}\right)^2} \int_0^1 \mathrm{d}\omega~
[\omega(1-\omega)]^{1-d/2}  ~,\label{E-A-2.0-25}
\end{equation}
where the conventional renormalization scale factor $(\mu/2)^{3-d}$ has been introduced as in eq~\eqref{E-A-2.0-18}.
The remaining integral has been already met in eq.~\eqref{E-A-2.0-20} and it can be evaluated exactly
in $d=2$ or expanded in powers of $3-d$ in the three-dimensional case. Plugging eq.~\eqref{E-A-2.0-17} into
eq.~\eqref{E-A-2.0-24}, it turns out that the integral is again divergent in the limit $d\rightarrow 3$ and
includes also a threefold PDS pole in $d\rightarrow 2$. Taking the former limit, the Gamma function can
be expanded in power series as before and the amplitude can be re-expressed as
\begin{equation}
\mathfrak{L}_2 = \cos\theta [D(E^*)]^2 \frac{\alpha M^2}{4\pi}\frac{\mathbf{p}^4}{4}\left[ \frac{1}{3-d}
  -\gamma_E + 2+ \mathrm{i}\pi  -\log \frac{\mathbf{p}^2}{8\pi\mu} \right] ~,
\label{E-A-2.0-26}
\end{equation}
where the ultraviolet divergence of the original integral is made explicit. Then, the application of the
PDS scheme to the $d\rightarrow 2$ pole yields the following correction,
\begin{equation}
\delta\mathfrak{L}_2^{\mathrm{PDS}} = -\alpha\cos\theta D_0^2 M^2 \frac{\mu\mathbf{p}^2}{8\pi}~.\label{E-A-2.0-27}
\end{equation}
Finally, we subtract the PDS contribution just determined to eq.~\eqref{E-A-2.0-26}, obtaining
\begin{equation}
\mathfrak{L}_2^{\mathrm{PDS}} =  \cos\theta [D(E^*)]^2 \frac{\alpha M^2}{4\pi}\frac{\mathbf{p}^{2}}{2}\Big\{
\frac{\mathbf{p}^2}{2}\left[\frac{1}{3-d}   -\gamma_E+ 2 + \mathrm{i}\pi -
\log\left(\frac{2\mathbf{p}^2}{\pi\mu}\right)\right]  + \mu\Big\} ~.\label{E-A-2.0-28}
\end{equation}
Now we collect the two results in eqs.~\eqref{E-A-2.0-21} and \eqref{E-A-2.0-28} and write the one-loop
scattering amplitude with one photon exchange in the power divergence subtraction scheme,
\begin{equation}
T_{\mathrm{SC}}^{\mathrm{1-loop}}(\mathbf{p},\mathbf{p}')\Big\lvert^{\mathrm{PDS}} = \cos\theta [D(E^*)]^2
\frac{\alpha M^2 }{4\pi} \Big\{ \frac{\mu \mathbf{p}^{2}}{4}  +  \frac{\mathbf{p}^{4}}{6}\left[\frac{1}{3-d}
-\gamma_E + \frac{7}{3}- \log\left(\frac{2\mathbf{p}^2}{\pi\mu}\right) + \mathrm{i}\pi\right] \Big\}~.
\label{E-A-2.0-29}
\end{equation}

\section{\textsf{Three dimensional integrations}}\label{S-A-3.0}

Here, we focus our attention on the computation of the leading order matrix element of the Coulomb-corrected 
strong fermion-fermion scattering amplitude in eq.~\eqref{E-2.2-02}. The process can be reduced to the
evaluation of only one of the two integrals presented in the first row of the last equation, by virtue
of the complex-conjugation property satisfied by the repulsive Coulomb wavefunctions,
$\psi_{\mathbf{p}'}^{(-)*} = \psi_{-\mathbf{p}'}^{(+)}$. In particular, we choose to concentrate on the
first term on the l.h.s. of eq.~\eqref{E-2.2-02}, 
\begin{equation}
 \mathfrak{F} \equiv \mathrm{i} \nabla \psi_{\mathbf{p}}^{(-)*}(\mathbf{r}) \Big\lvert_{\mathbf{r}=\mathbf{0}}
 = \mathrm{i}\int_{\mathbb{R}^3}\mathrm{d}^3 r'~\delta(\mathbf{r}')~\nabla' \psi^{(-)*}_{\mathbf{p}'}(\mathbf{r}')~.
 \label{E-A-3.0-01}
\end{equation}
Recalling the parity rule of the spherical harmonics (cf. sec.~$A_{VI}$ in ref.~\cite{CTD77-02}),
the repulsive Coulomb eigenstate turns out to be given by
\begin{equation}
\psi_{\mathbf{p}'}^{(-)*}(\mathbf{r}') = \frac{4\pi}{|\mathbf{p}'| r'} \sum_{\ell=0}^{+\infty}\sum_{m=-\ell}^{\ell}
(-\mathrm{i})^{\ell} e^{\mathrm{i}\sigma_{\ell}} \cdot Y_{\ell}^{m*}(\hat{\mathbf{p}'})  Y_{\ell}^{m}(\hat{\mathbf{r}'})
F_{\ell}(\eta,|\mathbf{p}'|r')~.\label{E-A-3.0-02}
\end{equation}
Now, we start by observing that the three-dimensional Dirac delta function peaked at the origin can be
rewritten  in spherical coordinates as 
\begin{equation}
\delta (\mathbf{r}) \equiv \delta(x)\delta(y)\delta(z) = \frac{\delta(r)}{4\pi r^2}~,\label{E-A-3.0-03}
\end{equation}
due to rotational invariance. The last equation can be proven by exploiting the integral representation of
$\delta (\mathbf{r})$ \cite{CTD77-01}, then expressing the exponential in terms of spherical
harmonics and spherical Bessel functions of the first kind and finally performing the angular and radial
integrations by means of the identity in sec.~11.2 of ref.~\cite{ArW12}. Besides, the rewriting in
eq.~\eqref{E-A-3.0-03} paves the way for the integration of the angular variables and on the radial distance
on which the Dirac delta effectively acts separately. Equipped with the last equation, we can split the
expression in eq.~\eqref{E-A-3.0-01} into two parts,
\begin{equation}
\mathfrak{F} = \mathfrak{F}_1 + \mathfrak{F}_2~,\label{E-A-3.0-04}
\end{equation}
where
\begin{equation}
\mathfrak{F}_1 \equiv \mathrm{i}\int_{0}^{2\pi} \mathrm{d}\varphi \int_{0}^{\pi} \mathrm{d}\theta' \int_{0}^{+\infty}
\mathrm{d}r' \delta(r')  \sum_{\ell=0}^{+\infty}\sum_{m=-\ell}^{\ell} (-\mathrm{i})^{\ell}  e^{\mathrm{i}\sigma_{\ell}}
Y_{\ell}^{m*}(\hat{\mathbf{p}'}) Y_{\ell}^{m}(\hat{\mathbf{r}'}) \nabla'\frac{F_{\ell}(\eta,|\mathbf{p}'|r')}
{|\mathbf{p}'|r'}~.\label{E-A-3.0-05}
\end{equation}
and
\begin{equation}
 \mathfrak{F}_2 \equiv \mathrm{i}\int_{0}^{2\pi} \mathrm{d}\varphi \int_{0}^{\pi} \mathrm{d}\theta' \int_{0}^{+\infty}
 \mathrm{d}r' \delta(r') \sum_{\ell=0}^{+\infty}\sum_{m=-\ell}^{\ell} (-\mathrm{i})^{\ell}  e^{\mathrm{i}\sigma_{\ell}}
 Y_{\ell}^{m*}(\hat{\mathbf{p}'}) \frac{F_{\ell}(\eta,|\mathbf{p}'|r')}{|\mathbf{p}'|r'} \nabla'
 Y_{\ell}^{m}(\hat{\mathbf{r}'})~.\label{E-A-3.0-06}
\end{equation}
Considering the explicit expression for the regular repulsive Coulomb eigenfunctions in eq.~\eqref{E-2.1-15},
we first concentrate on the application of the gradient to $F_{\ell}(\eta,|\mathbf{p}'|r')/|\mathbf{p}'|r'$.
Recalling the transformation property of the Kummer functions, i.e. the confluent Hypergeometric functions
${}_1\mathrm{F}_1$, under differentiation with respect to their third argument
\begin{equation}
\frac{\partial}{\partial z} M(a,b, z) = \frac{a}{b} M(a+1,b+1,z)\hspace{0.99cm}a,b,z \in \mathbb{C}~,
\label{E-A-3.0-07}
\end{equation}
we can rewrite the term of interest as 
\begin{equation}
\begin{gathered}
\nabla' \left(\frac{F_{\ell}(\eta,|\mathbf{p}'|r')}{|\mathbf{p}'|r'}\right) =|\mathbf{p}'|\hat{\mathbf{r}}'
\frac{2^{\ell}e^{-\pi\eta/2}| e^{\mathrm{i}|\mathbf{p}'|r}|\Gamma(\ell+1+\mathrm{i}\eta)|}{(2\ell+1)!}
(|\mathbf{p}'|r')^{\ell-1} \\ \cdot \left[ \frac{}{} (\mathrm{i}r'|\mathbf{p}'| +\ell)~M(\ell+1
+\mathrm{i}\eta,2\ell+2,-2\mathrm{i}|\mathbf{p}'|r')  - \mathrm{i}|\mathbf{p}'|r'
\frac{\ell+1+\mathrm{i}\eta}{\ell+1} M(\ell+2+\mathrm{i}\eta,2\ell+3,-2\mathrm{i}|\mathbf{p}'|r')\right]~.
\label{E-A-3.0-08}
\end{gathered}
\end{equation}
In the last equation, we note that the application of the gradient effectively reduces to the application
of the derivative with respect to the radial variable $r'$, therefore the resulting vector is parallel
to $\mathbf{r}'$. It is then convenient to exploit the expression of the latter vector in terms of
the spherical harmonics given in eq.~(5.24) and sec.~5.1 of ref.~\cite{MaG96}, in order to perform the
integration on the radial and the angular variables associated to $r'$ in eq.~\eqref{E-A-3.0-05} separately.
Making use of the complex conjugation (cf. eq.~(4.31) in ref.~\cite{Ros57}) and the orthonormality (cf.
chap.~VI of ref.~\cite{CTD77-01}) properties of spherical harmonics, the integral over the angular
variables can be carried out rapidly, obtaining
\begin{equation}
\begin{split}
\mathfrak{F}_1 = \sum_{\ell=0}^{+\infty}\sum_{m=-\ell}^{\ell} (-\mathrm{i})^{\ell}  2\mathrm{i}\sqrt{\frac{\pi}{3}}
\int_{0}^{+\infty} \mathrm{d}r'\delta(r')  e^{i\sigma_{\ell}} Y_{\ell}^{m*}(\hat{\mathbf{p}}') \\
\cdot \left( \delta_{m1}\delta_{\ell 1} \mathbf{e}_1 +\delta_{m0}\delta_{\ell 1} \mathbf{e}_{0} \right. & \left. +  \delta_{m-1}
\delta_{\ell 1} \mathbf{e}_{-1} \right) \frac{\partial}{\partial r'}\left[\frac{F_{\ell}(\eta,|\mathbf{p}'|r')}
{|\mathbf{p}'|r'}\right] ~,\label{E-A-3.0-09}
\end{split}
\end{equation}
where $\Omega'$ denotes collectively the angular variables associated to $\mathbf{r}'$. Recalling eq.~(5.24)
and sec.~5.1 in ref.~\cite{MaG96}, the remaining spherical harmonic on the second row of eq.~\eqref{E-A-3.0-09}
together with the round bracket with the Kronecker deltas can be identified as the unit-vector
parallel to $\mathbf{p}'$, up to a multiplication factor that cancels out with $2\sqrt{\pi/3}$ on the
left of the last integration sign. Then, exploiting eq.~\eqref{E-A-3.0-08} for $\ell=1$, eq.~\eqref{E-A-3.0-09}
can be rewritten as
\begin{equation}
\begin{gathered}
\mathfrak{F}_1 = \int_{0}^{+\infty} \mathrm{d}r'\delta(r')  e^{\mathrm{i}\sigma_{1}}\frac{\partial}{\partial r'}
\left[\frac{F_{1}(\eta,|\mathbf{p}'|r')}{r'}\right]\mathbf{p}'  = \frac{|\Gamma(2+\mathrm{i}\eta)|}
{3~e^{-\mathrm{i}\sigma_{1}} e^{\pi\eta/2}} \\ \cdot \lim_{r'\rightarrow 0} \left[\frac{}{} (1+\mathrm{i}|\mathbf{p}'|r)
M(2+\mathrm{i}\eta,4,-2\mathrm{i}|\mathbf{p}'|r')  -\mathrm{i}|\mathbf{p}'|r'
\frac{2+\mathrm{i}\eta}{2}  M(3+\mathrm{i}\eta,5,-2\mathrm{i}|\mathbf{p}'|r')\right]\mathbf{p}'~.
\label{E-A-3.0-10}
\end{gathered}
\end{equation}
The explicit evaluation of the limit yields immediately to the disappearance of the terms depending linearly
on the radial coordinate $r'$, since the Kummer functions are equal to unity for zero values of the third argument, 
\begin{equation}
\lim_{z\rightarrow 0} M(a,b,z) = \lim_{z\rightarrow 0} {}_1\mathrm{F}_1(a,b,z) = 1 \hspace{0.99cm}a,b,z
\in \mathbb{C}~.\label{E-A-3.0-11}
\end{equation}
Besides, by exploiting the fundamental property $\Gamma(z+1)=z\Gamma(z)$ of the Gamma functions,
the constants outside the limit in eq.~\eqref{E-A-3.0-10} can be rewritten in terms of the Sommerfeld
factor (cf. eq.~\eqref{E-2.1-16}), 
\begin{equation}
 \frac{|\Gamma(2+\mathrm{i}\eta)|}{e^{\pi\eta/2}} = \frac{\sqrt{1+\eta^2}}{ e^{\pi\eta/2}} |\Gamma(1+\mathrm{i}\eta)|
 = \sqrt{1+\eta^2} C_{\eta}\label{E-A-3.0-12}~, 
\end{equation}
thus recovering the polynomial on the r.h.s of the generalized effective range expansion (cf. eq.~\eqref{E-2.1-26}).
Equipped with the two last results we can, finally, obtain the desired expression for the r.h.s. of
eq.~\eqref{E-A-3.0-09},
\begin{equation}
\mathfrak{F}_1  = e^{\mathrm{i}\sigma_{1}} C_{\eta} \sqrt{1+\eta^2} \frac{\mathbf{p}'}{3}~.\label{E-A-3.0-13}
\end{equation}
Now, we can proceed with the application of the gradient to the spherical harmonics (cf. eq.~\eqref{E-A-3.0-06}).
From ref.~\cite{MaG96}, the result of the latter derivative can be rewritten as a linear combination of
shperical harmonics as in eqs.~(5.24) and (5.27) in ref.~\cite{MaG96}. In particular, by the introduction
of $1=\sqrt{4\pi}Y_0^{0*}(\theta,\varphi)$ in the relevant integral of eq.~\eqref{E-A-3.0-06}, we can
observe that the surface integrals over the $Y_{\ell+1}^{\mu}(\theta,\varphi)$ yield no contribution,
since $\ell$ cannot assume negative values. It follows that our term of interest becomes
\begin{equation}
\begin{split}
\mathfrak{F}_2 = -\sum_{\ell, m} (-\mathrm{i})^{\ell+1} \sqrt{\frac{4\pi\ell}{2\ell+1}}
\frac{(\ell+1)}{|\mathbf{p}'|}e^{\mathrm{i}\sigma_{\ell}}  Y_{\ell}^{m*}(\hat{\mathbf{p}'}) \\
\int_{0}^{+\infty} \mathrm{d}r'~\delta(r')\frac{F_{\ell}(\eta,|\mathbf{p}'|r')}{r^{'2}} & \sum_{\mu,\mu'}
\delta_{\ell 1}\delta_{\mu 0}(\ell-1 1 \ell|\mu\mu'm)  \mathbf{e}_{\mu'}~,\label{E-A-3.0-14}
\end{split}
\end{equation}
where the Clebsch-Gordan coefficients $(j_1 j_2 J|m_1 m_2 M) \equiv \langle J M, j_1 j_2 | j_1 m_1, j_2 m_2 \rangle$
vanish whenever $m \neq \mu +\mu'$. The evaluation of the latter in the second row of eq.~\eqref{E-A-3.0-14} leads to 
\begin{equation}
\begin{split}
\mathfrak{F}_2 =  \sqrt{\frac{4\pi}{3}} e^{\mathrm{i}\sigma_{1}} \int_{0}^{+\infty} \mathrm{d}r'~\delta(r')\frac{2}{r'}
\frac{F_{1}(\eta,|\mathbf{p}'|r')}{|\mathbf{p}'|r'}  \left(Y_{1}^{-1*} (\hat{\mathbf{p}'}) \mathbf{e}_{1}
+ Y_{1}^{0*}(\hat{\mathbf{p}'}) \mathbf{e}_{0} \right. \\ \left. + Y_{1}^{1*}(\hat{\mathbf{p}'}) \mathbf{e}_{-1}\right) 
=  2  \mathbf{p}' \frac{|\Gamma(2+\mathrm{i}\eta)|}{3~e^{\pi\eta/2} e^{-\mathrm{i}\sigma_{1}}} \lim_{r\rightarrow 0}
\left[ e^{\mathrm{i}|\mathbf{p}'|r'} \right. & \left. M(2+\mathrm{i}\eta,4,-2\mathrm{i}|\mathbf{p}'|r')\right]~.\label{E-A-3.0-15}
\end{split}
\end{equation}
Then, exploiting again the results in eqs.~\eqref{E-A-3.0-11} and \eqref{E-A-3.0-12}, the limit in the equation
can be evaluated and the expression simplified as eq.~\eqref{E-A-3.0-13}, giving
\begin{equation}
\mathfrak{F}_2 = e^{\mathrm{i}\sigma_{1}} C_{\eta} \sqrt{1+\eta^2} \frac{2\mathbf{p}'}{3}~.\label{E-A-3.0-16}
\end{equation}
Combining the last result together with the one in eq.~\eqref{E-A-3.0-13}, we obtain the expression of the
complete integral in eq.~\eqref{E-A-3.0-01},
\begin{equation}
\mathrm{i}\int_{\mathbb{R}^3}\mathrm{d}^3 r'~\delta(\mathbf{r}')~\nabla' \psi^{(-)*}_{\mathbf{p}'}(\mathbf{r}')
= e^{\mathrm{i}\sigma_{1}} C_{\eta} \sqrt{1+\eta^2} \mathbf{p}'~.\label{E-A-3.0-17}
\end{equation}
By exploiting the complex-conjugation property of the regular Coulomb repulsive wavefunction
$\psi_{\mathbf{p}}^{(-)}(\mathbf{r}) = \psi_{-\mathbf{p}}^{(+)*}(\mathbf{r})$, also the second integral in
eq.~\eqref{E-2.2-02} can be evaluated, by noting that the latter property implies only the disappearance
of a $(-1)^{\ell}$ factor in eq.~\eqref{E-A-3.0-02}, that for $\ell = 1$ is compensated by the overall
minus sign in front of the integral,
\begin{equation}
-\mathrm{i}\int_{\mathbb{R}^3}\mathrm{d}^3 r~\delta(\mathbf{r})~\nabla \psi^{(+)}_{\mathbf{p}}(\mathbf{r})
= e^{\mathrm{i}\sigma_{1}} C_{\eta} \sqrt{1+\eta^2} \mathbf{p}'~.\label{E-A-3.0-18}
\end{equation}
Therefore, the full leading order $T_{\mathrm{SC}}$ matrix element in eq.~\eqref{E-2.2-01} reads
\begin{equation}
\langle \psi^{(-)}_{\mathbf{p}'}|\hat{\mathcal{V}}^{(1)}| \psi_{\mathbf{p}}^{(+)}\rangle = e^{2\mathrm{i}\sigma_{1}}
C_{\eta}^2 D(E^*) (1+\eta^2) \mathbf{p}'\cdot\mathbf{p} = \cos\theta D(E^*)  C_{\eta}^2  e^{2\mathrm{i}\sigma_{1}}
(1  +\eta^2) \mathbf{p}^2~,\label{E-A-3.0-19}
\end{equation}
where $\theta$ is the scattering angle. 

\section{\textsf{Three-dimensional Riemann sums}}\label{S-A-5.0}

The derivation of the energy corrections from the finite volume ERE for the lowest-energy $T_1$ scattering
state implies the computation of the sum of the single and double three-dimensional Riemann series treated
in this appendix.

\subsection{\textsf{Single sums}}\label{S-A-5.1}

Let us begin with the derivation of the sum of $\mathcal{S}_0(\tilde{p}) \equiv \Upsilon$, 
\begin{equation}
\Upsilon = \sum_{\mathbf{n}}^{\Lambda_n} 1 -\frac{4\pi}{3} \Lambda_n^3 = 0\label{E-A-5.1-01}
\end{equation}
which can be carried out analytically. After rewriting the series over three-vectors of integers
in integral form,
\begin{equation}
\sum_{\mathbf{n}}^{\Lambda_n} 1 = \int_{\mathbb{R}^3} \mathrm{d}^3k~\sum_{\mathbf{n}}^{\Lambda_n}\delta(\mathbf{k}
-\mathbf{n}) = \int_{\Lambda_n} \mathrm{d}^3k~\sum_{\mathbf{n}}^{\infty}\delta(\mathbf{k}-\mathbf{n})~,
\label{E-A-5.1-02}
\end{equation}
the \textit{Poisson summation formula} for a three-dimensional Dirac delta function, 
\begin{equation}
\sum_{\mathbf{n}}^{\infty} \delta(\mathbf{k}-\mathbf{n}) = \sum_{\mathbf{n}}^{\infty} e^{-2\pi \mathrm{i} \mathbf{n}\cdot
\mathbf{k}}~,\label{E-A-5.1-03}
\end{equation}
can be directly applied, obtaining
\begin{equation}
\Upsilon = \sum_{\mathbf{n}}^{\Lambda_n} 1 -\frac{4\pi}{3} \Lambda_n^3 = \int_{\Lambda_n} \mathrm{d}^3k~
\sum_{\mathbf{n}}^{\infty}e^{-2\pi\mathrm{i} \mathbf{k}\cdot\mathbf{n}} -\frac{4\pi}{3} \Lambda_n^3~.\label{E-A-5.1-04}
\end{equation}
Then, the remaining integral can be computed by singling out the zero mode,
\begin{equation}
\int_{\Lambda_n} \mathrm{d}^3k~\sum_{\mathbf{n}}^{\infty}e^{-2\pi\mathrm{i} \mathbf{k}\cdot\mathbf{n}}
= \int_{\Lambda_n} \mathrm{d}^3k + \sum_{\mathbf{n} \neq \mathbf{0}}^{\infty} \int_{\Lambda_n} \mathrm{d}^3k~
e^{-2\pi\mathrm{i} \mathbf{k}\cdot\mathbf{n}}  = \frac{4\pi}{3}\Lambda_n^2 + \sum_{\mathbf{n} \neq \mathbf{0}}^{\infty}
\delta (2\pi \mathbf{n}) = \frac{4\pi}{3}\Lambda_n^2~,\label{E-A-5.1-05}
\end{equation}
where the second integral vanishes, since the sum excludes the null vector. Equipped with last result,
it immediately follows that
\begin{equation}
\Upsilon = \sum_{\mathbf{n}}^{\Lambda_n} 1 -\frac{4\pi}{3} \Lambda_n^3
= \frac{4\pi}{3}\Lambda_n^2 - \frac{4\pi}{3} \Lambda_n^3 = 0~.\label{E-A-5.1-06}
\end{equation}
Second, we report the sum of the series whose general term is given by the inverse of the norm of
the three vector of integers $\mathbf{n}$ \cite{DaS14}, 
\begin{equation}
\mathcal{G}^{(0)} = \sum_{\mathbf{n}\neq \mathbf{0}}^{\Lambda_n} \frac{1}{|\mathbf{n}|} - 2\pi\Lambda_n^2
= -2.8372~. \label{E-A-5.1-07}
\end{equation}
Even if it does not play any role in the expression of the finite volume energy corrections, the sum of the series
shares its asymptotic behaviour with the one of $\Qoppa_1$. A precise determination of the sum of
$\mathcal{G}^{(0)}$, thus, provides a benchmark test, which has to be passed successfully before
addressing the $\Qoppa_1$ calculation. Due to the rapid oscillation of the sum of the series for
similar values of the cutoff constant $\Lambda_n$, the original series in eq.~\eqref{E-A-5.1-07}
has been recast as
\begin{equation}
\mathcal{G}^{(0)} = \lim_{\varepsilon\rightarrow 0^+}\left[\sum_{\mathbf{n}}^{\Lambda_n} \frac{e^{-\varepsilon|\mathbf{n}|}}
{|\mathbf{n}|} - 4\pi \int_0^{\Lambda_n} \mathrm{d}n~ne^{-\varepsilon n} \right]~,\label{E-A-5.1-07}
\end{equation}
where $\epsilon$ is a small real constant and the exponential factor proves to quench the oscillations
of $\mathcal{G}^{(0)}$ for neighbouring values of $\Lambda_n$. Considering the interval $0.1 \leq \epsilon \leq 1$,
the sum of the series proves to decrease monotonically towards $\epsilon = 0$ n the $\Lambda_n \rightarrow
+\infty$ limit and the behaviour is linear with $\epsilon$, with small quadratic corrections. The subsequent
quadratic interpolation, in fact, returns a value of $\mathcal{G}^{(0)}$ compatible with the exact one
in literature (cf. eq.~\eqref{E-A-5.1-07} and ref.~\cite{DaS14}),
\begin{equation}
\mathcal{G}^{(0)} \approx -2.83739(11)~.\label{E-A-5.1-08}
\end{equation}
It follows that the chosen approach (cf. eq.~\eqref{E-A-5.1-07}) is successful in the evaluation of
the sum of the series and can be promoted to more involved cases. Moreover, exploring a larger
interval of $\varepsilon$ towards larger values, further deviations form linearity are likely to appear
in the fit. In particular, the following class of fitting functions,
\begin{equation}
f(\varepsilon) = \frac{a}{\varepsilon + b} + c\hspace{1.5cm}a,b,c \in \mathbb{R}~,\label{E-A-5.1-09}
\end{equation}
is expected to provide a satisfactory description of the behaviour of of $\mathcal{G}^{(0)}$ with
$\varepsilon$ both in the vicinity of zero and in the infinite $\varepsilon$ limit.

Correlated to $\mathcal{G}^{(0)}$  is the $\mathcal{I}^{(0)}$ series, which appears in both in the QED
leading order corrections to the scattering parameters (cf. eqs.~\eqref{E-3.0-05}-\eqref{E-3.0-10}) and in the
large binding momentum limit of the L\"uscher functions $\mathcal{S}_2(\mathrm{i}\tilde{\kappa})$ and
$\mathcal{S}_3(\mathrm{i}\tilde{\kappa})$, see eqs.~\eqref{E-3.3.2-03} and \eqref{E-3.3.2-04}. The sum of
this series is already known in literature \cite{DaS14,Lue86-02,BeS14} and is given by
\begin{equation}
 2\pi\mathcal{G}^{(0)} = \mathcal{I}^{(0)} = \sum_{\mathbf{n}\neq \mathbf{0}}^{\Lambda_n} \frac{1}{|\mathbf{n}|^2}
 - 4\pi \Lambda_n =  -8.9136~,\label{E-A-5.1-10}
\end{equation}
where the first equality is shown in tab.~1 and eq.~(2.61) of ref.~\cite{Lue86-02}. A precise evaluation
of $\mathcal{I}^{(0)}$ can be attained by isolating the cutoff-dependent part of the series via the
Poisson summation formula. In particular, the addition and subtraction of a $1/(|\mathbf{n}|^2+1)$ term
in the original series yields
\begin{equation}
\mathcal{I}^{(0)} = \sum_{\mathbf{n}\neq\mathbf{0}}^{\Lambda_n} \frac{1}{|\mathbf{n}|^2 (|\mathbf{n}|^2+1)}
+ \sum_{\mathbf{n}\neq\mathbf{0}}^{\Lambda_n}\frac{1}{|\mathbf{n}|^2+1} -4\pi\Lambda_n~,\label{E-A-5.1-11}
\end{equation}
where the first term on the r.h.s proves to converge in the $\Lambda \rightarrow +\infty$ limit as fast
as the $\mathcal{J} \equiv \mathcal{J}^{(0)}$ series in ref.~\cite{BeS14} and the linear divergence is
confined into the second summation. Exploiting the Poisson formula in eq.~\eqref{E-A-5.1-03}, the
non-regularized series can be evaluated as follows,
\begin{equation}
\sum_{\mathbf{n}\neq\mathbf{0}}^{\Lambda_n}\frac{1}{|\mathbf{n}|^2+1}  = - 1 +  \sum_{\mathbf{n}}^{\infty}
\int_{S^2(\Lambda_n)} \frac{\mathrm{d}^3k}{\mathbf{k}^2+1} \delta(\mathbf{n}-\mathbf{k}) 
= - 1 +  \sum_{\mathbf{n}}^{\infty} \int_{S^2(\Lambda_n)} \frac{\mathrm{d}^3k}{\mathbf{k}^2+1}
e^{-2\pi\mathrm{i} \mathbf{k}\cdot\mathbf{n}}~,\label{E-A-5.1-12}
\end{equation}
where the zero $\mathbf{n}$ term has been added to the sum and, then, the spherical cutoff has been moved
from the sum to the integral over the lattice momenta. Separating the zero modes from the others in the
result of eq.~\eqref{E-A-5.1-11}, we obtain 
\begin{equation}
\begin{split}
 -1 + \int_{S^2(\Lambda_n)}\frac{\mathrm{d}^3k}{\mathbf{k}^2+1} + \sum_{\mathbf{n} \neq \mathbf{0}}^{\infty}
 \int_{S^2(\Lambda_n)}\frac{\mathrm{d}^3k}{\mathbf{k}^2+1} e^{-2\pi\mathrm{i} \mathbf{k}\cdot\mathbf{n}} 
 = -1 \\ + 4\pi \Lambda_n & + 4\pi\sum_{\mathbf{n} \neq \mathbf{0}}^{\infty} \int_0^{\Lambda_n} \mathrm{d}k~
 \frac{k}{\mathrm{i}|\mathbf{n}|}\frac{e^{2\pi\mathrm{i} k |\mathbf{n}|} - e^{-2\pi\mathrm{i} k |\mathbf{n}|}}
 {|\mathbf{k}|^2+1}~,\label{E-A-5.1-13}
\end{split}
\end{equation}
where the third and the fourth term on the r.h.s. turn out to be finite in the infinite cutoff limit
and the last integral can be performed in the complex plane, by collecting the residues according to
Jordan's Lemma. Performing the remaining integration, in fact, the last equation becomes
\begin{equation}
-1 + 4\pi \Lambda_n - 2\pi^2 + \pi \sum_{\mathbf{n} \neq \mathbf{0}}^{\infty} \frac{e^{-2\pi|\mathbf{n}|}}{|\mathbf{n}|}~,
\label{E-A-5.1-14}
\end{equation}
an expression that can be directly plugged into eq.~\eqref{E-A-5.1-11}, obtaining the desired result,
\begin{equation}
\mathcal{I}^{(0)} = -1 - 2\pi^2 + \sum_{\mathbf{n}\neq\mathbf{0}}^{\infty} \frac{1}{|\mathbf{n}|^2 (|\mathbf{n}|^2+1)}
+ \pi \sum_{\mathbf{n} \neq \mathbf{0}}^{\infty} \frac{e^{-2\pi|\mathbf{n}|}}{|\mathbf{n}|}   \equiv -1 - 2\pi^2  
+ \gamma_1 + \pi \Sampi_0~,\label{E-A-5.1-15}
\end{equation}
As it can be observed, the cutoff dependent term in the original series has been removed and, at the same time,
two rapidly convergent sums, $\Sampi_0 = 0.0125$ and $\gamma_1$ (cf. eq.~\eqref{E-A-5.1-19}) replaced the
divergent one, thus reducing significantly the computational efforts. The procedure is completely analogous
to the one adopted in Appendix~B 1 of ref.~\cite{PWH19} and can be applied to other cutoff-regulated single sums.

At this stage, we focus on the single sums appearing as coefficients in the $\delta\tilde{p}^2$ expansion
of the L\"uscher functions $\mathcal{S}_1(\tilde{p})$, $\mathcal{S}_2(\tilde{p})$ and $\mathcal{S}_3(\tilde{p})$
around  $\tilde{p}^2 = 1$. Adopting a notation similar to the one used for eq.~\eqref{E-A-5.1-07} and
\eqref{E-A-5.1-10}, the sums of relevant three-dimensional Riemann series yield
\begin{equation}
\mathcal{I}^{(1)} = \sum_{\mathbf{n}\neq \mathbf{0}}^{\Lambda_n} \frac{1}{|\mathbf{n}|^2-1} - 4\pi \Lambda_n
=  -1.2113~, \hspace{1.0cm}
 \mathcal{J}^{(1)} = \sum_{\mathbf{n}\neq \mathbf{0}} \frac{1}{(|\mathbf{n}|^2-1)^2}  = 23.2432~,\nonumber
\end{equation}
\begin{equation}
\mathcal{K}^{(1)} = \sum_{\mathbf{n}\neq \mathbf{0}} \frac{1}{(|\mathbf{n}|^2-1)^3} = 13.0594~,\hspace{1.0cm}
\mathcal{L}^{(1)}  = \sum_{\mathbf{n} \neq \mathbf{0}} \frac{1}{(|\mathbf{n}|^2-1)^4} = 13.7312~,\nonumber
\end{equation}
\begin{equation}
\mathcal{O}^{(1)} = \sum_{\mathbf{n} \neq \mathbf{0}} \frac{1}{(|\mathbf{n}|^2-1)^{5}} = 11.3085~,\label{E-A-5.1-16}
\end{equation}
where the first three coincide respectively with $\mathcal{I}_1$, $\mathcal{J}_1$ and $\mathcal{K}_1$ in
Appendix~C of ref.~\cite{PWH19}. In particular, all the series not regulated by a cutoff can be
computed directly, without the need to resort to the techniques outlined above. On the other hand, the
sum of $\mathcal{I}^{(1)}$ can be obtained rapidly from the existing result for $\mathcal{I}^{(0)}$.
In fact, the addition and subtraction of a $1/|\mathbf{n}|^2$ term gives
\begin{equation}
\mathcal{I}^{(1)} = \sum_{|\mathbf{n}|\neq 0, 1}^{+\infty} \frac{1}{|\mathbf{n}|^2(|\mathbf{n}|^2 - 1)}
+ \mathcal{I}^{(0)} + 5 \equiv \digamma_1 + \mathcal{I}^{(0)} + 5~,\label{E-A-5.1-17}
\end{equation}
where in the first term on the r.h.s. the limit $\Lambda_n \rightarrow +\infty$ has been taken. Moreover,
replacing $\mathcal{I}^{(0)}$ with its expression given in eq.~\eqref{E-A-5.1-10}, the last formula can be
rewritten in a compact fashion as
\begin{equation}
\mathcal{I}^{(1)} = 9 - 2\pi^2 + 2\sum_{|\mathbf{n}|\neq 1}^{+\infty} \frac{1}{|\mathbf{n}|^2 + 1}
\frac{1}{|\mathbf{n}|^2 - 1}   + \pi \sum_{\mathbf{n} \neq \mathbf{0}}^{\infty} \frac{e^{-2\pi|\mathbf{n}|}}{|\mathbf{n}|}
\equiv 9 - 2\pi^2 + 2\gamma_2   + \pi \Sampi_0~,\label{E-A-5.1-18}
\end{equation}
so that $\mathcal{I}^{(1)}$ can be evaluated independently from $\mathcal{I}^{(0)}$. Finally, we conclude
the paragraph by enumerating the single series which do not appear directly in the $\delta\tilde{p}^2~\sim~0$
expansions of the L\"uscher functions but that play an ancillary role in the evaluation of the single
sums listed in eqs.~\eqref{E-A-5.1-15} and \eqref{E-A-5.1-18} (cf. $\gamma_1$, $\gamma_2$ and $\digamma_1$)
or in the double sums in eqs.~\eqref{E-A-5.2-04} and \eqref{E-A-5.2-05} (cf. $\digamma_1-\digamma_4$),
\begin{equation}
\gamma_1 = \sum_{\mathbf{n}\neq\mathbf{0}}^{\infty} \frac{1}{|\mathbf{n}|^2 (|\mathbf{n}|^2+1)} = 11.7861~,\hspace{1.0cm}
\gamma_2 = \sum_{|\mathbf{n}|\neq 1}^{+\infty} \frac{1}{|\mathbf{n}|^2 + 1}\frac{1}{|\mathbf{n}|^2 - 1}
= 10.7442~,\label{E-A-5.1-19}
\end{equation}
and
\begin{equation}
\digamma_1 = \sum_{|\mathbf{n}| > 1}^{\infty} \frac{1}{|\mathbf{n}|^2} \frac{1}{|\mathbf{n}|^2 - 1} = 14.7022~,\hspace{1.0cm}
\digamma_2 = \sum_{|\mathbf{n}>1}^{\infty} \frac{1}{|\mathbf{n}|^2}\frac{1}{(\mathbf{n}^2-1)^2} = 7.5410~,\nonumber
\end{equation}
\begin{equation}
\digamma_3 = \sum_{|\mathbf{n}| > 1}^{\infty} \frac{1}{|\mathbf{n}|^2} \frac{1}{(|\mathbf{n}|^2 - 1)^3} = 6.5185 ~,\hspace{1.0cm}
\digamma_4 = \sum_{|\mathbf{n}>1}^{\infty} \frac{1}{|\mathbf{n}|^2}\frac{1}{(\mathbf{n}^2-1)^4} = 6.2128~.
\label{E-A-5.1-20}
\end{equation}
Both the classes of series in eqs.~\eqref{E-A-5.1-17} and \eqref{E-A-5.1-18} do not display convergence issues and
can be directly evaluated. Note that $\digamma_1$ coincides with the sum listed as $\chi_3$ in
eq.~(A1) of ref.~\cite{BeS14}.

\subsection{\textsf{Double Sums}}\label{S-A-5.2}

Differently from their single counterparts, double sums appear only in the purely Coulombic contributions
in the $\ell=1$ ERE and arise from the $\tilde{p}^2 \rightarrow 1$ limit of the L\"uscher functions
$\mathcal{S}_2(\tilde{p})$ and $\mathcal{S}_3(\tilde{p})$. Furthermore, the $\tilde{p}^2 \approx 1$
expansion of the functions $\mathcal{S}_2(\tilde{p})$  and $\mathcal{S}_3(\tilde{p})$ in sec.~\ref{S-3.3.1}
generates two categories of double sums. The simplest of them consists in one three-dimensional Riemann sum
performed on $\mathbb{Z}^3$, followed by a sum over the six possible unit-vectors parallel to the axes of a
cubic lattice. Adopting the notation introduced for the derivation of the of the $\tilde{p}^2
\approx 1$ limit of $\mathcal{S}_2(\tilde{p})$ in sec.~III~D~2 of ref.~\cite{BeS14} , we write 
\begin{equation}
\chi_0 = \sum_{|\mathbf{n}| = 1} \sum_{\substack{|\mathbf{m}| = 1\\ \mathbf{n} \neq \mathbf{m} }}
\frac{1}{|\mathbf{m}-\mathbf{n}|^2} = \frac{27}{2}~,\hspace{0.6cm}
\chi_1 = \sum_{|\mathbf{n}|=1} \sum_{\substack{\mathbf{m} \neq \mathbf{n} \\ |\mathbf{m}| \neq  1}}^{\infty}
\frac{1}{|\mathbf{m}|^2 - 1} \frac{1}{|\mathbf{n}-\mathbf{m}|^2} = 86.1806~,\nonumber
\end{equation}
\begin{equation}
\chi_2 = \sum_{|\mathbf{n}|=1}\sum_{\substack{|\mathbf{m}| \neq 1 \\ \mathbf{m} \neq \mathbf{n}}}^{\infty}
\frac{1}{(|\mathbf{m}|^2 - 1)^2}\frac{1}{|\mathbf{n}-\mathbf{m}|^2} = 52.5687~,\hspace{0.4cm}
 \chi_3 = \sum_{|\mathbf{n}|=1}\sum_{\substack{|\mathbf{m}| \neq 1 \\ \mathbf{m} \neq \mathbf{n}}}^{\infty}
 \frac{1}{(|\mathbf{m}|^2 - 1)^3}\frac{1}{|\mathbf{n}-\mathbf{m}|^2} = 34.0562 ~,\nonumber
\end{equation}
\begin{equation}
 \chi_4 = \sum_{|\mathbf{n}|=1}\sum_{\substack{|\mathbf{m}| \neq 1 \\ \mathbf{m} \neq \mathbf{n}}}^{\infty}
 \frac{1}{(|\mathbf{m}|^2 - 1)^4}\frac{1}{|\mathbf{n}-\mathbf{m}|^2} =44.1196 ~.
 \label{E-A-5.2-01}
\end{equation}
The presence of the factor $1/|\mathbf{m}-\mathbf{n}|^2$ in the sums $\chi_1-\chi_4$ of eq.~\eqref{E-A-5.2-01}
ensures convergence without the need for the introduction of spherical cutoffs and regulators. Besides,
the sum $\chi_0$ is analytical and  denoted as $\chi_1$ in ref.~\cite{BeS14}, whereas the series $\chi_2$
and $\chi_3$ coincide with the $\chi_2$ and $\chi_5$, respectively, in the latter work, except for the
inclusion of the zero mode in the sum over $\mathbf{m}$.

Another group of series belonging to the same category is provided by the sums which are not present
in the expansions of the $\mathcal{S}_2(\tilde{p})$ and $\mathcal{S}_3(\tilde{p})$ L\"uscher functions,
but occur in the rewriting of certain double sums in terms of the already existing results in literature.
Due to the fast convergence, the evaluation such sums does not display difficulties and gives
\begin{equation}
 \sampi_1 = \sum_{|\mathbf{n}| = 1}\sum_{ |\mathbf{m}| > 1}^{\infty} \frac{1}{|\mathbf{m}|^2}
 \frac{1}{|\mathbf{m}-\mathbf{n}|^2} = 65.3498~,\hspace{1.0cm}
\sampi_2 = \sum_{|\mathbf{n}| = 1} \sum_{|\mathbf{n}| > 1}^{\infty} \frac{1}{|\mathbf{m}|^4}
\frac{1}{|\mathbf{m}-\mathbf{n} |^2} = 14.9350~,\nonumber
\end{equation}
\begin{equation}
 \sampi_3 = \sum_{|\mathbf{n}| = 1}\sum_{ |\mathbf{m}| > 1}^{\infty} \frac{1}{|\mathbf{m}|^6}
 \frac{1}{|\mathbf{m}-\mathbf{n}|^2} = 5.8426 ~,\hspace{1.0cm}
\sampi_4 = \sum_{|\mathbf{n}| = 1} \sum_{|\mathbf{n}| > 1}^{\infty} \frac{1}{|\mathbf{m}|^8}
\frac{1}{|\mathbf{m}-\mathbf{n} |^2} = 2.6217~,\label{E-A-5.2-02}
\end{equation}
thus permitting a fast evaluation of $\mathcal{R}^{(1)}$ and $\mathcal{R}_{2i~2j}^{(1)}$ in terms of the
existing results in ref.~\cite{BeS14}. Finally, of the same kind of the sums in eqs.~\eqref{E-A-5.2-01}
and \eqref{E-A-5.2-02} is the series $\Qoppa_0$ in eq.~\eqref{E-3.3.1-19}, which appears as a proportionality
constant in the $\mathcal{O}(1/\delta\tilde{p}^2)$ contributions to $\mathcal{S}_3$ (cf. eq.~\eqref{E-3.3.1-25})
and allows for a pairwise elementary numerical evaluation,
\begin{equation}
\Qoppa_0  = \sum_{|\mathbf{n}|=1}\sum_{|\mathbf{m}| \neq 1}^{\infty} \frac{ \mathbf{m}\cdot\mathbf{n}-1}{(|\mathbf{m}|^2
- 1)|\mathbf{m} - \mathbf{n}|^2} = -29.85670 (03) ~.\label{E-A-5.2-03}
\end{equation}

Now we switch to the second category of double sums, the one consisting of two three-dimensional sums
performed on $\mathbb{Z}^3$. First, we consider the series stemming from the $\tilde{p} \sim 1$
approximations of $\mathcal{S}_2(\tilde{p})$, see eq.~\eqref{E-3.3.1-09}. The latter sums, in fact,
are the counterpart of divergent double integrals contributing to the amplitudes of the relevant
two-particle scattering processes. Due to the large increase of the configuration space, for the
numerical calculation of the sum of such series it is advisable to parallelize the operations via the
development of GPU codes (e.g. in Cuda C++). The computational efforts can be significantly reduced by
subdividing the original double sum into an arbitrarily large finite number of single sums, characterized
by a three dimensional vector of integers. Then, assigning each of the outcoming single sums to a
different subunit of a graphic card, the sum of the original double series is derived by gathering the
results obtained simultaneously by each operating unit.

In particular, the series $\mathcal{R}^{(1)}$ in eq.~\eqref{E-3.3.1-07} can be expressed in terms
of the already known $\mathcal{R} \equiv \mathcal{R}^{(0)}$ in eq.~(30) of ref.~\cite{BeS14}. By adding and
subtracting $\mathcal{R}^{(0)}$ from $\mathcal{R}^{(1)}$ in eq.~ and performing few manipulations, the latter
series can be conveniently recast as
\begin{equation}
\mathcal{R}^{(1)} = \sum_{|\mathbf{n}| > 1}^{\infty}\sum_{\substack{|\mathbf{m}| > 1 \\ \mathbf{m} \neq \mathbf{n}}}^{\infty}
\frac{ |\mathbf{n}|^2 + |\mathbf{m}|^2 - 1}{|\mathbf{n}|^2(|\mathbf{n}|^2 - 1)|\mathbf{m}|^2(|\mathbf{m}|^2-1)|
\mathbf{m}-\mathbf{n}|^2} +  \mathcal{R}^{(0)}  - 2\chi_0 -2\sampi_1 -2\digamma_1 = -101.016  (11) ~.
\label{E-A-5.2-04}
\end{equation}
Once in this form, the sum of $\mathcal{R}^{(1)}$ can be obtained by exploiting the existing result for the
cutoff-regularized sum $\mathcal{R}^{(0)}$ in ref.~\cite{BeS14}, together with the single sums in
eqs.~\eqref{E-A-5.1-20}, \eqref{E-A-5.2-01} and \eqref{E-A-5.2-02}. The only additional computational
effort is given by the double sum explicitly shown on the r.h.s of eq.~\eqref{E-A-5.2-04}, which proves
to converge rapidly, differently from $\mathcal{R}^{(0)}$. For the last sum, in fact, an approach analogous
to the one shown in eq.~\eqref{E-A-5.1-07} or to the tail-singularity separation (TSS) in ref.~\cite{Tan08}
is recommendable. In the latter method, summarized in detail in ref.~\cite{BeS14}, a three-dimensional
Riemann sum is subdivided into an IR part, dominated by the the singularities of the summand and an UV
part, expressed in the form of a three-dimensional integral and describing the behaviour of the argument
of the original sum towards the infinity. As shown in the Appendix~A of ref.~\cite{Tan08} for the
sums $\theta_{As}$ and $\theta_{Bs}$, the TSS approach holds also for double sums regulated asymmetrically.

Second, we switch to the series of the kind $\mathcal{R}_{2i~2j}^{(1)}$ in eq.~\eqref{E-3.3.1-08}. Even if
the evaluation of such sums does not require stabilization techniques, we present for completeness the
expression of $\mathcal{R}_{2i~2j}^{(1)}$ in terms of $\mathcal{R}_{2i~2j}^{(0)}$ and ancillary single and double sums:
\begin{equation}
\begin{split}
\mathcal{R}_{2i~2j}^{(1)} = \sum_{|\mathbf{n}| > 1}^{\infty}\sum_{\substack{|\mathbf{m}|> 1 \\ \mathbf{m} \neq \mathbf{n}}}^{\infty}
\left[\frac{1}{(|\mathbf{n}|^2-1)^i} \frac{1}{(|\mathbf{m}|^2-1)^j} - \frac{1}{|\mathbf{n}|^{2i} |\mathbf{m}|^{2j}}
\right]\\ \cdot \frac{1}{|\mathbf{m}-\mathbf{n}|^2} +  \mathcal{R}_{2i~2j}^{(0)} - 2\chi_0 & - \sampi_i -
\sampi_j + (-1)^j \digamma_i + (-1)^i \digamma_j~.\label{E-A-5.2-05}
\end{split}
\end{equation}
As noticed in ref.~\cite{BeS14}, the only contribution of such sums in the expression of the finite volume
energy corrections for the lowest energy $T_1$ eigenstate (cf. eq.~\eqref{E-3.3.1-34} and \eqref{E-3.3.1-35})
is provided by $\mathcal{R}_{24}^{(1)} = \mathcal{R}_{42}^{(1)}$, whose explicit evaluation gives $-93.692 (10)$.
 
\begin{figure}[hb!]
\begin{center}
\includegraphics[width=0.95\columnwidth]{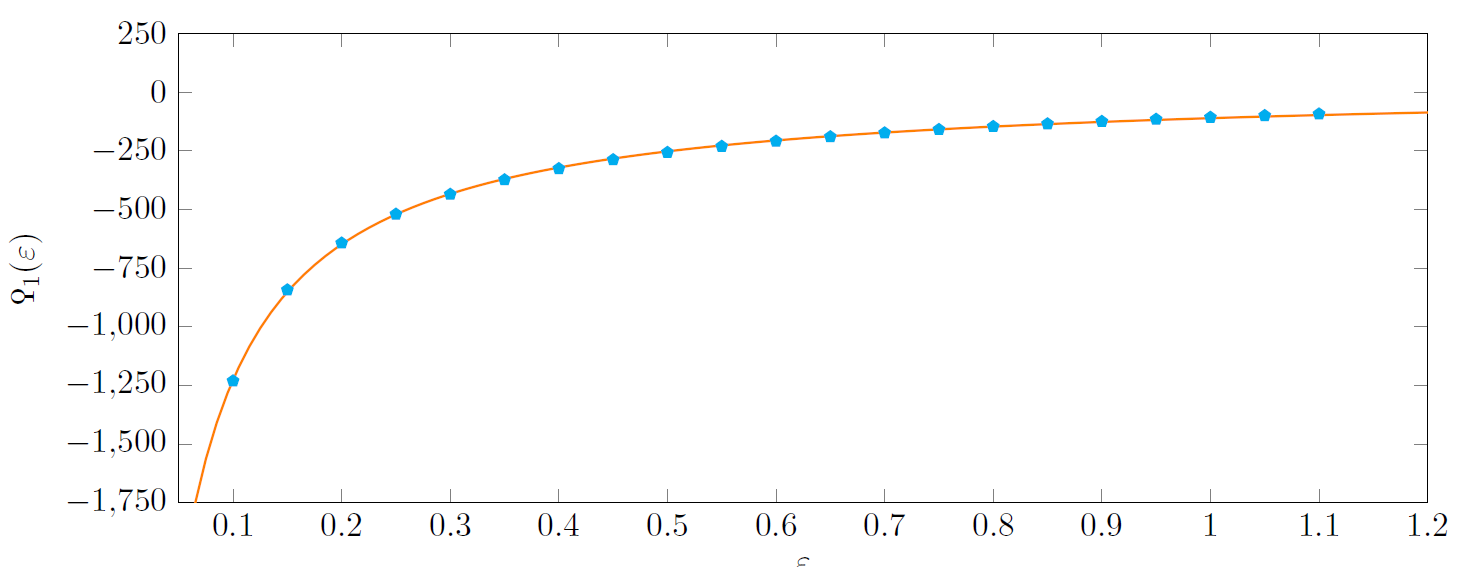}
\end{center}
\caption{Sum of the stabilized $\Qoppa_1$ series as a function of the parameter $\epsilon$. The fitting
model, marked by the continuous orange line, corresponds to $f(\varepsilon)$ in eq.~\eqref{E-A-5.1-09}
with $a = -149.59(47)$, $b = 0.0185484(64)$ and $c=36.5(1.4)$. The value of the parameter $b$ plays 
a crucial role in determining the sum of the series and maximizes the value of the
coefficient of determination ($R^2$) of the fit for the considered fitting function. In particular,
the R squared is equal to $0.999815$, thus ensuring the reliability of the interpolation.}\label{F-A-02}
\end{figure}

Finally, we present the double sums arising from the $\tilde{p} \sim 1$ approximation of
$\mathcal{S}_3(\tilde{p})$ in eq.~\eqref{E-3.3.1-25}, starting with the cutoff-regularized double sum
$\Qoppa_1$ in eq.~\eqref{E-3.3.1-20}. Since the argument of the sum is quadratically divergent with
the spherical cutoff, it is convenient to adopt a stabilization technique for the evaluation of the sum.
To this purpose, we chose to apply the approach in eq.~\eqref{E-A-5.1-07} to the cutoff-regularized
sum over $\mathbf{n}$, 
\begin{equation}
\Qoppa_1 = \lim_{\varepsilon\rightarrow 0^+}\left[\sum_{|\mathbf{n}|\neq 1}^{\Lambda_n} \sum_{\substack{\mathbf{m}\neq\mathbf{n} \\
|\mathbf{m}| \neq 1}}^{\infty} \frac{e^{-\varepsilon|\mathbf{n}|}}{|\mathbf{n}|^2 - 1}\frac{1}{|\mathbf{m}|^2 - 1} 
\frac{\mathbf{n}\cdot \mathbf{m} - 1}{|\mathbf{n}-\mathbf{m}|^2} - 2\pi^4 
\int_0^{\Lambda_n} \mathrm{d}n~ne^{-\varepsilon n} \right]~.\label{E-A-5.2-06}
\end{equation}
Since the $\varepsilon$-dependent $\Qoppa_1$ sums display a non-linear behaviour in the interval $0 \leq
\varepsilon \leq 1.1$, we choose to interpolate the data with the fitting function in eq.~\eqref{E-A-5.1-09}.
As shown in fig.~\ref{F-A-02}, $f(\varepsilon)$ describes the behaviour of the sum of the stabilized
series as a function of $\varepsilon$ satisfactorily, therefore the sum of the series becomes
\begin{equation}
\Qoppa_1  = \sum_{|\mathbf{n}|\neq 1}^{\Lambda_n} \sum_{\substack{\mathbf{m}\neq\mathbf{n} \\ |\mathbf{m}| \neq 1}}^{\infty}
\frac{1}{|\mathbf{n}|^2 - 1}\frac{1}{|\mathbf{m}|^2 - 1}  \frac{\mathbf{n}\cdot \mathbf{m} - 1}
{|\mathbf{n}-\mathbf{m}|^2}  - \pi^4\Lambda_n^2 = -8028.1  (24.2) ~.\label{E-A-5.2-07}
\end{equation}
Conversely, the double sum $\Qoppa_2$ (cf. eq.~\eqref{E-3.3.1-21}) appearing at order $\delta\tilde{p}^2$
in the power series expansion of $\mathcal{S}_3(\tilde{p})$ can be calculated efficiently even without
stabilization approaches, despite its sign-changing numerator. Its numerical evaluation yields 
\begin{equation}
\begin{split}
\Qoppa_2  =  \sum_{|\mathbf{n}|\neq 1}^{\infty} \sum_{\substack{\mathbf{m}\neq\mathbf{n} \\ |\mathbf{m}| \neq 1}}^{\infty} \frac{1
- |\mathbf{m}|^2 |\mathbf{n}|^2 + \mathbf{n}\cdot\mathbf{m}(|\mathbf{n}|^2 + |\mathbf{m}|^2 - 2)}
{(|\mathbf{m}|^2-1)^2(|\mathbf{n}|^2-1)^2 |\mathbf{m}-\mathbf{n}|^2} \\ = -315.981 (74) ~.\label{E-A-5.2-08}
\end{split}
\end{equation}
Analogous considerations hold for the $\Qoppa_3$ series (cf. eq.~\eqref{E-3.3.1-22}) emerging from the
$\delta\tilde{p}^4$ contributions to $\mathcal{S}_3(\tilde{p})$. For large values of $|\mathbf{n}|$ the
series proves to converge even more rapidly than $\Qoppa_2$, therefore the statistical errors associated
to the sum are smaller,
\begin{equation}
\begin{split}
\Qoppa_3 =  \sum_{|\mathbf{n}|\neq 1}^{\infty} \sum_{\substack{\mathbf{m}\neq\mathbf{n} \\ |\mathbf{m}| \neq 1}}^{\infty}
\frac{q_S(\mathbf{n},\mathbf{m}) + \mathbf{m}\cdot\mathbf{n}~q_X(\mathbf{n},\mathbf{m})}{(|\mathbf{m}|^2-1)^3
(|\mathbf{n}|^2-1)^3 |\mathbf{m}-\mathbf{n}|^2} \\ = -384.169 (03) &~.
\label{E-A-5.2-09}
\end{split}
\end{equation}
where the polynomials $q_S(\mathbf{n},\mathbf{m})$ and $q_X(\mathbf{n},\mathbf{m})$ (cf. eqs.~\eqref{E-3.3.1-23}
and \eqref{E-3.3.1-24}) in the numerator are given by
\begin{equation}
q_S(\mathbf{n},\mathbf{m}) = -1 + 3 |\mathbf{n}|^2 |\mathbf{m}|^2 - |\mathbf{m}|^2 |\mathbf{n}|^2
(|\mathbf{n}|^2+|\mathbf{m}|^2) \label{E-A-5.2-10}
\end{equation}
and
\begin{equation}
q_X(\mathbf{n},\mathbf{m}) = 3 - 3(|\mathbf{n}|^2 + |\mathbf{m}|^2) +  |\mathbf{m}|^2|\mathbf{n}|^2
+ |\mathbf{m}|^4 + |\mathbf{n}|^4~.\label{E-A-5.2-11}
\end{equation}

\section{\textsf{L\"uscher functions}}\label{S-A-6.0}

In this appendix, we concentrate on the derivation of the large imaginary momentum $\tilde{p}
= \mathrm{i}\tilde{\kappa}$ limit of the double sum $\mathcal{S}_3(\tilde{p})$ in eq.~\eqref{E-3.2-08},
encountered in the derivation of the finite volume energy corrections for the lowest bound $T_1$ state
in sec.~\ref{S-3.3.2}. To this aim, we start by splitting the original L\"uscher function
$\mathcal{S}_3(\mathrm{i}\tilde{\kappa})$ into three parts, 
\begin{equation}
\begin{split}
\mathcal{S}_3(\mathrm{i}\tilde{\kappa}) = \sum_{\mathbf{n}}^{\Lambda_n}\sum_{\mathbf{m}\neq\mathbf{n}}^{+\infty}
\frac{1}{|\mathbf{n}|^2 + \tilde{\kappa}^2}\frac{1}{|\mathbf{m}|^2 + \tilde{\kappa}^2}\frac{\mathbf{n}
\cdot\mathbf{m} + \tilde{\kappa}^2}{|\mathbf{n}-\mathbf{m}|^2} - \pi^4\Lambda_n^2 \\
= \lim_{\Lambda_m\rightarrow \infty} \Big\{ \mathfrak{S}_1^{\Lambda_m} + \mathfrak{S}_2^{\Lambda_m}
& + \mathfrak{S}_3^{\Lambda_m} \Big\}  - \pi^4\Lambda_n^2~,\label{E-A-6.0-01}
\end{split}
\end{equation}
where $\Lambda_m$ is a spherical cutoff accounting for the divergence of the following sums over $\mathbf{m}$,
\begin{equation}
\mathfrak{S}_1^{\Lambda_m} = \frac{1}{2} \sum_{\mathbf{n}}^{\Lambda_n}\sum_{\mathbf{m}\neq\mathbf{n}}^{\Lambda_m}
\frac{1}{|\mathbf{m}|^2 +\tilde{\kappa}^2}\frac{1}{|\mathbf{m}-\mathbf{n}|^2}~,\label{E-A-6.0-02}
\end{equation}
\begin{equation}
\mathfrak{S}_2^{\Lambda_m} = \frac{1}{2} \sum_{\mathbf{n}}^{\Lambda_n}\sum_{\mathbf{m}\neq\mathbf{n}}^{\Lambda_m}
\frac{1}{|\mathbf{n}|^2 + \tilde{\kappa}^2}\frac{1}{|\mathbf{m}-\mathbf{n}|^2}~,\label{E-A-6.0-03}
\end{equation}
\begin{equation}
\mathfrak{S}_3^{\Lambda_m} = - \frac{1}{2}\sum_{\mathbf{n}}^{\Lambda_n}\sum_{\mathbf{m}\neq\mathbf{n}}^{\Lambda_m}
\frac{1}{|\mathbf{n}|^2 + \tilde{\kappa}^2} \frac{1}{|\mathbf{m}|^2 + \tilde{\kappa}^2}~.\label{E-A-6.0-04}
\end{equation}
Due to the presence of a cutoff in the inner sum, the translation in momentum space operated in the
$\mathcal{S}_2(\mathrm{i}\tilde{\kappa})$ case (cf. Appendix~A3 in ref.~\cite{BeS14}) is no longer
allowed in the individual sums (cf. eqs.~\eqref{E-A-6.0-02}-\eqref{E-A-6.0-04}). Nevertheless, since the
purpose is the extraction of the finite and $\tilde{\kappa}$-dependent contributions from each of
the three double sums in eq.~\eqref{E-A-6.0-01}, terms depending on nonzero powers of $\Lambda_m$
and $\Lambda_n$ can be neglected without loss of information for the FVECs. Therefore, we assume
henceforth the limits $\Lambda_n$, $\Lambda_m \rightarrow +\infty$ and extract the finite parts from
the sums depending on the binding momentum. Now, we consider the first of the three double sums in the
second row of eq.~\eqref{E-A-6.0-01}. After the translation in the momenta $\mathbf{m} \mapsto \mathbf{m}
-\mathbf{n}\equiv \mathbf{p}$, we rewrite the argument of the inner sum in integral form,
\begin{equation}
\mathfrak{S}_1^{\infty} = \frac{1}{2} \sum_{\mathbf{p}\neq \mathbf{0}}^{\infty} \frac{1}{|\mathbf{p}|^2}
\int_{\mathbb{R}^3} \mathrm{d}^3\mathbf{q} \sum_{\mathbf{n}}^{\infty}\frac{\delta^3(\mathbf{q}-\mathbf{n})}{|\mathbf{q}
+ \mathbf{p}|^2 +\tilde{\kappa}^2}~.\label{E-A-6.0-05}
\end{equation}
Then, we apply the Poisson summation formula to the unconstrained sum over $\mathbf{n}$ and isolate
the zero modes from the nonzero ones,
\begin{equation}
\mathfrak{S}_1^{\infty} = \frac{1}{2} \sum_{\mathbf{p}\neq \mathbf{0}}^{\infty} \frac{1}{|\mathbf{p}|^2}
\int_{\mathbb{R}^3} \mathrm{d}^3\mathbf{q} \frac{1}{|\mathbf{q} + \mathbf{p}|^2 +\tilde{\kappa}^2} 
+ \frac{1}{2} \sum_{\mathbf{p}\neq \mathbf{0}}^{+\infty} \frac{1}{|\mathbf{p}|^2} \sum_{\mathbf{n} \neq \mathbf{0}}^{+\infty}
\int_{\mathbb{R}^3} \mathrm{d}^3\mathbf{q} \frac{e^{-\mathrm{i}2\pi \mathbf{q} \cdot \mathbf{n}}}{|\mathbf{q}
+ \mathbf{p}|^2 +\tilde{\kappa}^2}  ~.\label{E-A-6.0-06}
\end{equation}
Concentrating on the first term of eq.~\eqref{E-A-6.0-06}, we perform the translation in momentum
space $\mathbf{q} \mapsto \mathbf{q} - \mathbf{p} \equiv \mathbf{l}$ and integrate over the angular
variables associated to $\mathbf{l}$,
\begin{equation}
\frac{1}{2} \sum_{\mathbf{p}\neq \mathbf{0}}^{\infty} \frac{1}{|\mathbf{p}|^2} \int_{\mathbb{R}^3} \mathrm{d}^3\mathbf{q}
\frac{1}{|\mathbf{q} + \mathbf{p}|^2 +\tilde{\kappa}^2}  = 2\pi \sum_{\mathbf{p}\neq \mathbf{0}}^{+\infty}
\frac{1}{|\mathbf{p}|^2} \int_0^{+\infty} \mathrm{d}l  - 2\pi \sum_{\mathbf{p}\neq \mathbf{0}}^{+\infty}
\frac{1}{|\mathbf{p}|^2} \int_0^{+\infty} \mathrm{d}l \frac{\tilde{\kappa}^2}{\mathbf{l}^2 + \tilde{\kappa}^2}~.
\label{E-A-6.0-07}
\end{equation}
As it can be inferred, the first term on the r.h.s. of the last equation is independent on $\tilde{\kappa}$
and unbound, thus it can be neglected. On the other hand, the integral in the second term is finite and
generates linear contributions in the binding momentum,
\begin{equation}
- 2\pi \sum_{\mathbf{p}\neq \mathbf{0}}^{\infty} \frac{1}{|\mathbf{p}|^2} \int_0^{+\infty} \mathrm{d}l
\frac{\tilde{\kappa}^2}{\mathbf{l}^2 + \tilde{\kappa}^2}  = - 2\pi \tilde{\kappa}
\sum_{\mathbf{p}\neq \mathbf{0}}^{\infty} \frac{1}{|\mathbf{p}|^2} \mathrm{arctan} (x)\Big\lvert_0^{+\infty}
 = - \pi^2 \tilde{\kappa} \sum_{\mathbf{p}\neq \mathbf{0}}^{+\infty} \frac{1}{|\mathbf{p}|^2}~.\label{E-A-6.0-08}
\end{equation}
Moreover, the remaining sum in eq.~\eqref{E-A-6.0-07} can be evaluated by splitting it into two parts,
\begin{equation}
\begin{split}
- \pi^2 \tilde{\kappa} \sum_{\mathbf{p}\neq \mathbf{0}}^{+\infty} \frac{1}{|\mathbf{p}|^2}
= - \pi^2 \tilde{\kappa} \sum_{\mathbf{p}\neq \mathbf{0}}^{2\tilde{\kappa}}\frac{1}{\mathbf{p}^2}
- \pi^2 \tilde{\kappa} \sum_{|\mathbf{p}|> 2\tilde{\kappa} }^{+\infty}\frac{1}{\mathbf{p}^2} \\
\approx -\pi^2 \tilde{\kappa} (\mathcal{I}^{(0)} +  8\pi \tilde{\kappa} )  &  - 4\pi^3 \tilde{\kappa}
\int_{2\tilde{\kappa}}^{+\infty} \mathrm{d} p \rightarrow   -\pi^2 \mathcal{I}^{(0)} \tilde{\kappa}~,\label{E-A-6.0-09}
\end{split}
\end{equation}
where, in the last step, the sum has been approximated by an integral, since the binding momentum is expected to
be large, $\tilde{\kappa} \gg 1$. Additionally, the linearly divergent part of the radial integral has
been consistently discarded. At this stage, we focus on the nonzero modes, i.e. the second term in
eq.~\eqref{E-A-6.0-06}. After performing the translation $\mathbf{q} \mapsto \mathbf{q} - \mathbf{p}
\equiv \mathbf{l}$, we integrate over the angular variables associated to $\mathbf{l}$ and
simplify the expression as 
\begin{equation}
 \frac{1}{2} \sum_{\mathbf{p}\neq \mathbf{0}}^{+\infty} \frac{1}{|\mathbf{p}|^2} \sum_{\mathbf{n} \neq \mathbf{0}}^{\infty}
 \int_{\mathbb{R}^3} \mathrm{d}^3\mathbf{q} \frac{e^{-\mathrm{i}2\pi \mathbf{q} \cdot \mathbf{n}}}{|\mathbf{q}
 + \mathbf{p}|^2 +\tilde{\kappa}^2}  = \frac{1}{2}\sum_{\mathbf{p}\neq\mathbf{0}}^{\infty} \frac{1}{\mathbf{p}^2}
 \sum_{\mathbf{n}\neq\mathbf{0}}^{\infty} \int_0^{+\infty} \mathrm{d}l~\frac{l}{\mathrm{i}|\mathbf{n}|}
 \frac{e^{2\pi\mathrm{i} l |\mathbf{n}|} - e^{-2\pi\mathrm{i} l |\mathbf{n}|}}{l^2 +\tilde{\kappa}^2} ~,\label{E-A-6.0-10}
\end{equation}
where $l\equiv|\mathbf{l}|$ and the additional exponential factor $e^{2\pi\mathrm{i} \mathbf{p}\cdot\mathbf{n}}$
has been dropped, since for integer momenta it is equal to one. Furthermore, the argument of the radial
integral over the momenta $|\mathbf{l}|$ is even, thus the integration region can be extended to the
whole real axis. Additionally, the integrand tends uniformly to zero in the limit $l \rightarrow
\pm \infty$ and it is analytical all over the complex plane $l \in \mathbb{C}$, except for two simple
poles at $l_{\pm} = \pm \mathrm{i} \tilde{\kappa}$, located along the imaginary axis. It follows that
the integration region can be extended to an arbitrary large circular region about the origin of the
complex plane encompassing the two singularities. Moreover, the integrand can be split into two
functions of complex variable $l$
\begin{equation}
g_{\pm}(l) = \frac{l}{\mathrm{i}|\mathbf{n}|} \frac{\mathrm{e}^{\pm\mathrm{i}2\pi l |\mathbf{n}|} }
{(l - l_+)(l - l_-)}~,\label{E-A-6.0-11}
\end{equation}
so that $g_+(l)$ ($g_-(l)$) can be integrated in a semicircumference with arbitrarily large radius
about the origin in the upper (lower) part of the complex plane picking up the $l_+$ ($l_-$) singularity,
according to Jordan's Lemma. The residues associated to the two poles turn out to coincide and to
depend on $|\mathbf{n}|$ through negative exponentials. Observing again that $e^{-2\pi|\mathbf{n}|\tilde{\kappa}}
\leq e^{-\pi\tilde{\kappa}} e^{-\pi|\mathbf{n}|}$,  eq.~\eqref{E-A-6.0-10} becomes
\begin{equation}
\frac{1}{2} \sum_{\mathbf{p}\neq \mathbf{0}}^{\infty} \frac{1}{|\mathbf{p}|^2} \sum_{\mathbf{n}\neq \mathbf{0}}^{\infty}
\frac{\pi}{|\mathbf{n}|} e^{-2\pi|\mathbf{n}|\tilde{\kappa}} \leq \frac{1}{2} \sum_{\mathbf{p}\neq \mathbf{0}}^{\infty}
\frac{1}{|\mathbf{p}|^2} e^{-\pi\tilde{\kappa}}  \sum_{\mathbf{n}\neq \mathbf{0}}^{\infty}
\frac{\pi}{|\mathbf{n}|} e^{-\pi|\mathbf{n}|} = \Sampi_1 \frac{\pi}{2}
e^{-\pi\tilde{\kappa}} \sum_{\mathbf{p}\neq \mathbf{0}}^{\infty}  \frac{1}{|\mathbf{p}|^2}~,\label{E-A-6.0-12}
\end{equation}
where $\Sampi_1$ is a small constant equal to $0.400982(1)$. Finally, we evaluate the sum over $\mathbf{p}$
in the same fashion as eq.~\eqref{E-A-6.0-09} and we single out the non-divergent part, finding
 \begin{equation}
\begin{gathered}
\frac{1}{2} \sum_{\mathbf{p}\neq \mathbf{0}}^{+\infty} \frac{1}{|\mathbf{p}|^2} \sum_{\mathbf{n} \neq \mathbf{0}}^{+\infty}
\int_{\mathbb{R}^3} \mathrm{d}^3\mathbf{q} \frac{e^{-\mathrm{i}2\pi \mathbf{q} \cdot \mathbf{n}}}{|\mathbf{q} + \mathbf{p}|^2
+\tilde{\kappa}^2} \leq \Sampi_1 \frac{\pi}{2}  \sum_{\mathbf{p}\neq \mathbf{0}}^{\infty} \frac{e^{-\pi\tilde{\kappa}}}
{|\mathbf{p}|^2} =  \Sampi_1 \frac{\pi}{2} e^{-\pi\tilde{\kappa}} \sum_{\mathbf{p}\neq \mathbf{0}}^{2\tilde{\kappa}}
\frac{1}{|\mathbf{p}|^2}  \\ + \Sampi_1 \frac{\pi}{2} e^{-\pi\tilde{\kappa}} \sum_{|\mathbf{p}| > \tilde{\kappa}}^{\infty}
\frac{1}{|\mathbf{p}|^2} \approx \Sampi_1 \frac{\pi}{2} e^{-\pi\tilde{\kappa}}(\mathcal{I}^{(0)}+ 8\pi \tilde{\kappa})
+  \Sampi_1 2\pi^2 e^{-\pi\tilde{\kappa}} \int_{2\tilde{\kappa}}^{+\infty}\mathrm{d} p  \rightarrow \frac{\pi}{2}
\Sampi_1 \mathcal{I}^{(0)} e^{-\pi\tilde{\kappa}}~.\label{E-A-6.0-13}
\end{gathered}
\end{equation}
The finite contribution arising from the non-zero modes in eq.~\eqref{E-A-6.0-13} decays exponentially with
$\tilde{\kappa}$, thus it is negligible in the large binding momentum regime in comparison with the one
in eq.~\eqref{E-A-6.0-09}. Therefore, we retain only the latter and write
\begin{equation}
\mathfrak{S}_1^{\infty}  \rightarrow -\pi^2 \mathcal{I}^{(0)} \tilde{\kappa}~.\label{E-A-6.0-14}
\end{equation}
Now, we switch to the second term in the last row of eq.~\eqref{E-A-6.0-01} and we observe that, in the
$\Lambda_n, \Lambda_m \rightarrow +\infty$ limits, translational symmetry is restored and
$\mathfrak{S}_2^{\infty}$ coincides with $\mathfrak{S}_1^{\infty}$. As a consequence, we are allowed to write
\begin{equation}
\mathfrak{S}_2^{\infty}  = \frac{1}{2} \sum_{\mathbf{n}}^{\infty}\sum_{\mathbf{m}\neq\mathbf{n}}^{\infty}
\frac{1}{|\mathbf{n}|^2 + \tilde{\kappa}^2}\frac{1}{|\mathbf{m}-\mathbf{n}|^2}\rightarrow
-\pi^2 \mathcal{I}^{(0)} \tilde{\kappa}~.\label{E-A-6.0-15}
\end{equation}
At this stage, we can concentrate directly on the third term in the second row of eq.~\eqref{E-A-6.0-01},
which can be conveniently split as follows,
\begin{equation}
\mathfrak{S}_3^{\infty} = \frac{1}{2}\sum_{\mathbf{n}}^{\infty}\frac{1}{(|\mathbf{n}|^2 + \tilde{\kappa}^2)^2}
- \frac{1}{2}\sum_{\mathbf{n}}^{\infty}\sum_{\mathbf{m}}^{\infty}\frac{1}{|\mathbf{n}|^2 + \tilde{\kappa}^2}
\frac{1}{|\mathbf{m}|^2 + \tilde{\kappa}^2} ~,\label{E-A-6.0-16}
\end{equation}
so that the second term on the r.h.s. of the last equation is factorized. Since the two disentangled
sums in the product are identical, it is sufficient to evaluate only one of them and, then, to take the
square of the retained finite parts. As previously, we first rewrite the argument of the sum in integral
form and then exploit the Poisson summation formula,
\begin{equation}
\sum_{\mathbf{n}}^{\infty} \frac{1}{\mathbf{n}^2 +\tilde{\kappa}^2} = \sum_{\mathbf{n}}^{\infty} \int_{\mathbb{R}^3}
\mathrm{d}^3\mathbf{q} \frac{\delta^3(\mathbf{q}-\mathbf{n})}{\mathbf{q}^2 + \tilde{\kappa}^2} 
= \int_{\mathbb{R}^3} \mathrm{d}^3\mathbf{q}~ \frac{1}{\mathbf{q}^2 +\tilde{\kappa}^2}
+ \sum_{\mathbf{n}\neq\mathbf{0}}^{\infty}\int_{\mathbb{R}^3} \mathrm{d}^3\mathbf{q}~
\frac{e^{-2\pi\mathrm{i}\mathbf{n}\cdot\mathbf{q}}}{\mathbf{q}^2+\tilde{\kappa}^2}~,\label{E-A-6.0-17}
\end{equation}
where the zero modes have been isolated. As it can be observed, the expression on the r.h.s. of
eq.~\eqref{E-A-6.0-17} is identical to the one in eq.~\eqref{E-A-6.0-06} after the translation $\mathbf{q}
\mapsto \mathbf{q} - \mathbf{p} \equiv \mathbf{l}$ in momentum space, except for the outer sum over
$\mathbf{p}$ and the factor $1/2$. Therefore, we are allowed to exploit the results in
eq.~\eqref{E-A-6.0-08} and \eqref{E-A-6.0-12} for zero and non-zero modes respectively, obtaining
\begin{equation}
\int_{\mathbb{R}^3} \mathrm{d}^3\mathbf{q}~\frac{1}{\mathbf{q}^2 +\tilde{\kappa}^2}
+ \sum_{\mathbf{n}\neq\mathbf{0}}^{\infty}\int_{\mathbb{R}^3} \mathrm{d}^3\mathbf{q}~
\frac{e^{-2\pi\mathrm{i}\mathbf{n}\cdot\mathbf{q}}}{\mathbf{q}^2+\tilde{\kappa}^2} 
\rightarrow -2\pi^2 \tilde{\kappa}  + \Sampi_1 \pi e^{-\pi\tilde{\kappa}}~.\label{E-A-6.0-18}
\end{equation}
As a consequence the second term on the r.h.s. of eq.~\eqref{E-A-6.0-16} can be finally rewritten as
\begin{equation}
- \frac{1}{2}\sum_{\mathbf{n}}^{\infty}\sum_{\mathbf{m}}^{\infty}\frac{1}{|\mathbf{n}|^2 + \tilde{\kappa}^2}
\frac{1}{|\mathbf{m}|^2 + \tilde{\kappa}^2}  \rightarrow - 2\pi^4 \tilde{\kappa}^2
- \Sampi_1^2 \frac{\pi^2}{2} e^{-2\pi\tilde{\kappa}}  + 2\pi^3 \Sampi_1 \tilde{\kappa} e^{-\pi\tilde{\kappa}}~,
\label{E-A-6.0-19}
\end{equation}
where the last two terms on the r.h.s. are exponentially suppressed and they have to be neglected
for consistency. Second, we switch to the single sum on the r.h.s. of eq.~\eqref{E-A-6.0-16}. Introducing
the integral sign and exploiting again the Poisson summation formula, the last contribution to
$\mathcal{S}_3(\mathrm{i}\tilde{\kappa})$ becomes 
\begin{equation}
\frac{1}{2}\sum_{\mathbf{n}}^{\infty}\frac{1}{(|\mathbf{n}|^2 + \tilde{\kappa}^2)^2} = \frac{1}{2}
\int_{\mathbb{R}^3} \frac{1}{(\mathbf{q}^2 +\tilde{\kappa}^2)^2} 
+ \frac{1}{2}\sum_{\mathbf{n}\neq\mathbf{0}}^{\infty}\int_{\mathbb{R}^3} \mathrm{d}^3\mathbf{q}~
\frac{e^{-2\pi\mathrm{i}\mathbf{n}\cdot\mathbf{q}}}{(\mathbf{q}^2+\tilde{\kappa}^2)^2}~,\label{E-A-6.0-20}
\end{equation}
where the zero modes have been again isolated from the non-zero ones. The first integral in
eq.~\eqref{E-A-6.0-20} can be carried out after few manipulations,
\begin{equation}
\frac{1}{2} \int_{\mathbb{R}^3}\mathrm{d}^3\mathbf{q}~\frac{1}{(\mathbf{q}^2 +\tilde{\kappa}^2)^2}
= \frac{2\pi}{\tilde{\kappa}} \mathrm{arctan}(x)\Big\lvert_0^{+\infty} 
-  \frac{\pi x}{x^2 + \tilde{\kappa}^2}\Big\lvert_0^{+\infty} - \frac{\pi }{\tilde{\kappa}}
\mathrm{arctan}\left(\frac{x}{\tilde{\kappa}}\right)\Big\lvert_0^{+\infty}   =  \frac{\pi^2}{2\tilde{\kappa}} ~.
\label{E-A-6.0-21}
\end{equation}
Then, the integration over the angular variables associated to $\mathbf{q}$  in eq.~\eqref{E-A-6.0-21} gives
\begin{equation}
\frac{1}{2} \sum_{\mathbf{n}\neq\mathbf{0}}^{\infty}\int_{\mathbb{R}^3} \mathrm{d}^3\mathbf{q}~
\frac{e^{-2\pi\mathrm{i}\mathbf{n}\cdot\mathbf{q}}}{(\mathbf{q}^2+\tilde{\kappa}^2)^2} 
= \sum_{\mathbf{n}\neq\mathbf{0}}^{\infty} \int_0^{+\infty}\mathrm{d}q~\frac{q}{\mathrm{i}|\mathbf{n}|}
 \frac{e^{2\pi \mathrm{i} q |\mathbf{n}|} - e^{-2\pi \mathrm{i} q |\mathbf{n}|}}{(q^2+\tilde{\kappa}^2)^2}~.\label{E-A-6.0-22}
\end{equation}
where $q \equiv |\mathbf{q}|$. As in the case of eq.~\eqref{E-A-6.0-10}, the integrand is an even
function of $q$, thus the integration region can be extended to the whole real axis. Besides, the
integrand tends uniformly to zero in the limit $l \rightarrow \pm \infty$ and is analytical all over
the complex plane $l \in \mathbb{C}$, except for two double poles at $q_{\pm} = \pm \mathrm{i} \tilde{\kappa}$,
located along the imaginary axis. It follows that the integration region can be extended to an
arbitrary large circular region about the origin of the complex plane encompassing the two singularities.
Moreover, the integrand can be split into two functions of complex variable $q$
\begin{equation}
h_{\pm}(q) = \frac{q}{\mathrm{i}|\mathbf{n}|} \frac{\mathrm{e}^{\pm\mathrm{i}2\pi q |\mathbf{n}|} }{(q- q_+)^2(q- q_-)^2}~,
\label{E-A-6.0-23}
\end{equation}
so that $h_+(q)$ ($h_-(q)$) can be integrated in a semicircumference with arbitrarily large
radius about the origin in the upper (lower) part of the complex plane picking up the $q_+$ ($q_-$)
singularity, according to Jordan's Lemma. Again, the residues about the two double poles turn out
to coincide and to depend on $|\mathbf{n}|$ through negative exponentials, 
\begin{equation}
\sum_{\mathbf{n}\neq\mathbf{0}}^{\infty} \int_0^{+\infty}\mathrm{d}q~\frac{q}{\mathrm{i}|\mathbf{n}|}
\frac{e^{2\pi \mathrm{i} q |\mathbf{n}|} - e^{-2\pi \mathrm{i} q |\mathbf{n}|}}{q^2+\tilde{\kappa}^2}
= \frac{\pi^2}{2\tilde{\kappa}} \sum_{\mathbf{n}\neq\mathbf{0}}^{\infty} e^{-2\pi \tilde{\kappa} |\mathbf{n}|}~.
\label{E-A-6.0-24}
\end{equation}
Now, observing again that $e^{-2\pi|\mathbf{n}|\tilde{\kappa}} \leq e^{-\pi\tilde{\kappa}} e^{-\pi|\mathbf{n}|}$,
the sum in eq.~\eqref{E-A-6.0-24} can be bound from above
\begin{equation}
\frac{\pi^2}{2\tilde{\kappa}} \sum_{\mathbf{n}\neq\mathbf{0}}^{\infty} e^{-2\pi \tilde{\kappa} |\mathbf{n}|} \leq
\frac{\pi^2}{2\tilde{\kappa}} e^{-\pi \tilde{\kappa}} \sum_{\mathbf{n}\neq\mathbf{0}}^{\infty} e^{-\pi |\mathbf{n}|}
= \Sampi_2 \frac{\pi^2}{2\tilde{\kappa}} e^{-\pi \tilde{\kappa}}  ~,\label{E-A-6.0-25}
\end{equation}
where $\Sampi_2$ is a small constant equal to $0.485647(1)$. It follows that the contribution of the
nonzero modes associated to the single sum on the r.h.s. of eq.~\eqref{E-A-6.0-16} is exponentially
suppressed and can be neglected in the large binding momentum limit. Collecting all the results in
eqs.~\eqref{E-A-6.0-14}, \eqref{E-A-6.0-15}, \eqref{E-A-6.0-19} and \eqref{E-A-6.0-21}, the large
binding momentum limit of the double sum $\mathcal{S}_3(\mathrm{i}\tilde{\kappa})$ is found,
\begin{equation}
\mathcal{S}_3(\mathrm{i}\tilde{\kappa}) = \sum_{\mathbf{n}}^{\Lambda_n}\sum_{\mathbf{m}\neq\mathbf{n}}^{+\infty}
\frac{1}{|\mathbf{n}|^2 + \tilde{\kappa}^2}\frac{1}{|\mathbf{m}|^2 + \tilde{\kappa}^2}\frac{\mathbf{n}
\cdot\mathbf{m} + \tilde{\kappa}^2}{|\mathbf{n}-\mathbf{m}|^2}  - \pi^4\Lambda_n^2 \rightarrow
\frac{\pi^2}{2\tilde{\kappa}} - 2\pi^2 \mathcal{I}^{(0)}\tilde{\kappa}  - 2\pi^4\tilde{\kappa}^2~,
\label{E-A-6.0-26}
\end{equation}
where the ellipses include the cutoff-dependent divergent terms and functions of
$\tilde{\kappa}$ which are suppressed by negative exponentials. 

\end{appendices}

%\addto\captionsbritish{\renewcommand{\appendixname}{}}

\section*{\textsf{References}}

%\begingroup
%\renewcommand{\section}[2]{}

\end{document}